\documentclass[11pt]{article}
\usepackage{amssymb}
\usepackage{amsmath}
\usepackage{amsopn}
\usepackage{pifont}
\usepackage{color}
\usepackage{ulem}
\usepackage{mathrsfs}

\definecolor{andrey}{rgb}{0.7,0,0.3}

\definecolor{arthur}{rgb}{0.2,0.6,0.2}

\numberwithin{equation}{section}

\newtheorem{proposition}{Proposition}[section] 

\def\coun{\varepsilon}
\def\bn{\begin{equation}}
\def\ed{\end{equation}}

\newcommand{\so}{\scriptscriptstyle \rm I}
\newcommand{\st}{\scriptscriptstyle \rm I\hspace{-1pt}I}

\newcommand{\bu}{\bar u}
\newcommand{\bv}{\bar v}

\def\<{\langle}
\def\>{\rangle}
\newcommand{\CC}{{\mathbb C}}

\newcommand{\XX}{{\mathbb X}}
\newcommand{\ZZ}{{\mathbb Z}}
\newcommand{\YY}{{\mathbb Y}}
\newcommand{\DD}{{\mathbb D}}

\def\r#1{(\ref{#1})}

\def\ot{\otimes}

\def\sk#1{\left(#1\right)}

\def\Pep{{P}^+_e}
\def\Pem{{P}^-_e}

\def\Pfpm{{P}_f^\pm}
\def\Pfp{{P}^+_f}
\def\Pfm{{P}^-_f}
\def\hPfp{{\hat P}^+_f}
\def\hPfm{{\hat P}^-_f}
\def\hPep{{\hat P}^{+}_e}
\def\hPem{{\hat P}^{-}_e}
\def\hPfpm{{\hat P}^\pm_f}
\def\hPepm{{\hat P}^{\pm}_e}

\def\Uqsl2{U_q(\widehat{\mathfrak{sl}}_2)}

\def\EE{{\rm E}}\def\FF{{\rm F}}

\def\RR{{\rm R}}
\def\LL{{\rm T}}

\def\E{{\rm E}}
\def\F{{\mathcal{F}}}

\def\tSym{\overline{\rm Sym}}
\def\Sym{{\rm Sym}}

\def\rr{r}

\def\tFF{\hat{\rm F}}
\def\tk{\hat{k}}
\def\tEE{\hat{\rm E}}
\def\tiF{\hat{F}}
\def\tiE{\hat{E}}

\def\ticF{\hat{\mathcal{F}}}

\def\hU{\hat{U}}

\def\hF{\hat{\mathcal{F}}}

\def\qed{\hfill$\square$\medskip}
\def\prt#1{[#1]}
\def\DYglmn{DY(\mathfrak{gl}(m|n))}
\def\II{\mathbb{I}}
\def\cci#1{c_{\prt{#1}}}
\def\eF{\mathcal{F}}
\def\eE{\mathcal{E}}
\def\oU{\overline{U}}
\def\cA{\mathcal{A}}
\def\cB{\mathcal{B}}

\def\cB{\mathcal{B}}
\def\hcB{\hat{\mathcal{B}}}

\def\rvec{|0\rangle}
\def\lvec{\langle0|}
\def\BBB{\mathbb{B}}
\def\CCC{\mathbb{C}}
\def\Sc{\mathcal{S}}

\def\ep{\epsilon}
\def\Ra{\Rightarrow}

\def\el{\ell}
\def\fgf{\mathbb{G}}
\def\comb{\mathbb{C}}
\def\mode#1{(#1)}
\def\Fp{{\sf F}}

\def\Bp{{\sf B}}
\def\Cp{{\sf C}}
\def\hFp{\hat{\sf F}}

\def\gF{\mathscr{F}}

\def\cpr{\mu}
\def\Tpf#1#2#3{\Delta_{#1_{\prt{#2}}}(#3)}
\def\Tpfp#1#2#3{\Delta'_{#1_{\prt{#2}}}(#3)}
\def\Cau{\mathcal{C}}
\def\Cres{C}
\def\Fli#1{\gamma_{#1}}
\def\hFli#1{\hat\gamma_{#1}}
\def\sieq#1#2{\sim_{#1,#2}}

\def\proof{{\sl Proof.}}
\def\qed{$\scriptstyle{\blacksquare}$}
\def\mode#1{^{(#1)}}
\def\trf{{\mathfrak t}}

\def\vph{\varphi}
\newcommand{\wt}[1]{\widetilde{#1}}

\begin{document}

\begin{flushright}
LAPTH-068/16
\end{flushright}

\vspace{12pt}

\begin{center}
\begin{LARGE}
{\bf Current presentation for the double\\[2mm] super-Yangian $DY(\mathfrak{gl}(m|n))$ and Bethe vectors}
\end{LARGE}

\vspace{40pt}

\begin{large}
{A.~A.~Hutsalyuk${}^{a}$,  A.~N.~Liashyk${}^{b,c,d}$,
S.~Z.~Pakuliak${}^{a,e,f}$,\\ E.~Ragoucy${}^g$, N.~A.~Slavnov${}^h$\  \footnote{
hutsalyuk@gmail.com, a.liashyk@gmail.com, stanislav.pakuliak@jinr.ru, eric.ragoucy@lapth.cnrs.fr, nslavnov@mi.ras.ru}}
\end{large}

 \vspace{12mm}

${}^a$ {\it Moscow Institute of Physics and Technology,  Dolgoprudny, Moscow reg., Russia}

\vspace{4mm}

${}^b$ {\it Bogoliubov Institute for Theoretical Physics, NAS of Ukraine,  Kiev, Ukraine}

\vspace{4mm}

${}^c$ {\it National Research University Higher School of Economics, Faculty of Mathematics, Moscow, Russia}

\vspace{4mm}

${}^d$ {\it Skolkovo Institute of Science and Technology, Moscow, Russia}

\vspace{4mm}

${}^e$ {\it Laboratory of Theoretical Physics, JINR,  Dubna, Moscow reg., Russia}

\vspace{4mm}

${}^f$ {\it Institute for Theoretical and Experimental Physics, Moscow, Russia}

\vspace{4mm}

${}^g$ {\it Laboratoire de Physique Th\'eorique LAPTh, CNRS and USMB,\\
BP 110, 74941 Annecy-le-Vieux Cedex, France}

\vspace{4mm}

${}^h$ {\it Steklov Mathematical Institute of RAS, Moscow, Russia}

\end{center}

\vspace{4mm}


\begin{abstract}
We find Bethe vectors for quantum integrable models associated with the
supersymmetric Yangians $Y(\mathfrak{gl}(m|n)$  in terms of the current generators
of the Yangian double  $DY(\mathfrak{gl}(m|n))$. We
use the method of projections onto intersections of  different type Borel subalgebras in this infinite
dimensional algebra to construct the Bethe vectors. Calculating these projection the supersymmetric
Bethe vectors can be expressed through matrix elements of the universal monodromy matrix
elements. Using two different but isomorphic current realizations of the Yangian double
$DY(\mathfrak{gl}(m|n))$ we obtain two different presentations for the Bethe vectors.
These Bethe vectors are also shown to obey some recursion relations which prove their
equivalence.
\end{abstract}


\setcounter{footnote}{0}

\section{Introduction}

The problem of calculating  form factors and correlation functions in  quantum integrable models is
a cornerstone  in the field of exactly solvable models of  statistical physics and
low-dimensional quantum mechanics. A lot of results were obtained in this direction starting from the earlier
years of development of the Quantum Inverse Scattering Method  (QISM) \cite{TaFa79,SkTaFa79}.
For  models related to the different deformations of the affine algebra $\widehat{\mathfrak{gl}}(2)$
one of the most important results is a determinant presentation for a particular case of the scalar
product, in which one of vectors is an eigenstate of the transfer matrix
\cite{Sl89}. It allows one to address the problem of
calculating the correlation functions \cite{KoBoIz93}
of the local operators in solvable models  (see the review paper \cite{Sl07} and references therein).

One of the most important notions of QISM is {\it a Bethe vector}. In the $\widehat{\mathfrak{gl}}(2)$ based models
the Bethe vector is a monomial in the right-upper element of the monodromy matrix (creation operator) applied to a pseudo-vacuum vector. It depends
on a set of complex variables called Bethe parameters. The distinguishing feature of these vectors is that they become eigenvectors of
the transfer matrix provided the Bethe parameters satisfy a special system of equations (Bethe equations). In this case we call them
{\it on-shell Bethe vectors}. Otherwise, if the Bethe parameters are generic complex numbers, then the corresponding vectors are called
{\it off-shell Bethe vectors}, or simply Bethe vectors. In this paper we deal with the universal monodromy matrix. This means that it depends only on the underlying algebra generators. We  refer to the corresponding Bethe vectors as universal Bethe  vectors.

The main topic of this paper is to study Bethe vectors in the Yangian double
$DY(\mathfrak{gl}(m|n))$. Our first goal is to obtain explicit formulas for them.
The second goal is to derive formulas for the
action of the monodromy matrix entries onto the off-shell Bethe vectors. Achieving these two goals allows to put the problem
of calculating the scalar products of Bethe vectors, which in turn is necessary for the study of form factors and correlation
functions in integrable models with underlying ${\mathfrak{gl}}(m|n)$ supersymmetry.

For  models related to higher rank symmetries, QISM is based on the so-called
Nested Bethe Ansatz, which
was elaborated in the pioneering papers \cite{KulRes83,KulRes81,KulRes82}.  There a recursive procedure was developed  to
construct Bethe vectors corresponding to the  $\widehat{\mathfrak{gl}}(N)$ algebra from the known Bethe vectors of the
$\widehat{\mathfrak{gl}}(N-1)$ algebra. Formally, this method allows one to obtain explicit formulas for Bethe vectors
in terms of certain polynomials in the creation operators (upper-triangular entries of the monodromy matrix) acting onto
the pseudo-vacuum vector. However, the procedure is quite involved, therefore  no explicit representations in the mentioned early works were obtained,
with the exception of graphical representations found by N.~Reshetikhin in \cite{Res86} for models with $\widehat{\mathfrak{gl}}(3)$ algebra.
The use of this diagram technique yielded a formula for the scalar products of off-shell Bethe vectors  in terms of sums over partitions of the
sets of the Bethe parameters ({\it a sum formula}).

In papers \cite{VT,VTcom} the Bethe vectors for the integrable models associated with
deformed $\widehat{\mathfrak{gl}}(N)$ algebras were obtained as
traces of products of the monodromy matrices, $\RR$-matrices and certain projections.
These results were generalized to the supersymmetric algebras in \cite{BeRa08}.
This approach allows to calculate in some cases the norms of the nested Bethe
vectors, but not their scalar products.

An alternative approach to construct Bethe vectors was proposed in the paper \cite{KhP-Kyoto}.
This method explores the relation between two different realizations of the quantized Hopf algebra
$U_q(\widehat{\mathfrak{gl}}(N))$ associated with the
affine algebra $\widehat{\mathfrak{gl}}(N)$, one in terms of the universal monodromy
matrix $\LL(z)$ and $\RR\LL\LL$ commutation relations and second in terms of the
total currents, which are defined by the Gauss decomposition of the monodromy matrix
$\LL(z)$ \cite{DiFr93}. Further, it was shown in the paper \cite{OPS} that two different types of formulas
for the universal off-shell Bethe vectors (constructed from the monodromy matrix)
are related to the two different current
realizations of the quantum affine algebra  $U_q(\widehat{\mathfrak{gl}}(N))$ and their associated projections.

Moreover, the approach which uses the current generators  of the deformed current
algebras
allows to calculate the action of the monodromy matrix elements onto universal Bethe vectors.
These action formulas appeared to be very useful for calculating
form factors in the different quantum integrable models related to the rational and trigonometric deformations of the affine algebra  $\widehat{\mathfrak{gl}}(3)$
 \cite{BelPRS12c,BelPRS13a,BPRS-trigo}. Recently similar results were obtained for the models with $\widehat{\mathfrak{gl}}(1|2)$ and
$\widehat{\mathfrak{gl}}(2|1)$  superalgebras in \cite{HLPRS1,HLPRS2}. In these works the explicit formulas for the Bethe vectors
and the action formulas were essentially used \cite{HLPRS3,HLPRS4}.

In the present paper we use the approach of \cite{KhP-Kyoto}. In this framework the universal off-shell Bethe vector
is defined as a projection of a product of the total currents  applied to the peseudo-vacuum vector. We postpone the detailed definition
to section~\ref{projuwf}, because it requires the introduction of many new concepts and notation.
For the same reason, we postpone a description of our main results to section \ref{mainres}.
Here we would like to mention only that
we construct explicit formulas for the universal Bethe vectors
in terms of the current generators of the Yangian double $DY(\mathfrak{gl}(m|n))$ for
two different Gauss decompositions of the universal monodromy matrix and
two different current realizations of this algebra. These different Gauss decompositions
correspond to the embeddings of
$DY(\mathfrak{gl}(m-1|n))$ or $DY(\mathfrak{gl}(m|n-1))$ into $DY(\mathfrak{gl}(m|n))$. On the
level of the $\RR\LL\LL$ realization these embeddings are either in the upper-left corner
or in the lower-right corner of the universal monodromy matrix.
Using the first or the second type of these embeddings we obtain two different representations for the Bethe vectors,
which we  respectively  denote  by $\BBB(\bar t)$ and
$\hat\BBB(\bar t)$, where $\bar t$ is a set of Bethe parameters \r{Bpar}.
 We prove that these two representations are equivalent, i.e. $\BBB(\bar t)=\hat\BBB(\bar t)$.

\medskip

The paper is organized as follows. In section~\ref{Notations} we introduce the necessary
notation used for calculations in  graded vector spaces, as well as the $\RR\LL\LL$ and current
realizations of the algebra $DY(\mathfrak{gl}(m|n))$.
In  section~\ref{projuwf} we define  universal Bethe vectors using
the notion of projections onto intersections of different type Borel subalgebras.
As already mentioned, section~\ref{mainres}  contains the
main results obtained in this paper.
Section~\ref{gen-act}
collects calculations of the action of the monodromy matrix elements onto Bethe vectors in
the generic case of $DY(\mathfrak{gl}(m|n))$. It is proved there, using  these action formulas,
that the vectors we constructed become on-shell Bethe vectors if the supersymmetric Bethe
equations for the Bethe parameters are satisfied.
In the last section~\ref{BV-calc} we calculate projections of the product of the currents and
present explicit formulas for the off-shell Bethe vectors as sums over partitions
of the Bethe parameters.
Appendix~\ref{CCGC} introduces the notion
of the composed currents and the relation between them and the Gauss
coordinates of the universal monodromy matrix. Appendix~\ref{comPS} describes
important properties
of the projections. Appendix~\ref{appC} shows how the Izergin and
Cauchy determinants appear in the resolution of the hierarchical relations to obtain the explicit
formulas for the off-shell Bethe vectors.

\section{Universal monodromy matrix}
\label{Notations}

In this paper we adopt the following approach. We do not consider any specific supersymmetric
solvable model defined by a particular monodromy matrix $\LL(z)$ satisfying the standard $\RR\LL\LL$
relation. Instead, we  consider a $\LL$-operator
\r{L-oper} as the {\it universal} monodromy matrix whose matrix elements are generating
series of the full set of generators of the Yangian double $DY(\mathfrak{gl}(m|n))$ acting in a
generic representation space of this  algebra, which is a rational
deformation of the affine algebra $\widehat{\mathfrak{gl}}(m|n)$. These representations are
not specified, except the
existence of left and right pseudo-vacuum vectors
that ensures the applicability
of the algebraic Bethe ansatz. To construct Bethe vectors we will
use only one $\LL$-operator $\LL^+(z)$ picked up from the dual pair $\{\LL^+(z), \LL^-(z)\}$ which generates
the whole algebra  $DY(\mathfrak{gl}(m|n))$. The eigenvalues $\lambda_i(z)$ (see \r{rvec} and
\r{lvec}) of the diagonal matrix  elements onto the pseudo-vacuum vector  are free functional
parameters which possibly can be set equal to zero, if  necessary.

We first present a definition of  $\ZZ_2$-graded linear spaces and their multiplication rules, as well as the
 matrices acting in these spaces.

\subsection{$\ZZ_2$-graded linear space and notation}

Let $\CC^{m|n}$ be a $\ZZ_2$-graded space with a basis ${\rm e}_i$, $i=1,\ldots,m+n$.
We assume that the basis vectors $\{{\rm e}_1,{\rm e}_2,\ldots, {\rm e}_m\}$ are even
while $\{{\rm e}_{m+1},{\rm e}_{m+2},\ldots, {\rm e}_{m+n}\}$ are odd. We introduce the $\ZZ_2$-grading of the indices as
\begin{equation}\label{grade}
\prt{i}=0\quad\mbox{for}\quad i=1,2,\ldots,m\quad\mbox{and}\quad \prt{i}=1\quad\mbox{for}\quad i=m+1,m+2,\ldots,m+n.
\end{equation}

Let $\E_{ij}\in{{\textrm{End}}(\CC^{m|n})}$ be a matrix with the only nonzero entry
equal to $1$ at the intersection of the $i$-th row and $j$-th column.

The basis vectors ${\rm e}_i$ and the matrices $\E_{ij}$ have grade
\begin{equation*}
[{\rm e}_{i}]=\prt{i},\qquad [\E_{ij}]=\prt{i}+\prt{j} {\mod 2}.
\end{equation*}
The tensor product is also graded according to the rule
\begin{equation*}
(\E_{ij}\ot\E_{kl})\cdot (\E_{pq}\ot\E_{rs})=(-)^{(\prt{k}+\prt{l})(\prt{p}+\prt{q})} \E_{ij}\E_{pq}\ot \E_{kl}\E_{rs}\,.
\end{equation*}

Let ${\rm P}$ be a graded permutation operator acting in the tensor product $\CC^{m|n}\ot\CC^{m|n}$ as
\begin{equation*}
{\rm P}=\sum_{a,b}^{m+n}(-)^{\prt{b}}\ \E_{ab}\ot \E_{ba}.
\end{equation*}

Let
\begin{equation*}
g(u,v)=\frac{c}{u-v}
\end{equation*}
be a rational function of the spectral parameters $u$, $v$  and $c$ is a deformation parameter. By rescaling the spectral parameters
it is always possible to set $c=1$, but we will keep it  for later convenience.

We define  $\RR(u,v)\in{\textrm{End}}(\CC^{m|n}\ot\CC^{m|n})$ as a rational supersymmetric $\RR$-matrix associated with the vector
representation of $\mathfrak{gl}(m|n)$,
\begin{equation}\label{DYglmn}
\RR(u,v)\ =\  \mathbb{I}\ot\mathbb{I}+g(u,v){\rm P},
\end{equation}
where we introduced the unity matrix
in $\CC^{m|n}$
\begin{equation*}
\mathbb{I}=\sum_{i=1}^{m+n}\E_{ii}.
\end{equation*}

\subsection{Universal monodromy commutation relations}

The super-algebra $\DYglmn$  is a graded associative algebra with unit $\mathbf{1}$ and generated by the modes
$\LL_{i,j}\mode{\ell}$, $\ell\in\ZZ$, $1\leq i,j\leq N+1$ of the
$\LL$-operators
\begin{equation}\label{L-oper}
\LL^{\pm}(u)=
\II\ot\mathbf{1}+\sum_{\substack{\ell\geq 0\\ \ell<0}}
\sum_{i,j=1}^{N+1} \E_{ij}\otimes \LL_{i,j}\mode{\ell}u^{-\ell-1},
\end{equation}
where $\ell\geq 0$ (resp. $\ell<0$) refers to the '$+$' index (resp. '$-$' index) in $\LL^{\pm}(u)$
and $N=m+n-1$ is the number of simple root of the super-algebra $\mathfrak{gl}(m|n)$.
Monodromy matrix elements $\LL^\pm_{i,j}(u)$
subject to the relations
\begin{equation}\label{L-op-com}
\RR(u,v)\cdot (\LL^{\mu}(u)\ot \II)\cdot (\II\ot \LL^{\nu}(v))=
(\II\ot \LL^{\nu}(v))\cdot (\LL^{\mu}(u)\ot \II)\cdot \RR(u,v),
\end{equation}
where $\mu,\nu=\pm$.
For the monodromy matrix\footnote{We  use the
notation $\LL(u)$ to denote either $\LL^+(u)$ or $\LL^-(u)$ when both matrices
share the same property.}
$\LL(u)$ to be globally even, we fix the grading of the monodromy matrix
elements as
\begin{equation*}
[\LL_{i,j}(u)]=\prt{i}+\prt{j} {\mod 2}.
\end{equation*}
The tensor product between matrices and algebra generators is also graded, that is
\begin{equation*}
(\E_{ij}\ot \LL_{i,j}(u))\cdot (\E_{kl}\ot \LL_{k,l}({v}))=(-)^{(\prt{i}+\prt{j})(\prt{k}+\prt{l})}\E_{ij}\E_{kl}\ot \LL_{i,j}(u)\LL_{k,l}(v).
\end{equation*}

The subalgebras formed by the modes
$\LL\mode{\ell}_{i,j}$ for $\ell\geq 0$ and $\ell<0$ of the $\LL$-operators $\LL^\pm(u)$
are the standard  Borel subalgebras $U(\mathfrak{b}^\pm)\subset\DYglmn$.
These Borel subalgebras are Hopf subalgebras of $\DYglmn$.
Their coalgebraic structure  is given by the graded coproduct
\begin{equation}\label{coprL}
\Delta \sk{\LL^{\pm}_{i,j}(u)}=\sum_{k=1}^{n+m}\ (-)^{(\prt{i}+\prt{k})(\prt{k}+\prt{j})}\LL^{\pm}_{k,j}(u)\otimes
\LL^{\pm}_{i,k}(u)\,.
\end{equation}

Due to the commutation relations {\eqref{L-op-com}}, the universal transfer matrix $\trf(u)$ defined as
the supertrace of the universal monodromy matrix $\LL^+(u)$
\begin{equation}\label{strace}
\trf(u)=\mbox{str}\sk{\LL^+(u)} \equiv\sum_{i=1}^{n+m}(-)^{\prt{i}}\LL^+_{i,i}(u)
\end{equation}
commutes for arbitrary values of the spectral parameters
\begin{equation*}
[\trf(u),\trf(v)]=0.
\end{equation*}
Thus, it may be considered as a generating function of the commuting integrals
of motion in the corresponding supersymmetric quantum integrable model.

The whole commutation relations \r{L-op-com} can be rewritten in the following form
\begin{equation}\label{TM-1}
\begin{split}
[\LL_{i,j}^\mu(u),\LL_{k,l}^\nu(v)\}&\equiv \LL^\mu_{i,j}(u)\LL^\nu_{k,l}(v)
 -(-)^{(\prt{i}+\prt{j})(\prt{k}+\prt{l})}   \LL^\nu_{k,l}(v)  \LL^\mu_{i,j}(u)  \\
&=(-)^{\prt{i}(\prt{k}+\prt{l})+\prt{k}\prt{l}}g(u,v)\Big(\LL^\nu_{k,j}(v)
\LL^\mu_{i,l}(u)-\LL^\mu_{k,j}(u)\LL^\nu_{i,l}(v)\Big).
\end{split}
\end{equation}
Renaming in \r{TM-1} the indices and the spectral parameters as $i\leftrightarrow k$,
 $j\leftrightarrow l$, and  $u\leftrightarrow v$ we  obtain an equivalent relation
\begin{equation}\label{TM-2}
\begin{split}
[\LL^\mu_{i,j}(u),\LL^\nu_{k,l}(v)\}&=\LL^\mu_{i,j}(u)\LL^\nu_{k,l}(v)
 -(-)^{(\prt{i}+\prt{j})(\prt{k}+\prt{l})}   \LL^\nu_{k,l}(v)  \LL^\mu_{i,j}(u)   \\
&=(-)^{\prt{l}(\prt{i}+\prt{j})+\prt{i}\prt{j}}g(u,v)\Big(\LL^\mu_{i,l}(u)\LL^\nu_{k,j}(v)
-\LL^\nu_{i,l}(v)\LL^\mu_{k,j}(u)\Big),
\end{split}
\end{equation}
where $\mu,\nu=\pm$.

Note that according to the commutation relations \r{TM-1} and \r{TM-2} the odd matrix
elements of the monodromy matrix do not commute in contrast to the even ones
\begin{equation}\label{NCR-7}
\LL^\mu_{i,j}(u)\LL^\nu_{i,j}(v)=  \frac{h_{\prt{i}}(v,u)}{h_{\prt{j}}(v,u)}
\LL^\nu_{i,j}(v)\LL^\mu_{i,j}(u)\,.
\end{equation}
Here and below we  use the graded rational functions\footnote{We will keep usual notation
$f(u,v)=\frac{u-v+c}{u-v}$ and $h(u,v)=\frac{u-v+c}{c}$ and use them occasionally.}
\begin{equation*}
f_{\prt{i}}(u,v)=1+g_{\prt{i}}(u,v)=1+\frac{\cci{i}}{u-v}=\frac{u-v+\cci{i}}{u-v}\,,\qquad
 h_{\prt{i}}(u,v)=\frac{f_{\prt{i}}(u,v)}{g_{\prt{i}}(u,v)}.
\end{equation*}
and\footnote{Introducing this graded deformation parameters allows one to write many relations
systematically and this is a reason why we do not scale the deformation parameter $c$ to 1.}
\begin{equation*}
c_{\prt{i}}=(-)^{\prt{i}}c\,.
\end{equation*}

Further on we will use following combination of the Kronecker symbol
\begin{equation*}
\ep_{i,j}=1-\delta_{i,j}\,,
\end{equation*}
which is equal to zero for $i=j$ and to 1 otherwise.

\subsection{Morphism of $\DYglmn$, singular vectors and Gauss decompositions}

Since $\RR$-matrix \r{DYglmn} and universal monodromy matrix \r{L-oper} are
globally even, one can easily check that the mapping\footnote{We keep the superscripts $\pm$ to give the anti-morphism compatible with the inclusion of a central charge in the double Yangian.}
\begin{equation}\label{antimo}
\Psi:\ \LL^{\pm}_{ij}(u) \to (-)^{\prt{i}(\prt{j}+1)}\LL^{\mp}_{ji}(u)
\end{equation}
is an antimorphism of $\DYglmn$ which is a super or graded transposition compatible with
the notion of super-trace. This map satisfies a property
\begin{equation}\label{pro-anti}
\Psi(A\cdot B)=(-)^{\prt{A}\prt{B}}\Psi(B)\cdot\Psi(A)
\end{equation}
for arbitrary elements $A,B\in\DYglmn$ and
will be used to relate right and left states or equivalently, Bethe vectors and
the dual ones.

Let $\rvec$ and $\lvec$ be vectors which satisfy the following conditions:
\begin{equation}\label{rvec}
\LL^\pm_{i,j}(u)\rvec =0\,,\quad i>j\,,\quad \LL^\pm_{i,i}(u)\rvec=\lambda^\pm_{i}(u)\rvec\,,\quad {i}=1,\ldots,N+1
\end{equation}
and
\begin{equation}\label{lvec}
\lvec\LL^\pm_{i,j}(u) =0\,,\quad i<j\,,\quad \lvec\LL^\pm_{i,i}(u)=\lambda^\pm_{i}(u)\lvec\,,\quad {i}=1,\ldots,N+1\,,
\end{equation}
where in \r{rvec} the monodromy matrix elements are acting to the right, while in \r{lvec}
they are acting to the left. Such vectors, if they exist, are called {\it singular vectors}.
If the pseudo-vacuum vectors $\rvec$ and $\lvec$ belong to the finite-dimensional
representations of the Yangian double  $\DYglmn$, then functions $\lambda^\pm_{i}(u)$ are coinciding
rational functions of the spectral parameter \cite{CP94} expanded in the different domains:
the function $\lambda^+_{i}(u)$ is a series with respect to $u^{-1}$ and the same
function $\lambda^-_{i}(u)$ is a series with respect to $u$.
In what follows we will use the same notation $\lambda_{i}(u)$ for the functions
$\lambda^\pm_{i}(u)$.

For the $\LL$-operators fixed by the relations \r{L-op-com}  we have
two possibilities to introduce the Gauss coordinates. The first possibility is to introduce $\FF^{\pm}_{j,i}(u)$,
$\EE^{\pm}_{i,j}(u)$, $1\leq i<j\leq N+1$, and
$k^\pm_{\ell}(u)$, $\ell=1,\ldots,N+1$, such that
\begin{align}\label{GF2}
\LL^{\pm}_{i,j}(u)&=\FF^{\pm}_{j,i}(u)k^\pm_{i}(u)+\sum_{1\leq \ell< i}
\FF^{\pm}_{j,\ell}(u)k^\pm_{\ell}(u)\EE^{\pm}_{\ell,i}(u),\\
\label{GK2}
\LL^{\pm}_{i,i}(u)&=k^\pm_{i}(u) +\sum_{1\leq \ell< i} \FF^{\pm}_{i,\ell}(u)k^\pm_{\ell}(u)
\EE^{\pm}_{\ell,i}(u),\\
\label{GE2}
\LL^{\pm}_{j,i}(u)&=k^\pm_{i}(u)\EE^{\pm}_{i,j}(u)+\sum_{1\leq \ell<i}
\FF^{\pm}_{i,\ell}(u)k^\pm_{\ell}(u)\EE^{\pm}_{\ell,j}(u)\,.
\end{align}
In the second case we introduce $\tFF^{\pm}_{j,i}(u)$,
$\tEE^{\pm}_{i,j}(u)$, $1\leq i<j\leq N+1$, and
$\tk^\pm_{\ell}(u)$, $\ell=1,\ldots,N+1$, such that
\begin{align}\label{GF1}
\LL^{\pm}_{i,j}(u)&=\tFF^{\pm}_{j,i}(u)\tk^+_{j}(u)+\sum_{j<\ell\leq N+1} (-)^{(\prt{\ell}+\prt{i})(\prt{\ell}+\prt{j})}
\tFF^{\pm}_{\ell,i}(u)\tk^\pm_{\ell}(u)\tEE^{\pm}_{j,\ell}(u),\\
\label{GK1}
\LL^{\pm}_{j,j}(u)&=\tk^\pm_{j}(u) +\sum_{j<\ell\leq N+1}(-)^{(\prt{\ell}+\prt{j})} \tFF^{\pm}_{\ell,j}(u)\tk^\pm_{\ell}(u)
\tEE^{\pm}_{j,\ell}(u),\\
\label{GE1}
\LL^{\pm}_{j,i}(u)&=\tk^\pm_{j}(u)\tEE^{\pm}_{i,j}(u)+\sum_{j<\ell\leq N+1}(-)^{(\prt{\ell}+\prt{i})(\prt{\ell}+\prt{j})}
\tFF^{\pm}_{\ell,j}(u)\tk^\pm_{\ell}(u)\tEE^{\pm}_{i,\ell}(u).
\end{align}

One can verify that the antimorphism \r{antimo} and the Gauss decomposition \r{GF2}--\r{GE2}
imply the following formulas for the Gauss  coordinates
\begin{equation}\label{morGau}
\Psi\sk{\FF^\pm_{j,i}(u)}=(-)^{\prt{i}(\prt{j}+1)} \EE^\mp_{i,j}(u)\,,\quad
\Psi\sk{\EE^\pm_{i,j}(u)}=(-)^{\prt{j}(\prt{i}+1)} \FF^\mp_{j,i}(u)\,,\quad
\Psi\sk{k^\pm_{\ell}(u)}=k^\mp_{\ell}(u)\,.
\end{equation}
Analogously
\begin{equation*}
\Psi\sk{\tFF^\pm_{j,i}(u)}=(-)^{\prt{i}(\prt{j}+1)} \tEE^\mp_{i,j}(u)\,,\quad
\Psi\sk{\tEE^\pm_{i,j}(u)}=(-)^{\prt{j}(\prt{i}+1)} \tFF^\mp_{j,i}(u)\,,\quad
\Psi\sk{\tk^\pm_{\ell}(u)}=\tk^\mp_{\ell}(u)\,.
\end{equation*}

Gauss decomposition formulas  also yield
\begin{equation*}
\EE^\pm_{i,j}(u)\rvec =\tEE^\pm_{i,j}(u)\rvec=0\,,\quad i<j\,,\quad
k^\pm_{\ell}(u)\rvec=\tk^\pm_{\ell}(u)\rvec=\lambda^\pm_\ell(u)\rvec\,,
\end{equation*}
and
\begin{equation*}
\lvec\FF^\mp_{j,i}(u) = \lvec\tFF^\mp_{j,i}(u)=0\,,\quad i<j\,,\quad \lvec k^\mp_{{\ell}}(u)=\lvec\tk^\mp_{\ell}(u)=
\lambda^\mp_{\ell}(u)\lvec\,.
\end{equation*}

\subsection{Current realizations of $\DYglmn$}
\label{TCR}

Let
\begin{equation*}
F_i(u)=\FF^{+}_{i+1,i}(u)-\FF^{-}_{i+1,i}(u)\,,\quad
E_i(u)=\EE^{+}_{i,i+1}(u)-\EE^{-}_{i,i+1}(u)
\end{equation*}
be total currents \cite{Dr88}.
Note that according to the formulas \r{morGau} we have
\begin{equation}\label{mor-cur}
\begin{split}
\Psi\sk{F_i(u)}&=-(-)^{\prt{i}(\prt{i+1}+1)}E_i(u)=-E_i(u)\,,\\
\Psi\sk{E_i(u)}&=-(-)^{\prt{i+1}(\prt{i}+1)}F_i(u)=-(-)^{\delta_{i,m}}F_i(u)\,,\quad i=1,\ldots,N.
\end{split}
\end{equation}
This proves that  the graded transposition is idempotent of the order 4 and its square
counts the number of odd elements.

Using straightforward calculations \cite{DiFr93,Zh97} and the Gauss decomposition \r{GF2}--\r{GE2}
we can obtain the following nontrivial commutation relations in terms of the
 total currents
$F_i(t)$, $E_i(t)$ and the Cartan currents $k^\pm_i(t)$:
\begin{equation}\label{kiF}
\begin{split}
k^{\pm}_i(u)F_i(v)k^{\pm}_i(u)^{-1}&=
f_{\prt{i}}(v,u)
\ F_i(v),\\
k^{\pm}_{i+1}(u)F_i(v)k^{\pm}_{i+1}(u)^{-1}&=
f_{\prt{i+1}}(u,v)
\ F_i(v),
\end{split}
\end{equation}
\begin{equation}\label{kiE}
\begin{split}
k^{\pm}_i(u)^{-1}E_i(v)k^{\pm}_i(u)&=
f_{\prt{i}}(v,u)
\ E_i(v),\\
k^{\pm}_{i+1}(u)^{-1}E_i(v)k^{\pm}_{i+1}(u)&=
f_{\prt{i+1}}(u,v)
\ E_i(v),
\end{split}
\end{equation}
\begin{equation}\label{FiFi}
((u-v)\ep_{i,m}-\cci{i})\ F_i(u)F_i(v)= ((u-v)\ep_{i,m}+\cci{i})\  F_i(v)F_i(u),
\end{equation}
\begin{equation}\label{EiEi}
((u-v)\ep_{i,m}+\cci{i})\ E_i(u)E_i(v)= ((u-v)\ep_{i,m}-\cci{i})\  E_i(v)E_i(u), \\
\end{equation}
\begin{equation}\label{FiFii}
(u-v)\ F_i(u)F_{i+1}(v)= (u-v-\cci{i+1})\ F_{i+1}(v)F_i(u),
\end{equation}
\begin{equation}\label{EiEii}
(u-v-\cci{i+1})\ E_i(u)E_{i+1}(v)= (u-v)\  E_{i+1}(v)E_i(u),
\end{equation}
\begin{equation}\label{EF}
\begin{split}
[E_i(u),F_j(v)\}&=E_i(u)F_j(v)-(-)^{(\prt{i}+\prt{i+1})(\prt{j}+\prt{j+1})}F_j(v)E_i(u)\\
&=\delta_{i,j}\cci{i+1}\delta(u,v)\Big(k^-_{i+1}(u)\cdot k^-_{i}(u)^{-1}-k^+_{i+1}(v)\cdot k^+_{i}(v)^{-1}\Big)
\end{split}
\end{equation}
where  $\delta(u,v)$ is the rational $\delta$-function
given by the equality \r{del1}.
These calculations also lead to the Serre relations. For the simple root currents
$F_{i}(u)$, $i=1,\ldots,N$ they read:
\begin{equation}\label{serF1}
\begin{split}
{\rm Sym}_{u_1,u_2}\Big(((u_2-u_1)\delta_{i,m}&-\cci{i+1})
(F_i(u_1)F_i(u_2)F_{i+1}(v)\\
&{}-2F_i(u_1)F_{i+1}(v)F_i(u_2)+F_{i+1}(v)F_i(u_1)F_i(u_2))\Big)=0\,,
\end{split}
\end{equation}
\begin{equation}\label{serF2}
\begin{split}
{\rm Sym}_{u_1,u_2}\Big(((u_1-u_2)\delta_{i,m}&+\cci{i})
(F_i(u_1)F_i(u_2)F_{i-1}(v)\\
&{}-2F_i(u_1)F_{i-1}(v)F_i(u_2)+F_{i-1}(v)F_i(u_1)F_i(u_2))\Big)=0\,,
\end{split}
\end{equation}
\begin{equation}\label{serF3}
\begin{split}
{\rm Sym}_{u_1,u_2}\Big(
(u_1-u_2&+c)\big[F_m(u_1)F_m(u_2)F_{m-1}(v_1)F_{m+1}(v_2)\\
&-2  F_m(u_1)F_{m-1}(v_1)F_m(u_2)F_{m+1}(v_2)\big]\\
&+2c\ F_{m-1}(v_1)F_m(u_1)F_m(u_2)F_{m+1}(v_2)\\
+(u_2-u_1&+c)\big[F_{m-1}(v_1)F_{m+1}(v_2)F_m(u_1)F_m(u_2)\\
&-2  F_{m-1}(v_1)F_m(u_1)F_{m+1}(v_2)F_m(u_2)\big]
\Big)=0\,.
\end{split}
\end{equation}
Analogous formulas for the currents $E_i(u)$, $i=1,\ldots,N$, can be obtained by
applying  the antimorphism $\Psi$ to these relations. It amounts to do the
replacements $F_i(u)\to E_i(u)$ and $c\to -c$ in the formulas \r{serF1}--\r{serF3}.

The 'rational' or equivalently, 'additive' $\delta$-function used in the relation \r{EF}
can be presented as difference of two series
\begin{equation}\label{del1}
\delta(u,v)=\delta(v,u)=\frac{1}{(u-v)_>}-\frac{1}{(u-v)_<}=\sum_{n\in\ZZ}\frac{v^n}{u^{n+1}}\,,
\end{equation}
where
\begin{equation}\label{expan}
\frac{1}{(u-v)_>}= \frac{1}{u}\sum_{k\geq0}\sk{\frac{v}{u}}^k,\qquad
\frac{1}{(u-v)_<}= -\frac{1}{v}\sum_{k\geq0}\sk{\frac{u}{v}}^k\,.
\end{equation}
Here the symbol '$>$' in the rational function $\frac{1}{(u-v)_>}$ means that $|u|>|v|$ and
this rational function should be presented as the first series in \r{expan}. In its turn,
the symbol '$<$' in the rational function $\frac{1}{(u-v)_<}$ means that $|u|<|v|$ and
this rational function should be presented as the second series in \r{expan}.
Below we will also use notation  $\frac{1}{(u-v)_\lessgtr}$ to stress that one can use either one or another
decomposition into series \r{expan} of the rational function $\frac{1}{u-v}$.

It is known  \cite{OPS}
that another current realization of the Yangian double $\DYglmn$ can be obtained using
a different Gauss decomposition of the monodromy matrix, as in \r{GF1}--\r{GE1}. The commutation relations
between Cartan currents $\tk^\pm_i({u})$ and the simple root total currents $\tiF_i({u})$, $\tiE_i({u})$
\begin{equation}\label{DF-iso1}
\tiF_i({u})=\tFF^{+}_{i+1,i}({u})-\tFF^{-}_{i+1,i}({u})\,,\quad
\tiE_i({u})=\tEE^{+}_{i,i+1}({u})-\tEE^{-}_{i,i+1}({u})\,,
\end{equation}
are gathered below
\begin{equation}\label{tkiF}
\begin{split}
\tk^{\pm}_i(u)\tiF_i(v)\tk^{\pm}_i(u)^{-1}&=
f_{\prt{i}}(v,u)
\ \tiF_i(v),\\
\tk^{\pm}_{i+1}(u)\tiF_i(v)\tk^{\pm}_{i+1}(u)^{-1}&=
f_{\prt{i+1}}(u,v) {\tiF_i(v)},
\end{split}
\end{equation}
\begin{equation}\label{tkiE}
\begin{split}
\tk^{\pm}_i(u)^{-1}\tiE_i(v)\tk^{\pm}_i(u)&=
f_{\prt{i}}(v,u)
\ \tiE_i(v),\\
\tk^{\pm}_{i+1}(u)^{-1}\tiE_i(v)\tk^{\pm}_{i+1}(u)&=
f_{\prt{i+1}}(u,v)
\ \tiE_i(v),
\end{split}
\end{equation}
\begin{equation}\label{tFiFi}
((u-v)\ep_{i,m}+\cci{i})\ \tiF_i(u)\tiF_i(v)= ((u-v)\ep_{i,m}-\cci{i})\  \tiF_i(v)\tiF_i(u),
\end{equation}
\begin{equation}\label{tEiEi}
((u-v)\ep_{i,m}-\cci{i})\ \tiE_i(u)\tiE_i(v)= ((u-v)\ep_{i,m}+\cci{i})\  \tiE_i(v)\tiE_i(u),
\end{equation}
\begin{equation}\label{tFiFii}
(u-v-\cci{i+1})\ \tiF_i(u)\tiF_{i+1}(v)= (u-v)\ \tiF_{i+1}(v)\tiF_i(u),
\end{equation}
\begin{equation}\label{tEiEii}
(u-v)\ \tiE_i(u)\tiE_{i+1}(v)= (u-v-\cci{i+1})\  \tiE_{i+1}(v)\tiE_i(u),
\end{equation}
\begin{equation}\label{tEF}
\begin{split}
[\tiE_i(u),\tiF_j(v)\}=&\tiE_i(u)\tiF_j(v)-(-)^{(\prt{i}+\prt{i+1})(\prt{j}+\prt{j+1})}\tiF_j(v)\tiE_i(u)\\
=&\delta_{i,j}\cci{i+1}\delta(u,v)\Big(\tk^+_{i}(u)\cdot\tk^+_{i+1}(u)^{-1}-\tk^-_{i}(v)\cdot\tk^-_{i+1}(v)^{-1}\Big)\,.
\end{split}
\end{equation}
The Serre relations for the simple root currents
$\tiE_{i}(u)$, $i=1,\ldots,N$ now read
\begin{equation}\label{stE1}
\begin{split}
{\rm Sym}_{u_1,u_2}\Big(((u_2-u_1)\delta_{i,m}&-\cci{i+1})
(\tiE_i(u_1)\tiE_i(u_2)\tiE_{i+1}(v)\\
&-2\tiE_i(u_1)\tiE_{i+1}(v)\tiE_i(u_2)+\tiE_{i+1}(v)\tiE_i(u_1)\tiE_i(u_2))\Big)=0\,,
\end{split}
\end{equation}
\begin{equation}\label{stE2}
\begin{split}
{\rm Sym}_{u_1,u_2}\Big(((u_1-u_2)\delta_{i,m}&+\cci{i})
(\tiE_i(u_1)\tiE_i(u_2)\tiE_{i-1}(v)\\
&-2\tiE_i(u_1)\tiE_{i-1}(v)\tiE_i(u_2)+\tiE_{i-1}(v)\tiE_i(u_1)\tiE_i(u_2))\Big)=0\,,
\end{split}
\end{equation}
\begin{equation}\label{stE3}
\begin{split}
{\rm Sym}_{u_1,u_2}\Big(
(u_1-u_2&+c)\big[\tiE_m(u_1)\tiE_m(u_2)\tiE_{m-1}(v_1)\tiE_{m+1}(v_2)\\
&-2  \tiE_m(u_1)\tiE_{m-1}(v_1)\tiE_m(u_2)\tiE_{m+1}(v_2)\big]\\
&+2c\ \tiE_{m-1}(v_1)\tiE_m(u_1)\tiE_m(u_2)\tiE_{m+1}(v_2)\\
+(u_2-u_1&+c)\big[\tiE_{m-1}(v_1)\tiE_{m+1}(v_2)\tiE_m(u_1)\tiE_m(u_2)\\
&-2  \tiE_{m-1}(v_1)\tiE_m(u_1)\tiE_{m+1}(v_2)\tiE_m(u_2)\big]
\Big)=0\,.
\end{split}
\end{equation}
Due to the antimorphism $\Psi$, analogous relations occur for the currents $\tiF_i(u)$, $i=1,\ldots,N$ with the
replacements $\tiE_i(u)\to \tiF_i(u)$ and $c\to -c$ in the formulas \r{stE1}--\r{stE3}.
The action of the antimorphism \r{antimo} onto currents $\tiF_i(u)$, $\tiE_i(u)$ and $\tk_\ell(u)$
is given by the same formulas as in \r{mor-cur}.

Note that in  the  commutation relations \r{FiFi}, \r{EiEi},
\r{tFiFi} and \r{tEiEi}, one can replace $\cci{i}$ by $\cci{i+1}$. Indeed, when $i\neq m$, $\cci{i}=\cci{i+1}$, while
for $i=m$, the factor $(u-v)\ep_{i,m}$ vanishes and it does not matter to use either  $\cci{i}$ or $\cci{i+1}$.

\section{Universal Bethe vectors}
\label{projuwf}

It follows from the commutation relations \r{L-op-com} that the subalgebras $U^\pm$ generated
by the modes of the $\LL$-operators $\LL_{ij}\mode{n}$ form two Borel subalgebras
in $DY(\mathfrak{gl}(m|n))$. Moreover due to \r{coprL} they are Hopf subalgebras.
 We call the subalgebras $U^\pm$ the {\it standard Borel subalgebras} in
 the  Yangian double $DY(\mathfrak{gl}(m|n))$.

As we already mentioned, the universal Bethe vectors are constructed from the matrix elements of one
universal monodromy matrix $\LL^+_{ij}$ and belong to the standard 'positive' Borel subalgebra
$U^+$.  The goal of this section is to express the universal Bethe vectors through the current
generators of the Yangian double $DY(\mathfrak{gl}(m|n))$  using the approach
developed in the papers \cite{KhP-Kyoto,OPS,EKhP}.

In this paper we  consider formulas for the Bethe vectors compatible with two different
possibilities to embed an algebra of smaller rank  into the one of bigger rank. Namely, from the explicit
formulas for the right Bethe vectors $\BBB(\bar t)$ (see eq. \r{hr77}), one may deduce
that the Bethe vector $\BBB(\bar t)$
is obtained by resolving the hierarchical relations based on the embedding
of the Yangian double $DY(\mathfrak{gl}(m-1|n))$ into the bigger algebra
$DY(\mathfrak{gl}(m|n))$. Similarly, it follows from the eq. \r{hr77h}, the Bethe vector $\hat\BBB(\bar t)$ is obtained
by resolving the hierarchical relations based on the embedding
of the Yangian double $DY(\mathfrak{gl}(m|n-1))$ into
$DY(\mathfrak{gl}(m|n))$. To express Bethe vectors $\BBB(\bar t)$ and  $\hat\BBB(\bar t)$
in terms of the current generators we will use two different types of  Gauss
decompositions of the monodromy matrix elements and the corresponding current generators
\cite{OPS}.

The general theory of the relation between Bethe vectors and currents was developed
in the paper \cite{EKhP} and then applied in the papers \cite{KhP-Kyoto,OPS}
to the construction of the hierarchical Bethe vectors for
 quantum integrable models based on the quantum affine algebra
$U_q(\widehat{\mathfrak{gl}}(N))$. The main tool used in these papers was the language of projections
onto intersections of  different type Borel subalgebras.

To describe Bethe vectors
$\BBB(\bar t)$   and $\CCC(\bar t)$ we will  use the current Borel subalgebras associated with
the Gauss decomposition \r{GF2}--\r{GE2} and the antimorphism \r{antimo}.
For the Bethe vectors  $\hat\BBB(\bar t)$ and   $\hat\CCC(\bar t)$ we
will use the same antimorphism and the current Borel subalgebras associated with the second
Gauss decomposition \r{GF1}--\r{GE1}.

\subsection{Notation and conventions}
As in \r{Bpar} we  denote sets of the parameters by bar: $\bu$, $\bv$ and so on.
To simplify further formulas we use a shorthand notation for the products of functions depending on one
or two variables. Namely, whenever a function $\lambda_j$ depends on a set of variables $\lambda_j(\bu)$, it
stands for the product of $\lambda_j(u_\el)$ over the set $\bu$. Similarly, the notation
$f_{\prt{i}}(\bu,\bv)$ or
$g_{\prt{i}}(\bu,\bv)$ or $h_{\prt{i}}(\bu,\bv)$  means the double  product of these functions over
corresponding sets. For example,
\begin{equation*}
\lambda_j(\bu)=\prod_{u_\el\in\bu}  \lambda_j(u_\el), \qquad f_{\prt{i}}(\bu,\bv)=\prod_{u_\el\in\bu\,,\ v_{\el'}\in\bv}
f_{\prt{i}}(u_\el,v_{\el'}).
\end{equation*}
Moreover, we use the same convention when considering products of
commuting operators. For example,
\begin{equation*}
\LL_{i,j}(\bu)=\prod_\el \LL_{i,j}(u_\el), \qquad\text{for}\qquad \prt{i}+\prt{j}=0\mod 2.
\end{equation*}

We also introduce several  rational functions
which will appear in the text below.  First,
for any function $x(u_1,u_2)$, we set
\begin{equation*}
\Delta_x(\bu)=\prod_{1\leq\el<\el'\leq a} x(u_{\el'},u_\el)\quad \mbox{and}\quad
\Delta'_x(\bu)=\prod_{1\leq\el<\el'\leq a} x(u_{\el},u_{\el'}),
\end{equation*}
where $a=\#\bu$.

Second,  for arbitrary sets of the parameters
$\bu$ and $\bv$ we define
\begin{equation}\label{dc9}
\Fli{i}(\bu)=\frac{\Delta_{f_{\prt{i}}}(\bu)}{\Delta_{h}(\bu)^{\delta_{i,m}}}\quad\mbox{and}\quad
\Fli{i}(\bu,\bv)=\frac{f_{\prt{i}}(\bu,\bv)}{h(\bu,\bv)^{\delta_{i,m}}}\,.
\end{equation}
The first function coincides with the function $\Delta_{f_{\prt{i}}}(\bu)$ for $i\not=m$ and with
$\Delta_{g}(\bu)$ for $i=m$.
The second function coincides with the function $f_{\prt{i}}(\bu,\bv)$ for $i\not=m$ and with
$g(\bu,\bv)$ for $i=m$.
Analogously, we define
\begin{equation*}
\hFli{i}(\bu)=\frac{\Delta_{f_{\prt{i+1}}}(\bu)}{\Delta'_{h}(\bu)^{\delta_{i,m}}}\quad\mbox{and}\quad
\hFli{i}(\bu,\bv)=\frac{f_{\prt{i+1}}(\bu,\bv)}{h(\bv,\bu)^{\delta_{i,m}}}\,.
\end{equation*}
For $i\not=m$, $\hFli{i}(\bu)=\Delta_{f_{\prt{i+1}}}(\bu)$ and $\hFli{i}(\bu,\bv)=
f_{\prt{i+1}}(\bu,\bv)$, while for $i=m$, $\hFli{m}(\bu)=\Delta'_{g}(\bu)$ and
$\hFli{m}(\bu,\bv)= g(\bv,\bu)$. Note that function $\Fli{m}(\bu)$
differs from the function  $\hFli{m}(\bu)$ by the factor $(-)^{\#\bu(\#\bu-1)/2}$. Similarly,
\begin{equation}\label{equ}
\Fli{m}(\bu,\bv)=(-)^{\#\bu\cdot\#\bv} \hFli{m}(\bu,\bv)\,.
\end{equation}
Also note that $\Fli{i}(\bu) = \hFli{i}(\bu)$ and $\Fli{i}(\bu,\bv) = \hFli{i}(\bu,\bv)$ for $i\not=m$.

\subsection{Deformed symmetrization}

For any formal series $G(\bar t)$ depending on the set
of the variables $\bar t$ \r{Bpar}
we define {\it a deformed symmetrization} (or $c$-symmetrization)
as a summation\footnote{Recall that $N=m+n-1$ denotes the number of  simple roots of the
superalgebra $\mathfrak{gl}(m|n)$.}
\begin{equation}\label{sym}
\tSym_{\,\bar t}\ G(\bar t)=
\sum_{\sigma\in S_{\bar\rr}}
\prod_{s=1}^{N}\!\!
\prod_{\substack{\ell<\ell'\\ \sigma^s(\ell)>\sigma^s(\ell')}}
\frac{(t^s_{\sigma^s(\ell')}-t^s_{\sigma^s(\ell)})\ep_{s,m}+\cci{s}}
{(t^s_{\sigma^s(\ell')}-t^s_{\sigma^s(\ell)})\ep_{s,m}-\cci{s}} \
G(^\sigma\bar t)\,,
\end{equation}
where
 $S_{\bar\rr}=S_{\rr_1}\times\cdots\times S_{\rr_N}$ is the direct product of the groups
$S_{\rr_s}$ that permute the integers $1,\ldots,\rr_s$, $s=1,\ldots,N$
and $^\sigma\bar t$ is the corresponding permuted set of Bethe parameters \r{Bpar}.
Due to the argument given in the end of  section~\ref{TCR}
the formula of deformed symmetrization can be equally written as
\begin{equation}\label{sym1}
\tSym_{\,\bar t}\ G(\bar t)=
\sum_{\sigma\in S_{\bar\rr}}
\prod_{s=1}^{N}\!\!
\prod_{\substack{\ell<\ell'\\ \sigma^s(\ell)>\sigma^s(\ell')}}
\frac{(t^s_{\sigma^s(\ell')}-t^s_{\sigma^s(\ell)})\ep_{s,m}+\cci{s+1}}
{(t^s_{\sigma^s(\ell')}-t^s_{\sigma^s(\ell)})\ep_{s,m}-\cci{s+1}} \
G(^\sigma\bar t).
\end{equation}
In what follows we will use either formula \r{sym} or formula \r{sym1} depending on the situation.

We say that a series $Q(\bar t)$ is {\it $c$-symmetric} if
\begin{equation*}
\tSym_{\,\bar t}\ Q(\bar t)= \Big(\prod_{s=1}^N\rr_s!\Big)\
Q(\bar t)\,.
\end{equation*}

Note that for $s=m$, the product over $\ell$ and $\ell'$ is equal to $(-)^{P(\sigma^m)}$, where $P(\sigma^m)$
is a parity of the permutation $\sigma^m$ and the summation over all permutations $\sigma^m$ is nothing else
but anti-symmetrization over set $\bar t^m$.

\subsection{Bethe vector $\BBB(\bar t)$ and the dual Bethe vector $\CCC(\bar t)$}
\label{Bvsec}

We first  explain the relation between the Bethe vector $\BBB(\bar t)$
and the current presentation \r{kiF}--\r{EF}.

Let $U_F\subset DY(\mathfrak{gl}(m|n))$ be the $DY(\mathfrak{gl}(m|n))$ subalgebra
generated by the modes of the simple roots currents
$F_i\mode{\ell}$; $i=1,\ldots,N$; $\ell\in\ZZ$ and by the modes of the `positive' Cartan currents
$k_j\mode{\ell'}$;
$j=1,\ldots,N+1$; $\ell'\geq0$. In the framework of the quantum double construction,
the subalgebra $U_E\subset DY(\mathfrak{gl}(m|n))$, dual to $U_F$, is  generated
by the modes of the simple roots currents
$E_i\mode{\ell}$; $i=1,\ldots,N$; $\ell\in\ZZ$ and by the modes of the `negative' Cartan currents
$k_j\mode{\ell'}$;
$j=1,\ldots,N+1$; $\ell'<0$.

We  call the subalgebras $U_F$ and $U_E$ the {\it current  Borel subalgebras}.  These
subalgebras are Hopf subalgebras in $DY(\mathfrak{gl}(m|n))$ with respect to the so-called
Drinfeld coproduct
\begin{equation}\label{Dcopr}
\begin{split}
\Delta^{(D)}(F_i(z))&=F_i(z)\ot\mathbf{1}+k^+_{i+1}(z)\sk{k^+_{i}(z)}^{-1}\ot F_i(z),\\
\Delta^{(D)}(k^\pm_j(z))&=k^\pm_j(z)\ot k^\pm_{j}(z),\\
\Delta^{(D)}(E_i(z))&=\mathbf{1}\ot E_i(z)+ E_i(z)\ot k^-_{i+1}(z)\sk{k^-_{i}(z)}^{-1},
\end{split}
\end{equation}
which obviously differs from the coproduct given by the relation \r{coprL}.

In order to express Bethe vectors $\BBB(\bar t)$ in terms of the current generators
we need only one current Borel subalgebra $U_F$ and its coalgebraic property given by
the first two formulas in \r{Dcopr}.  Let us consider the following intersections of this current Borel
subalgebra with the standard Borel subalgebras $U^\pm$:
\begin{equation}\label{inter}
U_F^-=U_F\cap U^-\,,\qquad U_F^+=U_F\cap U^+\,.
\end{equation}
Each of these intersections is a subalgebra in $DY(\mathfrak{gl}(m|n))$ \cite{EKhP} and they are coideals
with respect to the coproduct \r{Dcopr}
\begin{equation}\label{coid}
\Delta^{(D)}(U_F^+)=U_F^+\ot U_F\,,\qquad \Delta^{(D)}(U_F^-)=U_F\ot U_F^-\,.
\end{equation}
To see this we introduce the expansion of the combination of  Cartan currents
\begin{equation*}
k^+_{i+1}(z)\sk{k^+_{i}(z)}^{-1}=\mathbf{1}+\sum_{\ell\geq 0}
\kappa_i\mode{\ell}z^{-\ell-1}\,.
\end{equation*}
Then,  the coproduct \r{Dcopr} maps the modes $F_i\mode{\ell}$ of the currents  $F_i(z)$  to
\begin{equation}\label{Dcmod}
\Delta^{(D)}\sk{F_i\mode{\ell}}=F_i\mode{\ell}\ot\mathbf{1} + \mathbf{1} \ot F_i\mode{\ell} +\sum_{\ell'\geq 0}
\kappa_i\mode{\ell'}\ot F_i\mode{\ell-\ell'-1}\,.\end{equation}
Properties \r{coid} become obvious from the relations \r{Dcmod}.

According to the Cartan--Weyl construction of the
Yangian double we have to find a
 global ordering on the generators in this algebra. There are two different choices
for this ordering. We choose the ordering such that elements in the subalgebra
$U^-_F$ precede  elements from the subalgebra $U^+_F$ \cite{EKhP,FKPR}.
We say that an arbitrary element $\eF\in U_F$
is ordered if it is presented in the form
\begin{equation*}
\eF=\eF_-\cdot \eF_+\,,
\end{equation*}
where $\eF_\pm\in U_F^\pm$.

According to the general theory \cite{EKhP} one may define the projections of any ordered elements from the
subalgebras $U_F$  onto subalgebras \r{inter} using  the formulas
\begin{equation}\label{proj1}
\Pfp(\eF_-\cdot\eF_+)=\varepsilon(\eF_-)\,\eF_+\,,
\quad \Pfm(\eF_-\cdot \eF_+)=\eF_-\,\varepsilon(\eF_+)\,,\quad \eF_\pm\in U_F^\pm\,,
\end{equation}
where the counit mapping $\varepsilon: U_F\to \CC$ is defined by the rules
\begin{equation*}
\varepsilon(F_i\mode{\ell})=0 \quad\mbox{and}\quad \varepsilon(k_j\mode{\ell})=0\,.
\end{equation*}

Let $\oU_F$ be the closure of the algebra $U_F$ formed by infinite sums of monomials
that are ordered products $\cA_{i_1}\mode{\ell_1}\cdots \cA_{i_a}\mode{\ell_a}$ with
$\ell_1\leq\cdots\leq\ell_a$,
where $\cA_{i_l} \mode{\ell_l}$ is either $F_{i_l}\mode{\ell_l}$ or $k_{i_l}\mode{\ell_l}$.
It can be proved \cite{EKhP} that
\begin{enumerate}
\item[(1)] the action of the projections \r{proj1} extends to the algebra $\oU_F$;
\item[(2)] for any $\eF\in\oU_F$ with $\Delta^{(D)}(\eF)=\eF'\otimes\eF''$ we have
\begin{equation}\label{ordF}
\eF=\Pfm\sk{\eF'}\cdot \Pfp\sk{\eF''}\,.
\end{equation}
\end{enumerate}

Formula \r{ordF} is an important tool for calculating  the
universal Bethe vectors. It allows one to present  an
arbitrary product of currents in the ordered form using
simple formulas for the  Drinfeld current coproducts.

Now we are in position to give a definition of the universal Bethe vector. Let
\begin{equation}\label{Bpar}
\bar{t}\
=\{t^{1}_{1},\ldots,t^{1}_{\rr_{1}};
t^{2}_{1},\ldots,t^{2}_{\rr_{2}};\ldots;
t^{N}_{1},\ldots,t^{N}_{\rr_{N}} \}
\end{equation}
be a set of parameters. The superscript indicates the type of parameters corresponding
to simple roots and the subscript counts the number of parameters of a given type.  There are
$\rr_\el$ Bethe parameters of the type $\el=1,\ldots,N$.

Define an ordered product of total currents
\begin{equation}\label{FF-ff}
\eF(\bar t)=\!\!\prod_{1\leq a\leq N}
^{\longrightarrow} \sk{\prod_{1\leq\ell\leq\rr_a}^{\longrightarrow} F_{a}(t^a_{\ell})},
\end{equation}
where the
symbol $\mathop{\prod}\limits^{\longleftarrow}_a A_a$ (resp.
$\mathop{\prod}\limits^{\longrightarrow}_a A_a$)
stands for the  ordered product of
noncommutative operators $A_a$, such that $A_\el$ is on the right (resp. on the left)
 of $A_{\el'}$ for {$\el' \geq \el$}:
\begin{equation*}
\mathop{\prod}\limits^{\longleftarrow}_{j\geq a\geq i} A_a = A_j\,A_{j-1}\,
\cdots\, A_{i+1}\,A_i\quad\mbox{and}\quad
\mathop{\prod}\limits^{\longrightarrow}_{i\leq a\leq j} A_a = A_i\,A_{i+1}\,
\cdots\, A_{j-1}\,A_j\,.
\end{equation*}

The product of  currents $\eF(\bar t)$
is a formal series over the ratios $t^b_k/t^c_l$ with $b>c$ and $t^a_i/t^a_j$ with $i>j$
 taking values in the completions
$\oU_F$ (see \cite{EKhP}).
 The product \r{FF-ff} has poles
for some values of the ratios $t^b_k/t^c_l$ and $t^a_i/t^a_j$. The operator valued
coefficients at these poles take values in the completions $\oU_F$
 and can be identified with composed root currents (see Appendix~\ref{CCGC}).
Note also that due to the commutation relations between currents, the product  \r{FF-ff}
as well as its projection  are all $c$-symmetric.

Let us introduce the normalized product of  currents
\begin{equation}\label{FFnor}
\Fp(\bar t)=\frac{
\prod_ {\el=1}^{N} \Fli{\el}(\bar t^\el) }
{\prod_ {\el=1}^{N-1} f_{\prt{\el+1}}(\bar t^{\el+1},\bar t^\el)}\
\eF(\bar t),
\end{equation}
where $\Fli{\el}$ is given by \eqref{dc9}.
Then the
universal off-shell Bethe vector $\BBB(\bar t)$ is defined
as the action of the projections onto this normalized product and applied to the singular vector $\rvec$
\begin{equation}\label{uBV1}
\BBB(\bar{t})=\Pfp\left(\Fp(\bar t)
\right)\prod_{s=1}^{N}\lambda_s(\bar t^s)\rvec\,.
\end{equation}
Note that due to the commutation relations between currents \r{FiFi} and \r{FiFii} the
normalized product of the currents \r{FFnor} is symmetric with respect to
permutations of
 Bethe parameters of the same type.

The normalization of the universal off-shell Bethe vector is chosen in such a way
that it removes all zeros and poles originating from the products of  currents.
  For example, according to the commutation
relations \r{FiFi},  the products of the currents
$\F_\el(\bar t^\el)$  have poles when $t^\ell_j-t^\ell_i+\cci{\ell}=0$ for $j>i$ and $\el\not=m$
and zeros for all $\el$ when $t^\ell_j-t^\ell_i=0$.
The potential singularities are
compensated by the rational functions in the prefactor numerator in the formulas \r{FFnor}.
On the other hand, the products of the currents $\F_\el(\bar t^\el)\F_{\el+1}(\bar t^{\el+1})$
 have poles when
$t^{\ell+1}_j-t^\ell_i=0$ and zeros when  $t^{\ell+1}_j-t^\ell_i+\cci{\el+1}=0$ for all $i,j$.
These possible singularities are
compensated by the product of the rational functions
$f_{\prt{\el+1}}(\bar t^{\el+1},\bar t^\el)^{-1}$
in the denominator of the prefactor in the formula \r{FFnor}.

Our strategy is to calculate first the projection in \r{uBV1} and then to rewrite the result of this
calculation as certain polynomial in
the monodromy matrix elements. This will be done in  section~\ref{BV-calc}.
Then we define the dual Bethe vector $\CCC(\bar t)$ by the formula
 \begin{equation}\label{dual1}
 \CCC(\bar t)=\Psi\bigl(\BBB(\bar t)\bigr),
 \end{equation}
 where the antimorphism \r{antimo} is extended from the
 algebra to vectors of the representation of this algebra  using  $\Psi(\rvec)=\lvec$ and $\Psi(\lvec)=\rvec$.

Alternatively, the formula for the dual Bethe vector can be found via the  projection method  and
another choice of
current Borel subalgebras,  Drinfeld coproduct,  and associated projections
from the ordered product of the currents
\begin{equation*}
\eE(\bar t)=\!\!\prod_{N\geq a\geq
1}^{\longleftarrow} \sk{\prod_{\rr_a\geq\ell\geq1 }^{\longleftarrow} E_{a}(t^a_{\ell})}\,.
\end{equation*}
We do not perform these calculations in this paper.

\subsection{Bethe vector $\hat\BBB(\bar t)$ and dual Bethe vector $\hat\CCC(\bar t)$}

For the Bethe vector $\hat\BBB(\bar t)$ and dual Bethe vector $\hat\CCC(\bar t)$
one has to explore  the second current realization \r{tkiF}--\r{tEF} of the Yangian
double $DY(\mathfrak{gl}(m|n))$
given by the currents $\tiF_i(z)$, $\tiE_i(z)$ and $\tk^\pm_j(z)$ which are related to the
monodromy matrix elements through the Gauss decomposition \r{GF1}--\r{GE1} and
Frenkel-Ding formulas \r{DF-iso1}.

As in the previous sections, to describe Bethe vector $\hat\BBB(\bar t)$
we define a Borel subalgebra $\hU_F$
 such that 'positive' Cartan currents $\tk^+_j(z)$ are in $\hU_F$ with coalgebraic properties
\begin{equation}\label{Dcopr1}
\begin{split}
\hat\Delta^{(D)}(\tiF_i(z))&=\mathbf{1}\ot\tiF_i(z)+\tiF_i(z)\ot\tk^+_{i}(z)\sk{\tk^+_{i+1}(z)}^{-1},\\
\hat\Delta^{(D)}(\tk^+_j(z))&=\tk^+_j(z)\ot \tk^+_{j}(z).
\end{split}
\end{equation}

We again consider the intersections of this current Borel subalgebra
with the standard Borel subalgebras $U^\pm$
\begin{equation}\label{inter1}
\begin{split}
\hU_F^-=\hU_F\cap \hU^-\,,&\qquad \hU_F^+=\hU_F\cap \hU^+\,,
\end{split}
\end{equation}
and check the coideal properties of these intersections
\begin{equation*}
\begin{split}
\hat\Delta^{(D)}(\hU_F^+)=\hU_F\ot \hU^+_F\,,&\qquad \hat\Delta^{(D)}(\hU_F^-)=\hU^-_F\ot \hU_F\,,
\end{split}
\end{equation*}
with respect to coproduct \r{Dcopr1}.

Using the same cycling ordering for the Cartan--Weyl generators of $\hU_F$ as we used
for ordering elements in $U_F$, we  say that an arbitrary element $\hat\eF\in \hU_F$
is ordered if
\begin{equation*}
\hat\eF=\hat\eF_-\cdot \hat\eF_+\,,
\end{equation*}
where $\hat\eF_\pm\in \hU_F^\pm$.

Again according to the general theory formulated in \cite{EKhP} one can define the projections of any ordered elements from the
subalgebras $\hU_F$ and $\hU_E$ onto subalgebras \r{inter1} using  the formulas
\begin{equation}\label{proj}
\begin{split}
\hPfp(\hat\eF_-\cdot\hat\eF_+)=\hat\varepsilon(\hat\eF_-)\,\hat\eF_+\,,
\quad \hPfm(\hat\eF_-\cdot \hat\eF_+)=\hat\eF_-\,\hat\varepsilon(\hat\eF_+)\,,\quad \hat\eF_\pm\in \hU_F^\pm\,,
\end{split}
\end{equation}
where the counit mapping $\hat\varepsilon: DY(\mathfrak{gl}(m|n))\to \CC$ is defined by the rules
\begin{equation*}
\hat\varepsilon(\tiF_i\mode{\ell})=0\,,\quad \varepsilon(\tk_j\mode{\ell})=0\,,
\end{equation*}
and $\tiF_i\mode{\ell}$ and $\tk_j\mode{\ell}$ are modes of the currents
$\tiF_i(z)$ and $\tk^+_i(z)$ in the second current realization of the Yangian double
 $DY(\mathfrak{gl}(m|n))$.

 Defining the completion $\hat\oU_F$  we can check \cite{EKhP} that:
 \begin{enumerate}
\item[(1)] the action of the projections \r{proj} extends to the algebras $\hat\oU_F$;
\item[(2)] for any $\hat\eF\in\hat\oU_F$ with $\hat\Delta^{(D)}(\hat\eF)=\hat\eF'\otimes\hat\eF''$ we have
\begin{equation}\label{ordF1}
\hat\eF=\hPfm\sk{\hat\eF''}\cdot \hPfp\sk{\hat\eF'}\,.
\end{equation}
\end{enumerate}

For the set \r{Bpar} of  Bethe parameters we consider the normalized
ordered products of  currents
\begin{equation}\label{hFFnor}
\hat\Fp(\bar t)=\frac{
\prod_ {\el=1}^{N} \hFli{\el}(\bar t^\el) }
{\prod_ {\el=1}^{N-1} f_{\prt{\el+1}}(\bar t^{\el+1},\bar t^\el)}\
\hF(\bar t)\,,
\end{equation}
where
\begin{equation}\label{tFF-EE}
\hF(\bar t)=\!\!\prod_{N\geq a\geq
1}^{\longleftarrow} \sk{\prod_{\rr_a\geq\ell\geq1 }
^{\longleftarrow} \tiF_{a}(t^a_{\ell})}.
\end{equation}

The universal off-shell Bethe vectors associated with the second current realization of the
Yangian double $\DYglmn$ are defined
through the action of the above projections on the singular vector $\rvec$
as follows
\begin{equation}\label{bv}
\hat\BBB(\bar t)=\hPfp\sk{\hat\Fp(\bar t)}\prod_{s=1}^{N}\lambda_{s+1}(\bar t^s)\rvec\,.
\end{equation}
The normalization of these universal off-shell Bethe vector is chosen again in such a way
to remove all zeros and poles originating from the products of  currents.

The dual Bethe vector $\hat\CCC(\bar t)$ is defined by antimorphism \r{antimo}:
\begin{equation}\label{dbv}
\hat\CCC(\bar t)=\Psi\bigl(\hat\BBB(\bar t)\bigr)\,.
\end{equation}

\subsection{Main results}
\label{mainres}

\medskip

In this paper we claim the following statements:
\begin{itemize}\item
We prove that the two different ways to construct Bethe vectors lead eventually to the same result, that is
\begin{equation}\label{conj}
\BBB(\bar t)=\hat\BBB(\bar t)\quad\mbox{and}\quad
\CCC(\bar t)=\hat\CCC(\bar t).
\end{equation}
In section~\ref{gen-act} we will give the proof of this statement for the
Bethe vectors $\BBB(\bar t)$ and $\hat\BBB(\bar t)$ only. The proof for the dual
vectors $\CCC(\bar t)$ and $\hat\CCC(\bar t)$ follows from application of antimorphism
$\Psi$ to the first equality in \r{conj}.

\item Bethe vectors become on-shell or equivalently, become eigenvectors of the supersymmetric
transfer matrix  $\trf(z)$ \eqref{strace}  with the eigenvalue \r{be8}
if the Bethe equations \r{be4} for the parameters \r{Bpar}
are satisfied.

\item Explicit formulas for the Bethe vectors in terms of the monodromy matrix elements
are given by the formulas \r{hr77} and \r{hr77h}. Explicit formulas for the dual vectors
can be obtained using antimorphism \r{antimo}.

\item The coproduct properties for the Bethe vectors are given in relations \eqref{BB1} and \eqref{BB2}.
They express the coproduct of a Bethe vector in term of Bethe vectors belonging to two copies of
$DY(\mathfrak{gl}(m|n))$ arising under application of the coproduct.

\end{itemize}

\section{{Monodromy matrix elements} action formulas}
\label{gen-act}

The goal of the present section is to prove that
the  Bethe vectors $\BBB(\bar t)$ and $\hat\BBB(\bar t)$
coincide.
Restoring formulas for the universal off-shell Bethe vectors in terms of  matrix elements
of the monodromy matrix  (see section~\ref{BV-calc}) we will see that the
problem to prove \r{conj}
is a rather complicated combinatorial problem. Instead, we will prove it
by checking that both of these vectors satisfy the same recurrence relations
with respect to the action of the upper triangular and diagonal monodromy matrix elements onto
these vectors. To check this statement it is not necessary to get explicit formulas
for the universal off-shell Bethe  vectors in terms of the monodromy matrix elements.
 Before starting this analysis we show that
Bethe vectors $\BBB(\bar t)$ and $\hat\BBB(\bar t)$
satisfy the same coproduct properties coming from the coproduct of
the monodromy matrix \r{coprL}.

\subsection{Coproduct properties of the Bethe vectors}

Calculating the coproduct of the product of  currents $F_i(t)$ using the first formula
in \r{Dcopr} we obtain that the Drinfeld coproduct of the ordered product of
simple root currents $\F(\bar t)$ is
\begin{equation}\label{FFor-cop}
\begin{split}
\Delta^{(D)}\sk{\eF(\bar t)}&=\sum_{0\leq s_{1}\leq \rr_1}\cdots \sum_{0\leq s_{N}\leq \rr_N}
\prod_{\el=1}^N\frac{1}{s_\el!(\rr_\el-s_\el)!} \\
&\qquad\times  \tSym_{\,\bar t}\sk{Z_{\bar s}(\bar t)\
\eF(\bar t')\prod_{s=1}^N \prod_{\el=s_\el+1}^{\rr_\el}k^+_{s+1}(t^s_{\el})k^+_{s}
(t^s_{\el})^{-1} \ot \eF(\bar t'')},
\end{split}
\end{equation}
where the sets $\bar t'$ and $\bar t''$ are
\begin{equation*}
\begin{split}
\bar t' \ &=\{t^{1}_{1},\ldots,t^{1}_{s_{1}};
t^{2}_{1},\ldots,t^{2}_{s_{2}};\ldots;
t^{N}_{1},\ldots,t^{N}_{s_{N}} \},\\
\bar t'' \ &=\{t^{1}_{s_1+1},\ldots,t^{1}_{\rr_{1}};
t^{2}_{s_2+1},\ldots,t^{2}_{\rr_{2}};\ldots;
t^{N}_{s_N+1},\ldots,t^{N}_{\rr_{N}} \},
\end{split}
\end{equation*}
and $Z_{\bar s}(\bar t)$ is the rational function
\begin{equation*}
Z_{\bar s}(\bar t)=\prod_{a=1}^{N-1}\prod_{\substack{s_a<\ell\leq \rr_a\\
0<\ell'\leq s_{a+1} }}\frac{t^a_{\ell}-t^{a+1}_{\ell'}-\cci{a+1}}{t^a_{\ell}-t^{a+1}_{\ell'}}
=\prod_{a=1}^{N-1}\prod_{\substack{s_a<\ell\leq \rr_a\\
0<\ell'\leq s_{a+1} }} f_{\prt{a+1}}(t^{a+1}_{\ell'},t^a_{\ell}).
\end{equation*}

Formula \r{FFor-cop} allows one to obtain the coalgebraic properties
of the normalized product of the currents \r{FFnor}
 with respect to the Drinfeld coproduct.
  Indeed, the $c$-symmetrization  can be transformed into the usual symmetrization over the set $\{\bar t^s\}$ due to  the property
\begin{equation}\label{sy-p}
\Fli{s}(\bar t^s)\
\tSym_{\,\bar t^s}\sk{G(\bar t^s)}=
\Sym_{\,\bar t^s}\sk{\Fli{s}(\bar t^s) G(\bar t^s)}.
\end{equation}
Then the symmetrization can be replaced by the sum over partitions and symmetrizations
over parted subsets:
\begin{equation}\label{sy-d}
\Sym_{\,\bar t^s}(\cdot) =\sum_{\bar t^s\Ra\{\bar t^s_{\so},\bar t^s_{\st}\}}\
\Sym_{\,\bar t^s_{\so}}\ \Sym_{\,\bar t^s_{\st}}(\cdot)\,.
\end{equation}
Here the summation goes over partitions of the set $\{\bar t^s\}$ onto two non-intersecting
subsets $\{\bar t^s_{\so}\}$ and $\{\bar t^s_{\st}\}$ with cardinalities
$\#\bar t^s_{\so}+ \#\bar t^s_{\st}=\#\bar t^s$, where
 \begin{equation}\label{par1}
\bar t=\{\bar t^1,\ldots,\bar t^N\}\Ra \bar t_{\so} \cup \bar t_{\st}\,,
\end{equation}
and
\begin{equation}\label{par2}
\bar t_{\so}=\{\bar t^1_{\so},\ldots,\bar t^N_{\so}\}\qquad
\bar t_{\st}=\{\bar t^1_{\st},\ldots,\bar t^N_{\st}\}.
\end{equation}

Using \r{sy-p} and \r{sy-d} and the fact that the normalized product of the currents $\Fp(\bar t)$
is symmetric with respect to permutations in  each set of the Bethe parameters $\bar t^\el$,
$\el=1,\ldots,N$ we can transform equation \r{FFor-cop} into the sum over partitions given by \r{par1} and
\r{par2}
\begin{equation}\label{FFcop}
\Delta^{(D)}\sk{\Fp(\bar t)}=\sum_{\rm part}
\frac{\prod_ {s=1}^{N}  \Fli{s}(\bar t^s_{\st},\bar t^s_{\so})  }
{\prod_ {s=1}^{N-1} f_{\prt{s+1}}(\bar t^{s+1}_{\st},\bar t^s_{\so})}\
\Fp(\bar t_{\so}) \prod_{s=1}^N k^+_{s+1}(\bar t^s_{\st})k^+_{s}
(\bar t^s_{\st})^{-1}\otimes \Fp(\bar t_{\st}).
\end{equation}
Using the Drinfeld coproduct
\r{Dcopr1} for the second current realization of $DY(\mathfrak{gl}(m|n))$ we can show that
the coproduct of the normalized product of the currents \r{hFFnor} is given by
\begin{equation}\label{hFFcop}
\hat\Delta^{(D)}\sk{\hFp(\bar t)}=\sum_{\rm part}
\frac{(-)^{\#\bar t^m_{\so}\cdot\#\bar t^m_{\st}}
\prod_ {s=1}^{N}  \hFli{s}(\bar t^s_{\st},\bar t^s_{\so})  }
{ \prod_ {s=1}^{N-1} f_{\prt{s+1}}(\bar t^{s+1}_{\st},\bar t^s_{\so})}
\ \hFp(\bar t_{\so})\otimes \hFp(\bar t_{\st}) \prod_{s=1}^N
\tk^+_{s}(\bar t^s_{\so})\tk^+_{s+1}(\bar t^s_{\so})^{-1},
\end{equation}
where summation goes over non-intersecting subsets defined by the equalities \r{par1} and \r{par2}.

We can use formulas \r{FFcop} and \r{hFFcop} to establish the coproduct properties of the
universal Bethe vectors \r{uBV1} and \r{bv}.
It was proved in \cite{EKhP} that for any elements $\F\in\oU_F$ and
$\hat\F\in\hat{\oU}_F$ { the following equations hold:}
\begin{equation*}
\begin{split}
\Delta\sk{\Pfp\sk{\F}}&\equiv \sk{\Pfp\otimes\Pfp}\sk{\Delta^{(D)}\sk{\F}}\quad\mbox{mod}\quad
U^+_F\otimes J,\\
\Delta\sk{\hPfp\sk{\hat\F}}&\equiv \sk{\hPfp\otimes\hPfp}\sk{\hat\Delta^{(D)}\sk{\hat\F}}\quad\mbox{mod}\quad
\hat U^+_F\otimes \hat J,
\end{split}
\end{equation*}
where $J$ and $\hat J$ are ideals in the corresponding subalgebras  killing the  singular vectors $\rvec$.
A proper definition of these ideals is given in the beginning of the next subsection.
Using these equalities and formulas \r{FFcop}, \r{hFFcop} we obtain
\begin{equation}\label{BB1}
\BBB(\bar t)=\sum_{\rm part}
\frac{\prod_ {s=1}^{N}  \Fli{s}(\bar t^s_{\st},\bar t^s_{\so})  }
{ \prod_ {s=1}^{N-1} f_{\prt{s+1}}(\bar t^{s+1}_{\st},\bar t^s_{\so})}\
\BBB^{(1)}(\bar t_{\so}) \prod_{s=1}^N \lambda^{(1)}_{s+1}(\bar t^s_{\st})\otimes
\BBB^{(2)}(\bar t_{\st}) \prod_{s=1}^N \lambda^{(2)}_{s}(\bar t^s_{\so})
\end{equation}
and
\begin{equation}\label{BB2}
\hat\BBB(\bar t)=\sum_{\rm part}
\frac{(-)^{\#\bar t^m_{\so}\cdot\#\bar t^m_{\st}}\prod_ {s=1}^{N}  \hFli{s}(\bar t^s_{\st},\bar t^s_{\so})  }
{\prod_ {s=1}^{N-1} f_{\prt{s+1}}(\bar t^{s+1}_{\st},\bar t^s_{\so})}\
 \hat\BBB^{(1)}(\bar t_{\so}) \prod_{s=1}^N \lambda^{(1)}_{s+1}(\bar t^s_{\st})\otimes
\hat\BBB^{(2)}(\bar t_{\st}) \prod_{s=1}^N \lambda^{(2)}_{s}(\bar t^s_{\so}).
\end{equation}
Taking into account \r{equ} we conclude that universal Bethe vectors
$\BBB(\bar t)$ and $\hat\BBB(\bar t)$ satisfy the same
coproduct properties, which signify that they may coincide. Below we show  that these vectors satisfy the same recurrence relations, thus proving that they coincide.

Coproduct formulas \r{FFor-cop} and \r{FFcop} are very powerful tools for the calculation
of the projection of the product of the currents. Indeed, using the fundamental property of the
projection $\Pfpm$ given by the equality \r{ordF}  we obtain from \r{FFor-cop}
\begin{equation}\label{FFor}
\eF(\bar t)=\sum_{0\leq s_{1}\leq \rr_1}\cdots \sum_{0\leq s_{N}\leq \rr_N}
\prod_{\el=1}^N\frac{1}{s_\el!(\rr_\el-s_\el)!} \ \tSym_{\,\bar t}\sk{Z_{\bar s}(\bar t)
\Pfm\sk{\eF(\bar t')}\Pfp\sk{\eF(\bar t'')}},
\end{equation}
and from \r{FFcop}
\begin{equation}\label{FFpar}
\Fp(\bar t)=\sum_{\bar t^s\Ra\{\bar t^s_{\so},\bar t^s_{\st}\}}\ \
\frac{ \prod_ {s=1}^{N}  \Fli{s}(\bar t^s_{\st},\bar t^s_{\so})  }
{ \prod_ {s=1}^{N-1} f_{\prt{s+1}}(\bar t^{s+1}_{\st},\bar t^s_{\so})}\
\Pfm\sk{\Fp(\bar t_{\so})}\cdot \Pfp\sk{\Fp(\bar t_{\st})}\,.
\end{equation}
This equality and analogous equality for the product of the currents $\tiF_i(t)$
 will be used in the section~\ref{BV-calc} to resolve the hierarchical relations for the nested Bethe
vectors  and to obtain the explicit formulas for the Bethe vectors in terms of the monodromy matrix
elements. This will be achieved by the explicit calculation of the projection of the  corresponding
products of the currents, which is reduced to the calculation presented in  Appendix~\ref{appC}.

\subsection{Ideals of the Yangian double and presentation of the projections}

For the calculation  of the monodromy matrix elements actions
onto Bethe vectors we have to formulate an important  auxiliary statement about the action
of the monodromy matrix elements $\LL^+_{i,j}(z)$ onto 'negative' projections of the
composed currents  $\Pfm\sk{F_{k,l}(w)}$ and
$\hPfm\sk{\tiF_{k,l}(w)}$ modulo certain ideals. The proof of this statement can be
performed along the route used in the paper \cite{FKPR} for the quantum affine
algebra $U_q(\widehat{\mathfrak{gl}}(N))$, so that we just sketch it below.

Let $U^\pm_F$ and $U^\pm_E$ be intersections of the standard Borel subalgebras
$U^\pm$ and current Borel subalgebras $U_F$ and $U_E$ used in  section~\ref{Bvsec}.
Let $I\subset DY(\mathfrak{gl}(m|n))$ be the ideal in the
Yangian double  generated by the elements of the form ${\cal F}_-\cdot {\cal F}$
such that ${\cal F}_-\in U_F^-$, ${\cal F}\in U_F$ and
$\coun({\cal F}_-)=0$.
Here and below $\coun$ is a counit in the Hopf algebra
$DY(\mathfrak{gl}(m|n))$.  It is clear from the definition \r{proj1} of the projection $\Pfp$
that the whole  ideal $I$ is annihilated by this projection: $\Pfp(I)=0$.
 Let $K\in DY(\mathfrak{gl}(m|n))$ be the ideal
generated by the  elements which contain any combination of the 'negative' Cartan
currents $k^-_j(u)$. Due to the commutation relations in $DY(\mathfrak{gl}(m|n))$,
$K$ is  an ideal because the 'negative' Cartan
currents cannot be 'killed' by any of the commutation relations in $DY(\mathfrak{gl}(m|n))$.
Let $J \subset DY(\mathfrak{gl}(m|n))$ be the ideal in the Yangian double generated by the elements
of the form ${\cal F}\cdot {\cal E}_+$ such that
${\cal E}_+\in U_E^+$, ${\cal F}\in U_F^+$ and
$\coun({\cal E}_+)=0$. By definition of this ideal, any element in $J$ annihilates
the right vacuum vector: $J\rvec=0$.
Further on we will
use the symbols $\sim_I$, $\sim_K$, and $\sim_J$ for equalities in the
Yangian double $DY(\mathfrak{gl}(m|n))$ modulo terms
from the corresponding ideals $I$, $K$, and $J$. Analogously, from the current Borel
subalgebras $\hat U_F$ and $\hat U_E$, we define the ideals $\hat I$,
$\hat K$, and $\hat J$ and the equivalence relations $\sim_{\hat I}$,
$\sim_{\hat K}$, and $\sim_{\hat J}$.

Since off-shell Bethe vectors defined  in \r{uBV1} and \r{bv}
obviously do not belong to the ideals $I$ and $K$ or the ideals $\hat I$ and $\hat K$
we can compute the action of the monodromy matrix elements on these Bethe
vectors modulo these ideals.
Moreover, since the ideals $J$ and
$\hat J$ kill the vacuum vector $\rvec$ we can also skip the terms from these ideals
when calculating the action of monodromy on the projections of the currents.

Using the commutation relations between the Gauss coordinates of the
'positive' and 'negative' monodromy matrices \r{TM-1} and \r{TM-2},  as well as the
relations between the 'negative' projections of the composed currents and the
Gauss coordinates \r{nrca3} and \r{nrca4} we can prove

\begin{proposition} \cite{FKPR}
\begin{eqnarray}\label{idca3}
&&\LL^+_{i,j}(z)\cdot \Pfm\sk{F_{k,l}(w)}\ \sim_{I,K}\
-\ \phi_k\ c_{[l,k]}\ \delta_{jl}\ g(z,w)\ \LL^+_{i,k}(z),
\\
\label{idca4}
&&\LL^+_{i,j}(z)\cdot \hPfm\sk{\tiF_{k,l}(w)}\ \sim_{{\hat I},{\hat K}}\
-\ \hat{\phi}_l\ c_{[l,k]}\
\delta_{ik}\ g(z,w)\ \LL^+_{l,j}(z),
\end{eqnarray}
where $c_{[l,k]}$ is given by the relation \r{ccc},
and\;\footnote{The asymmetry in the signs $\phi_k$ and $\hat\phi_l$ is related to
the asymmetry in the different Gauss decompositions. }
\begin{equation}\label{phases}
\begin{split}
\phi_k&=(-)^{(\prt{i}+\prt{j})\prt{k}+\prt{i}\prt{j}}\quad\mbox{for}\quad k>j\,, \\
\hat{\phi}_l&=(-)^{1+\prt{i}}\quad\mbox{for}\quad l<i\,.
\end{split}
\end{equation}
We extend the values of the indices
$k$ and $l$ in \r{phases} to the values $k=j$ and $l=i$:
\begin{equation*}
\phi_k=1\quad\mbox{for}\quad k=j\quad \mbox{and}\quad
\hat{\phi}_l=1\quad\mbox{for}\quad l=i\,.
\end{equation*}
This extension will be justified later on, see Proposition~\ref{prop74}.
\end{proposition}

{\sl Sketch of the proof.}
The appearance of the Kroneker symbols $\delta_{jl}$ and $\delta_{ik}$  respectively in \r{idca3} and \r{idca4}
were proved in \cite{FKPR}. Let us
give arguments which fix the rest in the right hand side of equations
\r{idca3} and \r{idca4}  including the phases \r{phases}.
In order to do this we consider equations \r{idca3} and \r{idca4} applied to
the right singular vectors.

It is clear from the Gauss decompositions \r{GF2} that
\begin{equation*}
\FF^-_{k,l}(w)\rvec = \LL^-_{l,k}(w)\LL^-_{l,l}(w)^{-1} \rvec.
\end{equation*}
Then the equation \r{TM-2} can be interpreted as
\begin{equation}\label{apA10}
\LL^+_{i,j}(z)\LL^-_{l,k}(w)\LL^-_{l,l}(w)^{-1}\ \sim_{I,K}\ (-)^{\prt{k}(\prt{i}+\prt{j})+\prt{i}\prt{j}}
g(z,w)\LL^+_{i,k}(z)\LL^-_{l,j}(w)\LL^-_{l,l}(w)^{-1},
\end{equation}
and due to the Kroneker symbol $\delta_{jl}$ in the right hand side of \r{idca3}
the 'negative' monodromy matrix elements in the right hand side of \r{apA10} cancel each other.
Taking into account the relation \r{nrca3} we obtain that $\phi_k=
(-)^{\prt{k}(\prt{i}+\prt{j})+\prt{i}\prt{j}}$.

Similarly, it follows from the  Gauss decomposition \r{GF1} that
\begin{equation*}
\tFF^-_{k,l}(w)\rvec = \LL^-_{l,k}(w)\LL^-_{k,k}(w)^{-1} \rvec.
\end{equation*}
Then the equation \r{TM-1} can be interpreted as
\begin{equation}\label{apA12}
\LL^+_{i,j}(z)\LL^-_{l,k}(w)\LL^-_{k,k}(w)^{-1}\
\sim_{\hat I,\hat K}\ (-)^{1+\prt{i}(\prt{k}+\prt{l})+\prt{k}\prt{l}}
g(z,w)\LL^+_{l,j}(z)\LL^-_{i,k}(w)\LL^-_{k,k}(w)^{-1},
\end{equation}
and due to the Kroneker symbol $\delta_{ik}$ in the right hand side of \r{idca4}
the 'negative' monodromy matrix elements in the right hand side of \r{apA12} disappear,  leading to $\hat\phi_l=(-)^{1+\prt{i}}$.

\hfill\qed

We conclude this subsection with formulation the following
\begin{proposition}\label{main-prop}
The off-shell Bethe vectors given by the formulas \r{uBV1} and \r{bv} satisfy the same
recurrence relations following from the action by the upper-triangular monodromy
matrix elements $\LL_{i,j}(z)$, $i\leq j$ onto these vectors. This implies that
the Bethe vectors coincide
\begin{equation*}
\BBB(\bar t)=\hat\BBB(\bar t)\,.
\end{equation*}
\end{proposition}
The proof of this proposition will be given in the next two sections~\ref{ssec71} and \ref{ssec72}.

\subsection{Auxiliary presentations for the projections}\label{ssec71}

In order to calculate the action of the upper triangular and diagonal  monodromy matrix  elements
onto Bethe vectors \r{uBV1} and \r{bv} we have to obtain a special presentation for the
projections for the products of the total simple root currents.
The systematic way to get such
 presentation is based on the techniques elaborated in the paper
\cite{KhP-series}. Below we use results of this paper adapting them
to the case under consideration.

\begin{proposition}
We have following series identities for $i<j$
\begin{eqnarray}\label{pmau1}
\Pfm\sk{F_i(t^i)\cdots F_j(t^j)}&=&\sum_{\el=0}^{j-i}
c_{[i,i+\el+1]}^{-1}\prod_{s=i}^{i+\el-1}g_{\prt{s+1}}(t^{s+1},t^s)\Pfm\sk{F_{i+\el+1,i}(t^{i+\el})}\nonumber \\
&&\times\
\Pfm\sk{F_{i+\el+1}(t^{i+\el+1})\cdots F_j(t^j)},
\\
\label{pmau2}
\hPfm\sk{\tiF_j(t^j)\cdots \tiF_i(t^i)}&=&\sum_{\el=0}^{j-i}
c_{[j-\el,j+1]}^{-1}\prod_{s=j-\el}^{j+1}g_{\prt{s+1}}(t^{s+1},t^{s})\hPfm\sk{\tiF_{j+1,j-\el}(t^{j-\el})}\nonumber \\
&&\times\
\hPfm\sk{\tiF_{j-\el-1}(t^{j-\el-1})\cdots \tiF_i(t^i)}.
\end{eqnarray}
\end{proposition}

\proof\  Both equalities can be proved analogously using the definition of the projections.
Therefore, we give a detailed proof only for  \r{pmau1}.
We start from the definition
\begin{equation}\label{s7a1}
\begin{split}
\Pfm\sk{F_i(t^i)\cdots F_j(t^j)}&=\Pfm\sk{F_i(t^i)}\cdot \Pfm\sk{F_{i+1}(t^{i+1})\cdots F_j(t^j)}\\
&+\Pfm\sk{F^{(+)}_i(t^i)F_{i+1}(t^{i+1})F_{i+2}(t^{i+2})\cdots F_j(t^j)}.
\end{split}
\end{equation}
Using the definition
\begin{equation*}
F^{(+)}_i(t^i)=\int dw\  \frac{F_i(w)}{t^i-w}
\end{equation*}
and the commutation relation
\begin{equation}\label{cc12}
F_i(u)F_{i+1}(v)=\frac{u-v-\cci{i+1}}{(u-v)_<}F_{i+1}(v)F_i(u)-\delta(u,v)F_{i+2,i}(v),
\end{equation}
which is a particular case of the definition of the composed current \r{F-com} or
\r{iF},
 we obtain
\begin{equation}\label{s7a2}
\begin{split}
F^{(+)}_i(t^i)F_{i+1}(t^{i+1})&=f_{\prt{i+1}}(t^{i+1},t^i)F_{i+1}(t^{i+1})F^{(+)}_i(t^i;t^{i+1})\\
&+\cci{i+1}^{-1}g_{\prt{i+1}}(t^{i+1},t^i)F_{i+2,i}(t^{i+1}),
\end{split}
\end{equation}
where
$F^{(+)}_i(t^i;t^{i+1})=F^{(+)}_i(t^i)-\frac{\cci{i+1}}{t^{i+1}-t^i+\cci{i+1}}F^{(+)}_i(t^{i+1})$.
Because of the commutativity of the current $F_i(t)$ with the currents $F_{i+2}(t^{i+2})\cdots F_j(t^j)$
the first term in \r{s7a2} vanishes under 'negative' projections in the second term
of \r{s7a1}. On the other hand, due to the second relation in \r{iF1}
\begin{equation}\label{s7a3}
F_{i+2,i}(t^{i+1})=-\Sc_{F_i\mode{0}}\sk{F_{i+1}(t^{i+1})}+\cci{i+1}F_{i+1}(t^{i+1})F^{(+)}_i(t^{i+1})
\end{equation}
where the operators $\Sc_{F_i\mode{0}}(\cdot)$ are called screening operators and are defined
by equations \r{sc1}.
The second term in the right hand side of \r{s7a3}  also vanishes under the 'negative' projection
in the second line of \r{s7a1}. Thus, \r{s7a1} turns into
\begin{equation}\label{s7a4}
\begin{split}
&\Pfm\sk{F_i(t^i)\cdots F_j(t^j)}=\Pfm\sk{F_i(t^i)}\cdot \Pfm\sk{F_{i+1}(t^{i+1})\cdots F_j(t^j)}\\
&\qquad-\cci{i+1}^{-1}g_{\prt{i+1}}(t^{i+1},t^i)
\Sc_{F_i\mode{0}}\sk{\Pfm\sk{F_{i+1}(t^{i+1})\cdots F_j(t^j)}}.
\end{split}
\end{equation}

In the second line of \r{s7a4} we obtain the 'negative' projection of the product of
currents $F_{i+1}(t^{i+1})\cdots F_j(t^j)$. Therefore, we can use this equality recursively to obtain
at the first step
\begin{equation*}
\begin{split}
&\Pfm\sk{F_i(t^i)\cdots F_j(t^j)}=\Pfm\sk{F_i(t^i)}\cdot \Pfm\sk{F_{i+1}(t^{i+1})\cdots F_j(t^j)}\\
&+\cci{i,i+2}^{-1}g_{\prt{i+1}}(t^{i+1},t^i)\Pfm\sk{F_{i+2,i}(t^{i+1})}
\Pfm\sk{F_{i+2}(t^{2+1})\cdots F_j(t^j)}\\
&+\cci{i,i+3}^{-1}g_{\prt{i+1}}(t^{i+1},t^i)g_{\prt{i+2}}(t^{i+2},t^{i+1})
\Sc_{F_i\mode{0}}\sk{\Sc_{F_{i+1}\mode{0}}\sk{\Pfm\sk{F_{i+2}(t^{i+2})\cdots F_j(t^j)}}},
\end{split}
\end{equation*}
where we again used  \r{s7a3} and the commutativity of the screening operators
and projections (see Appendix~\ref{comPS}).
Continuing this recursion we prove \r{pmau1}. Equation \r{pmau2} can be proved
similarly starting from the commutation relations
\begin{equation*}
\tiF_{i+1}(u)\tiF_{i}(v)=\frac{u-v+\cci{i+1}}{(u-v)_<}\tiF_{i}(v)\tiF_{i+1}(u)+\delta(u,v)\tiF_{i+2,i}(v)
\end{equation*}
 and using
the first equation in  \r{itF1}. \hfill\qed

\bigskip

For each simple root index $i=1,\ldots,N$
  we introduce the following notation for  ordered products of the
currents
\begin{equation*}
\F_i(\bar t^i)=F_i(t^i_1)\cdots F_i(t^i_{r_i})\quad\mbox{and}\quad
\ticF_i(\bar t^i)=\tiF_i(t^i_{r_i})\cdots \tiF_i(t^i_1).
\end{equation*}

Using the normal ordering relation \r{ordF} (in the form \r{FFor}) and  \r{pmau1} we can prove
following
\begin{proposition}\label{prop44}
There is an equality
\begin{equation}\label{usre1}
\begin{split}
&\Pfp\sk{\F_1(\bar t^1)\cdots \F_N(\bar t^N)}= \Pfp\sk{\F_1(\bar t^1)\cdots \F_{N-1}(\bar t^{N-1})}\cdot
\F_N(\bar t^N)\\
&\quad -\sum_{\el=1}^N \   \comb_\el\ \tSym_{\,\bar t^\el,\ldots,\bar t^N}\left[
\fgf_\ell(\bar t^{\el-1},\ldots,\bar t^N)\   \cci{\el,N+1}^{-1}  \Pfm\sk{F_{N+1,\el}(t^N_1)}
\right.\\
&\left.\quad\times\
\Pfp\sk{\F_1(\bar t^1)\cdots\F_{\el-1}(\bar t^{\el-1})\F_{\el}(\bar t_1^{\el})\cdots \F_{N-1}(\bar t_1^{N-1})}
\cdot \F_N(\bar t_1^N)\right]+\mathbb{W},
\end{split}
\end{equation}
where for any  $1\leq \el \leq N$  we have introduced rational functions
\begin{equation}\label{zz1}
\fgf_\ell(\bar t^{\el-1},\ldots,\bar t^N) =
 f_{\prt{\el}}(t^\el_1,\bar t^{\el-1})
\prod_{s=\el}^{N-1}g_{\prt{s+1}}(t_1^{s+1},t_1^{s}) f_{\prt{s+1}}(t^{s+1}_1,\bar t_1^{s})\,,
\end{equation}
 and a combinatorial factor
\begin{equation}\label{comb1}
\comb_\el=\prod_{s=\el}^{N} \frac{1}{(r_s-1)!}\,.
\end{equation}
In \eqref{usre1}, $\mathbb{W}$ denotes terms which have the structure $\Pfm\sk{F_{j_1,i_1}(w_1)}
\Pfm\sk{F_{j_2,i_2}(w_2)}\F$ with $j_1\geq j_2$ and some element $\F\in\oU_F$.
\end{proposition}
In \r{usre1} we used a shorthand notation for the set $\bar t^\ell_i$ (for $i=1,\ldots,\rr_\ell$)
\begin{equation*}
\bar t^\ell_i=\{t^\ell_1,\ldots,t^\ell_{i-1},t^\ell_{i+1},\ldots,t^\ell_{\rr_\ell}\}\,,
\end{equation*}
where the Bethe parameter $t^\ell_i$ is omitted from the set $\bar t^\ell$, $\ell=1,\ldots,N$.

 Due to \r{idca3}, the action of any monodromy matrix
element $\LL^+_{i,j}(z)$ onto terms $\mathbb{W}$ belongs to the ideal $I$, except for
 terms which  are proportional to $\delta_{j,i_1}\delta_{j_1,i_2}$. These terms are irrelevant
due to the condition $j_1\geq j_2>i_2$.

{\sl Proof.}
It was proved in \cite{KhP-series} that the projection $\Pfp\sk{\F_1(\bar t^1)\cdots \F_N(\bar t^N)}$
can be presented in the form\footnote{In fact, this statement was proved in \cite{KhP-series} for the case of the
currents $\tiF_\ell$, but it can be repeated easily for the currents $F_\ell$, leading to
\r{ser-res}.}
\begin{equation}\label{ser-res}
\Pfp\sk{\F_1\cdots \F_N}=\sum
{\cal P}\sk{\Pfm\sk{F_{N+1,\ell}}}\cdot \Pfp\sk{\F_1\cdots \F_{N-1}}\cdot \F_N\,,
\end{equation}
where ${\cal P}\sk{\Pfm\sk{F_{N+1,\ell}}}$ is a certain polynomial with rational coefficients of the
'negative' projections of the composed currents $F_{N+1,\ell}$, $\ell=1,\ldots,N$, and $\F_\ell$
are products of the currents corresponding to the simple root $\ell$. To be compact we didn't write the argument of the currents in \eqref{ser-res}.

It was shown in \cite{KhP-series} that only 'negative' projections of
the currents $\Pfm\sk{F_{N+1,\ell}(t)}$ appear in the right hand side of \r{ser-res}. Other
'negative' projections of the currents  $\Pfm\sk{F_{\ell',\ell}(t)}$ with $N\geq \ell'>\ell$ do not arise.
The main reason for such a phenomena is the factorisation  of projections
of the products of currents. We will demonstrate this phenomena below in the
simplest non-trivial case of $N=2$ using the normal ordering relation \r{FFor}.

Moreover,
due to  equation \r{idca3}, it is enough to keep in the formula \r{ser-res}
only the first order polynomials with respect to the 'negative' projections of the composed
currents. Indeed, after the action of the monodromy matrix element $\LL^+_{i,j}(z)$
onto the product of two 'negative' projections of the
composed currents $\Pfm\sk{F_{N+1,\ell_1}(t)}\cdot \Pfm\sk{F_{N+1,\ell_2}(t)}$,
the terms which are not in the ideal $I$ and $K$
are proportional to $\delta_{j,\ell_1}\delta_{N+1,\ell_2}$ and they vanish
due to $\ell_2<N+1$.

Let us show how  the relations of the type
 \r{ser-res} arise in the simple case of $m=2$ and $n=1$.
We rename the sets of parameters $\bar t^1\equiv \bu$
and $\bar t^2\equiv \bv$ with cardinalities $\#\bu=a$ and $\#\bv=b$
 to simplify formulas below.
Formula \eqref{FFor} in this case can be rewritten as
\begin{equation}\label{pf1}
\begin{split}
&\Pfp\left(F_{1}(u_{1})\cdots F_{1}(u_a)\cdot
F_{2}(v_1)\cdots F_{2}(v_b)\right)=F_{1}(u_{1})\cdots F_{1}(u_a)\cdot
F_{2}(v_1)\cdots F_{2}(v_b)\\
&\quad -\tSym_{\,\bu}\frac{1}{(a-1)!}\Pfm\sk{F_1(u_1)}\Pfp\sk{F_{1}(u_{2})\cdots F_{1}(u_a)\cdot
F_{2}(v_1)\cdots F_{2}(v_b)}\\
&\quad -\tSym_{\,\bv}\frac{f(v_1,\bu)}{(b-1)!}\Pfm\sk{F_2(v_1)}\Pfp\sk{
F_{1}(u_{1})\cdots F_{1}(u_a)\cdot
F_{2}(v_2)\cdots F_{2}(v_b)}\\
&\quad -\tSym_{\,\bu,\bv}\frac{f(v_1,\bu_1)}{(a-1)!(b-1)!}
\Pfm\sk{F_1(u_1)F_2(v_1)}\Pfp\sk{F_{1}(u_{2})\cdots F_{1}(u_a)\cdot
F_{2}(v_2)\cdots F_{2}(v_b)}\\
&\qquad +\mathbb{W}.
\end{split}
\end{equation}
We  keep the double symmetrized term in \r{pf1} because it is a source of the 'negative' projection
of the composed currents $\Pfm\sk{F_{3,1}(v)}$ (see \r{decomp} below), while
the quadratic terms from $\Pfm\sk{F_1(u_1)F_2(v_1)}$  disappears at the next step of the
recursion.

Next we apply this formula recursively in the right hand side of \r{pf1}
in order to replace 'positive' projections
by the corresponding products of  total currents. Using equality
\begin{equation}\label{decomp}
\Pfm\sk{F_1(u)F_2(v)}=\Pfm\sk{F_1(u)}\Pfm\sk{F_2(v)} +c^{-1} g(v,u)\Pfm\sk{F_{3,1}(v)},
\end{equation}
which is a direct consequence of the relation \r{cc12},
we obtain instead of \r{pf1} a formal series equality
(recall that $F_{i+1,i}(t)\equiv F_i(t)$)
\begin{equation*}
\begin{split}
&\Pfp\left(F_{1}(u_{1})\cdots F_{1}(u_a)\cdot
F_{2}(v_1)\cdots F_{2}(v_b)\right)=F_{1}(u_{1})\cdots F_{1}(u_a)\cdot
F_{2}(v_1)\cdots F_{2}(v_b)\\
&\quad -\tSym_{\,\bu}\frac{1}{(a-1)!}\Pfm\sk{F_{2,1}(u_1)}F_{1}(u_{2})\cdots F_{1}(u_a)\cdot
F_{2}(v_1)\cdots F_{2}(v_b)\\
&\quad -\tSym_{\,\bv}\frac{f(v_1,\bu)}{(b-1)!}\Pfm\sk{F_{3,2}(v_1)}
F_{1}(u_{1})\cdots F_{1}(u_a)\cdot
F_{2}(v_2)\cdots F_{2}(v_b)\\
&\quad -\tSym_{\,\bu,\bv}\frac{c^{-1}g(v_1,u_1)f(v_1,\bu_1)}{(a-1)!(b-1)!}
\Pfm\sk{F_{3,1}(v_1)}F_{1}(u_{2})\cdots F_{1}(u_a)\cdot
F_{2}(v_2)\cdots F_{2}(v_b)+\mathbb{W},
\end{split}
\end{equation*}
where again the terms denoted by $\mathbb{W}$  belong to the ideal $I$ after action of any
monodromy matrix element. Finally, using the normal ordering rule \r{FFor} for the
product of currents $\F_1$ we can replace these products by their
``positive'' projections to obtain
\begin{equation}\label{pf112}
\begin{split}
&\Pfp\left(F_{1}(u_{1})\cdots F_{1}(u_a)\cdot
F_{2}(v_1)\cdots F_{2}(v_b)\right)=\Pfp\sk{F_{1}(u_{1})\cdots F_{1}(u_a)}\cdot
F_{2}(v_1)\cdots F_{2}(v_b)\\
&\quad -\tSym_{\,\bv}\frac{f(v_1,\bu)}{(b-1)!}\Pfm\sk{F_{3,2}(v_1)}
\Pfp\sk{F_{1}(u_{1})\cdots F_{1}(u_a)}\cdot
F_{2}(v_2)\cdots F_{2}(v_b)\\
&\quad -\tSym_{\,\bu,\bv}\frac{c^{-1}g(v_1,u_1)f(v_1,\bu_1)}{(a-1)!(b-1)!}
\Pfm\sk{F_{3,1}(v_1)}\Pfp\sk{F_{1}(u_{2})\cdots F_{1}(u_a)}\cdot
F_{2}(v_2)\cdots F_{2}(v_b)+\mathbb{W},
\end{split}
\end{equation}
and we see that the terms containing the 'negative' projection of the
current $\Pfm\sk{F_{2,1}(u_1)}$ disappear from the final formula \r{pf112}.

 Now we prove the statement of  Proposition~\ref{prop44}
 in the general case using the normal ordering relation \r{FFor}. Taking into account the arguments above we write
 \begin{equation}\label{gcp441}
\begin{split}
&\Pfp\left(\F_1(\bar t^1)\cdots \F_{N-1}(\bar t^{N-1})\F_{N}(\bar t^{N})\right)=
\F_1(\bar t^1)\cdots \F_{N-1}(\bar t^{N-1})\F_{N}(\bar t^{N})\\
&\quad -\sum_{\ell=1}^N \comb_\ell\
\tSym_{\,\bar t^\ell,\cdots,\bar t^N} \left[f_{\prt{\ell}}(t^\ell_1,\bar t^{\ell-1})
\prod_{s=\ell}^{N-1} f_{\prt{s+1}}(t^{s+1}_1,\bar t^{s}_1)
\Pfm\left(F_\ell(t^\ell_1)\cdots F_N(t^N_1)\right)\right.\\
&\quad \left.\phantom{\prod_{s=\ell}^{N-1}}\times
\Pfp\left(\F_1(\bar t^1)\cdots \F_{\ell-1}(\bar t^{\ell-1})\cdot
\F_{\ell}(\bar t^{\ell}_1)\cdots
\F_{N}(\bar t^{N}_1)\right)\right]+\mathbb W,
\end{split}
\end{equation}
where we keep only the terms which contains $\Pfm\left(F_\ell(t^\ell_1)\cdots F_N(t^N_1)\right)$
as a source of the 'negative'  projection of the composed current
$\Pfm\left(F_{N+1,\ell}(t^N_1)\right)$, and $\mathbb W$ are the terms
which produce the elements of the ideal $I$ after action of any monodromy
matrix element $\LL^+_{i,j}(z)$. Using  \r{pmau1} we can replace \r{gcp441} by
\begin{equation}\label{gcp442}
\begin{split}
&\Pfp\left(\F_1(\bar t^1)\cdots \F_{N-1}(\bar t^{N-1})\F_{N}(\bar t^{N})\right)=
\F_1(\bar t^1)\cdots \F_{N-1}(\bar t^{N-1})\F_{N}(\bar t^{N})\\
&\quad -\sum_{\ell=1}^N  \comb_\el\ \tSym_{\,\bar t^\el,\ldots,\bar t^N}\left[
\fgf_\ell(\bar t^{\el-1},\ldots,\bar t^N)\   \cci{\el,N+1}^{-1}  \Pfm\sk{F_{N+1,\el}(t^N_1)}
\right.\\
&\quad \left.\phantom{\prod}\times
\Pfp\left(\F_1(\bar t^1)\cdots \F_{\ell-1}(\bar t^{\ell-1})\cdot
\F_{\ell}(\bar t^{\ell}_1)\cdots
\F_{N}(\bar t^{N}_1)\right)\right]+\mathbb W.
\end{split}
\end{equation}

Now we can use the result of the paper \cite{KhP-series} which states that
 only 'negative' projections of the
composed currents  $\Pfm\left(F_{N+1,\ell}(t^N_1)\right)$, $\ell=1,\ldots,N$ appear in the right hand side of the equation  \r{ser-res}. This allows
us to replace the first term $\F_1(\bar t^1)\cdots \F_{N-1}(\bar t^{N-1})\F_{N}(\bar t^{N})$
in the right hand side of \r{gcp442} by the term
 $\Pfp\sk{\F_1(\bar t^1)\cdots \F_{N-1}(\bar t^{N-1})}\F_{N}(\bar t^{N})$.

 Analogously, positive ``projections'' of the products of the composed currents
 \begin{equation*}
 \Pfp\left(\F_1(\bar t^1)\cdots \F_{\ell-1}(\bar t^{\ell-1})\cdot
\F_{\ell}(\bar t^{\ell}_1)\cdots
\F_{N}(\bar t^{N}_1)\right)
\end{equation*}
under the sum in \r{gcp442} can be replaced
by
\begin{equation*}
\Pfp\left(\F_1(\bar t^1)\cdots \F_{\ell-1}(\bar t^{\ell-1})\cdot
\F_{\ell}(\bar t^{\ell}_1)\cdots
\F_{N-1}(\bar t^{N-1}_1)\right)\cdot \F_{N}(\bar t^{N}_1),
\end{equation*}
and this replacement changes only the structure of the elements $\mathbb W$.
This finishes the proof of the Proposition~\ref{prop44}. \hfill\qed

Similarly,  using \r{ordF1} and \r{pmau2} we can prove following
\begin{proposition}
\begin{equation}\label{usre2}
\begin{split}
&\hPfp\sk{\ticF_N(\bar t^N) \cdots \ticF_1(\bar t^1)}= \Pfp\sk{\ticF_{N}(\bar t^{N}) \cdots
\ticF_2(\bar t^2)}\cdot
\ticF_1(\bar t^1)\\
&\quad -\sum_{\el=1}^{N} \  \tSym_{\,\bar t^1,\ldots,\bar t^{\el}}\left[
\hat\comb_{\el}\
  \hat\fgf_\el(\bar t^1,\ldots,\bar t^{\el+1})\  c_{\prt{1,\el+1}}^{-1} \hPfm\sk{\tiF_{\el+1,1}(t^1_{\rr_1})}
\right.\\
&\left.\quad\times\
\hPfp\sk{\ticF_N(\bar t^N)\cdots\ticF_{\el+1}(\bar t^{\el+1})
\ticF_{\el}(\bar t_{\rr_{\el}}^{\el})\cdots \ticF_{2}(\bar t_{\rr_2}^{2})}
\cdot \ticF_1(\bar t_{\rr_1}^1)\right]+\hat{\mathbb{W}},
\end{split}
\end{equation}
where for any $1\leq \el\leq N$ we introduce a rational function
\begin{equation}\label{zz2}
\hat\fgf_{\el}(\bar t^1,\ldots,\bar t^{\el+1}) = f_{\prt{\el+1}}(\bar t^{\el+1},t^{\el}_{\rr_{\el}})
\prod_{s=1}^{\el-1}g_{\prt{s+1}}(t_{\rr_{s+1}}^{s+1},t_{\rr_s}^{s})
f_{\prt{s+1}}(\bar t^{s+1}_{\rr_{s+1}},t_{\rr_s}^{s})\,,
\end{equation}
and a combinatorial factor
\begin{equation}\label{comb2}
\hat\comb_{\el}=\prod^{\el}_{s=1} \frac{1}{(r_s-1)!}\,.
\end{equation}
A symbol\ \
 $\hat{\mathbb{W}}$ denotes the terms which have the structure $\Pfm\sk{\tiF_{j_1,1}(w_1)}
\Pfm\sk{\tiF_{j_2,1}(w_2)}$.
\end{proposition}

Again, the action of any monodromy matrix
element $\LL^+_{i,j}(z)$ onto terms $\hat{\mathbb{W}}$
belongs to the ideal $\hat I$ due to \r{idca4}. The terms
which would not be in this ideal  are proportional to $\delta_{i,j_1}\delta_{1,j_2}$ and they vanish
due to the condition $1< j_2$.

\subsection{Action of the monodromy matrix element $\LL^+_{i,j}(z)$}
\label{ssec72}

Let us apply from the left the monodromy matrix element $\LL^+_{i,j}(z)$ to  equations
\r{usre1} and \r{usre2}.
As one can easily verify, the structure of the action formulas  differs significantly for the cases
$i\leq j$ and $i>j$.

The action of the monodromy matrix elements $\LL_{i,j}(z)$
for $i<j$  lead
to  recursions which relate the Bethe vectors with smaller number of Bethe
parameters with the ones with bigger numbers of these parameters.
If we prove that the action formulas for $i<j$ are the same for the Bethe vectors
$\BBB(\bar t)$ and $\hat\BBB(\bar t)$ this will provide the same recurrence relations
for these vectors. As a result we will prove that the vectors
$\BBB(\bar t)$ and $\hat\BBB(\bar t)$ coincide.

The action formulas for the diagonal monodromy
matrix elements $\LL_{i,i}(z)$ lead to the Bethe equations. They prove that Bethe vectors become
eigenvectors of the transfer matrix provided Bethe equations are satisfied.

Finally, the action formulas of
the monodromy matrix elements $\LL_{i,j}(z)$ with $i>j$ are necessary for calculating
the scalar products of the Bethe vectors.  This last problem is beyond the scope of the present paper, and  we will consider the general action formulas in this case in a separate publication.  From now on,  we restrict ourselves
to the action of the monodromy matrix elements $\LL_{i,j}(z)$ with $i\leq j$.

We introduce the shorthand notation
\begin{equation*}
\F_\el\equiv \F(\bar t^\el)\,,\quad
\F'_\el\equiv \F(\bar t_1^\el)\,,\quad
\F''_\el\equiv \F(\bar t_{\rr_\el}^\el)
\end{equation*}
and a similar notation for
\begin{equation*}
\ticF_\el\equiv \ticF(\bar t^\el)\,,\quad
\ticF'_\el\equiv \ticF(\bar t_1^\el)\,,\quad
\ticF''_\el\equiv \ticF(\bar t_{\rr_\el}^\el)\,,
\end{equation*}
where $\bar t^\el_{1}=\{\bar t^\el\}\setminus\{t^\el_1\}$ and
 $\bar t^\el_{\rr_\el}=\{\bar t^\el\}\setminus\{t^\el_{\rr_\el}\}$,
   are the sets of
Bethe parameters of the same type with  either the first or the last element omitted.

For any $\el$, $1\leq \el\leq N$ we introduce
two collections of  rational functions
\begin{equation}\label{zz11}
\begin{split}
\fgf^q_\el(\bar t^{\el-1},\ldots,\bar t^q) &=
 f_{\prt{\el}}(t^{\el}_1,\bar t^{\el-1})
\prod_{s=\el}^{q-1}g_{\prt{s+1}}(t_1^{s+1},t_1^{s}) f_{\prt{s+1}}(t^{s+1}_1,\bar t_1^{s})\,,
\quad \ell\leq q\leq N\,,\\
\hat\fgf^p_{\el}(\bar t^p,\ldots,\bar t^{\el+1}) &= f_{\prt{\el+1}}(\bar t^{\el+1},t^{\el}_{\rr_{\el}})
\prod_{s=p}^{\el-1}g_{\prt{s+1}}(t_{\rr_{s+1}}^{s+1},t_{\rr_s}^{s})
f_{\prt{s+1}}(\bar t^{s+1}_{\rr_{s+1}},t_{\rr_s}^{s})\,,
\quad 1\leq p\leq\ell.
\end{split}
\end{equation}
The rational functions in \r{zz1} and \r{zz2} are particular cases of the functions
\r{zz11}:
\begin{equation*}
\fgf_\el(\bar t)\equiv \fgf^N_\el(\bar t)\quad\mbox{and}\quad
\hat\fgf(\bar t)\equiv \hat\fgf^1_\el(\bar t)\,.
\end{equation*}

For $q=j+1,\ldots,N+1$ and $p=1,\ldots,i-1$
we also define the rational functions
\begin{equation}\label{ZZ-fun}
\ZZ_j^q(z;\bar t)=g(z,t_1^{q-1})\ \fgf_{j}^{q-1}(\bar t)\quad\mbox{and}\quad
\hat{\ZZ}_{i}^{p}(z;\bar t)=g(z,t_{\rr_{p}}^{p})\ \hat{\fgf}_{i-1}^{p}(\bar t).
\end{equation}
We extend these definition
 to $q=j$ and $p=i$ by setting
 $\ZZ_j^j(z;\bar t)=\hat{\ZZ}_{i}^{i}(z;\bar t)\equiv1$. Finally, let
\begin{equation*}
\comb^{\el'}_\el=\prod_{s=\el}^{\el'} \frac{1}{(r_s-1)!}\;.
\end{equation*}
Then the combinatorial factors given by the equations \r{comb1} and \r{comb2}
are
\begin{equation*}
\comb_\el\equiv\comb^N_\el\quad\mbox{and}\quad
\hat\comb_\el\equiv\comb^\el_1\,.
\end{equation*}

We have following
\begin{proposition}\label{prop74}
\begin{equation}\label{gc44}
\begin{split}
\LL^+_{i,j}(z) \cdot \Pfp\sk{\F_1\cdots \F_N}
 &\sim_{I,K} \sum_{q=j}^{N+1} \tSym_{\,\bar t^j,\ldots,\bar t^{q-1}}\left[
\phi_q\ \comb^{q-1}_j\ \ZZ^{q}_j(z;\bar t)\right. \\
&\times\left. \LL^+_{i,q}(z)\cdot
\F_1\cdots\F_{j-1}\F'_{j}\cdots \F'_{q-1}\right]\cdot \F_q\cdots \F_N
\end{split}
\end{equation}
 and
 \begin{equation}\label{gc66}
 \begin{split}
\LL^+_{i,j}(z) \cdot \hPfp\sk{\ticF_N\cdots \ticF_1}&\sim_{\hat{I},\hat{K}}
 \sum_{p=1}^{i} \tSym_{\,\bar t^{p},\ldots,\bar t^{i-1}}\left[
\hat\phi_p\ \comb^p_{i-1}\ \hat{\ZZ}_{i}^{p}(z;\bar t)\right.\\
& \times \left. \LL^+_{p,j}(z)\cdot
\ticF_N\cdots\ticF_{i}\ticF''_{i-1}\cdots \ticF''_{p}\right]\cdot \ticF_{p-1}\cdots \ticF_1,
\end{split}
\end{equation}
where the sign factors $\phi_q$ for $q=j+1,\ldots,N+1$ and $\hat\phi_p$ for
$p=1,\ldots,i-1$ are given by  \r{phases}, and we set
$\phi_j=\hat\phi_i\equiv1$.
\end{proposition}

\proof\ We begin the proof with the equation \r{usre1}.
Assume that $j=N+1$. Then due to \r{idca3} the sum over $\el$ in the right
hand side of \r{usre1}  as well as the terms  $\mathbb{W}$  generate
the elements from the ideal $I$ under the  action of
$\LL^+_{i,N+1}(z)$. As the result we obtain
\begin{equation}\label{Np1}
\LL^+_{i,N+1}(z)\cdot  \Pfp\sk{\F_1(\bar t^1)\cdots \F_N(\bar t^N)}\ \sim_{I,K}\
\LL^+_{i,N+1}(z)\cdot \Pfp\sk{\F_1(\bar t^1)\cdots \F_{N-1}(\bar t^{N-1})}\cdot
\F_N(\bar t^N).
\end{equation}
Using again \r{usre1} for the projection $\Pfp\sk{\F_1(\bar t^1)\cdots \F_{N-1}(\bar t^{N-1})}$  we can continue to obtain
\begin{equation}\label{Np2}
\LL^+_{i,N+1}(z)\cdot  \Pfp\sk{\F_1(\bar t^1)\cdots \F_N(\bar t^N)}\ \sim_{I,K}\
\LL^+_{i,N+1}(z)\cdot \F_1(\bar t^1)\cdots
\F_N(\bar t^N).
\end{equation}

Assume now that $j\leq N$. Then besides first terms as in \r{Np1} and due to \r{idca3}
there will be  a contribution of one term from the sum in the right hand side of  \r{usre1}
corresponding to  $\ell=j$, so that
\begin{equation}\label{gc1}
\begin{split}
&\LL^+_{i,j}(z) \cdot \Pfp\sk{\F_1\cdots \F_N}\sim_{I,K}
\LL^+_{i,j}(z) \cdot \Pfp\sk{\F_1\cdots \F_{N-1}}\cdot
\F_N\ \\
&+  \tSym_{\,\bar t^j,\ldots,\bar t^N}\left[
\phi_{N+1}g(z,t_1^N)\ \comb^N_j\ \fgf_{j}^{N}(\bar t) \ \LL^+_{i,N+1}(z)\cdot
\F_1\cdots\F_{j-1}\F'_{j}\cdots \F'_{N-1}
\cdot \F'_N\right].
\end{split}
\end{equation}
Observe that due to \r{Np2} and \r{idca3} we can relax the projection on the product of the
currents $ \F_1\cdots\F_{j-1}\F'_{j}\cdots \F'_{N-1}$.

We leave the second term in the right hand side of \r{gc1} as it is and consider the
first one. In the first term we have the projection $\Pfp\sk{\F_1\cdots \F_{N-1}}$
and we can use again the presentation \r{usre1} for the product of  currents
for the smaller rank algebra $\mathfrak{gl}(m|n-1)$. Again the only contribution is
a regular term and one term from the sum over $\el$ with $\el=j$. We obtain
\begin{equation*}
\begin{split}
&\LL^+_{i,j}(z) \cdot \Pfp\sk{\F_1\cdots \F_N}\sim_{I,K}
\LL^+_{i,j}(z) \cdot \Pfp\sk{\F_1\cdots \F_{N-2}}\cdot \F_{N-1}
\F_N\ \\
&+  \tSym_{\,\bar t^j,\ldots,\bar t^{N-1}}\left[
\phi_N\ g(z,t_1^{N-1})\ \comb^{N-1}_j\ \fgf_{j}^{N-1}(\bar t) \ \LL^+_{i,N}(z)\cdot
\F_1\cdots\F_{j-1}\F'_{j}\cdots \F'_{N-1}\right]
\cdot \F_N\\
& +  \tSym_{\,\bar t^j,\ldots,\bar t^N}\left[
\phi_{N+1}\ g(z,t_1^N)\ \comb^N_j\ \fgf_{j}^{N}(\bar t) \ \LL^+_{i,N+1}(z)\cdot
\F_1\cdots\F_{j-1}\F'_{j}\cdots
\cdot \F'_N\right].
\end{split}
\end{equation*}
Continuing this process we conclude that the action of the monodromy matrix
element $\LL^+_{i,j}(z)$ onto the projection $\Pfp\sk{\F_1\cdots \F_N}$
modulo elements from the ideals $I$ and $K$ is given by
\begin{equation}\label{gc4}
\begin{split}
\LL^+_{i,j}(z) \cdot \Pfp\sk{\F_1\cdots \F_N}&\sim_{I,K}
\LL^+_{i,j}(z) \cdot \F_1\cdots  \F_N\ \\
&+ \sum_{q=j+1}^{N+1} \tSym_{\,\bar t^j,\ldots,\bar t^{q-1}}\left[
\phi_q\ g(z,t_1^{q-1})\ \comb^{q-1}_j\ \fgf_{j}^{q-1}(\bar t)\right.\\
& \times \left. \LL^+_{i,q}(z)\cdot
\F_1\cdots\F_{j-1}\F'_{j}\cdots \F'_{q-1}\right]\cdot \F_q\cdots \F_N.
\end{split}
\end{equation}

Using relation \r{usre2} and similar argument as above we obtain
\begin{equation}\label{gc6}
\begin{split}
\LL^+_{i,j}(z) \cdot \hPfp\sk{\ticF_N\cdots \ticF_1}&\sim_{\hat{I},\hat{K}}
\LL^+_{i,j}(z) \cdot \ticF_N\cdots  \ticF_1\ \\
& + \sum_{p=1}^{i-1} \tSym_{\,\bar t^{p},\ldots,\bar t^{i-1}}\left[
\hat\phi_p\ g(z,t_{\rr_{p}}^{p})\ \hat\comb^p_{i-1}\
\hat{\fgf}_{i-1}^{p}(\bar t)\right.\\
& \times\left. \LL^+_{p,j}(z)\cdot
\ticF_N\cdots\ticF_{i}\ticF''_{i-1}\cdots \ticF''_{p}\right]\cdot \ticF_{p-1}\cdots \ticF_1 .
\end{split}
\end{equation}
With the notation \r{ZZ-fun} formulas \r{gc4} and
\r{gc6} are equivalent to the statement of the Proposition~\ref{prop74}.
\hfill\qed

The next step is to use the explicit expressions for the monodromy matrix element
$\LL^+_{i,j}(z)$ in terms of the Gauss coordinates \r{GF2}
\begin{equation}\label{gde1}
\LL^{+}_{i,q}(z)=\sum_{1\leq p\leq i}
\FF^{+}_{q,p}(z)k^+_{p}(z)\EE^{+}_{p,i}(z)
\end{equation}
and in terms of the 'hatted' Gauss coordinates \r{GF1}
\begin{equation}\label{gde2}
\LL^{+}_{p,j}(z)=\sum_{j\leq q\leq N+1} (-)^{(\prt{q}+\prt{p})(\prt{q}+\prt{j})}
\tFF^{+}_{q,p}(z)\tk^+_{q}(z)\tEE^{+}_{j,q}(z),
\end{equation}
where we formally have set the Gauss coordinates
$\FF^{+}_{i,i}(z)=\tFF^{+}_{j,j}(z)=\EE^{+}_{i,i}(z)=\tEE^{+}_{j,j}(z)\equiv 1$.
These representations give us a possibility to move the Gauss coordinates $\EE^{+}_{p,i}(z)$ and
$\tEE^{+}_{j,q}(z)$ through the corresponding products of the currents.

 As will be demonstrated below,
for $i\leq j$ these commutations  transform the product of the currents  in \r{gc44}
into
\begin{equation}\label{gc7}
\F_1\cdots\F_{p-1}\cdot\F''_{p}\cdots\F''_{i-1}\cdot\F_i\cdots \F_{j-1}\cdot \F'_j\cdots
\F'_{q-1}\cdot\F_{q}\cdots \F_{N},
\end{equation}
and the product of the currents in \r{gc66} into
\begin{equation}\label{gc8}
\ticF_N\cdots\ticF_{q}\cdot\ticF'_{q-1}\cdots\ticF'_{j}\cdot\ticF_{j-1}\cdots
\ticF_{i}\cdot \ticF''_{i-1}\cdots\ticF''_{p}\cdot\ticF_{p-1}\cdots \ticF_1,
\end{equation}
for $p=1,\ldots,i$ and
$q=j,\ldots,N+1$.

The Gauss coordinates  $\FF^{+}_{q,p}(z)$ and $\tFF^{+}_{q,p}(z)$ can
be replaced by the total composed currents $F_{q,p}(z)$ and $\tiF_{q,p}(z)$
according to \r{Fzm2} and \r{tFzm2} modulo terms from the ideal $I$.
Then, due to \r{cFr} and \r{tcFr} the products of the
currents in \r{gc7} and \r{gc8}
\begin{equation*}
\F''_{p}\cdots\F''_{i-1}\cdot\F_i\cdots \F_{j-1}\cdot \F'_j\cdots
\F'_{q-1}
\end{equation*}
and
\begin{equation}\label{gc88}
\ticF'_{q-1}\cdots\ticF'_{j}\cdot\ticF_{j-1}\cdots
\ticF_{i}\cdot \ticF''_{i-1}\cdots\ticF''_{p}
\end{equation}
 will be extended by the simple root currents depending on the auxiliary parameter $z$.

 This consideration shows that the action of the monodromy matrix element $\LL^+_{i,j}(z)$
 onto the projections of currents $\Pfp\sk{\F_1\cdots \F_N}$ and
$ \hPfp\sk{\ticF_N\cdots \ticF_1}$ have a similar structure.  This is the first sign that
the recursions for the Bethe vectors \r{uBV1} and \r{bv}  coincide.

Let us be more precise. Due to \r{Ezm2} the Gauss coordinate $\EE^+_{p,i}(z)$
commute with all the products of currents $\F_q\cdots\F_{j-1}$ except $\F_p\cdots\F_{i-1}$.
This is because  the Gauss coordinate $\EE^+_{p,i}(z)$ is constructed from the modes
of the currents $E_p(z),E_{p+1}(z),\ldots,E_{i-1}(z)$ due to \r{Ezm2}.
From the commutation relation \r{EF} for the simple root total currents we  obtain
the commutation relations of the simple root Gauss coordinates
\begin{equation*}
\begin{split}
[ \EE^+_{i,i+1}(v),\FF^+_{i+1,i}(u)\}&
=\frac{c_{\prt{i+1}}}{(v-u)_\lessgtr}\sk{k^+_{i+1}(v)k^+_i(v)^{-1}- k^+_{i+1}(u)k^+_i(u)^{-1}},\\
[ \EE^+_{i,i+1}(v),\FF^-_{i+1,i}(u)\}&
=\frac{c_{\prt{i+1}}}{(v-u)_>}\sk{k^+_{i+1}(v)k^+_i(v)^{-1}- k^-_{i+1}(u)k^-_i(u)^{-1}},
\end{split}
\end{equation*}
which also follows from \r{TM-2}. Then  we conclude that
\begin{equation*}
[\EE^+_{p,p+1}(z),F_p(t)\}\ \sim_{K}\ g_{\prt{p+1}}(t,z)\psi^+_{p}(t),
\end{equation*}
where
\begin{equation*}
\psi^+_{p}(t)=k^+_{p+1}(t)k^+_{p}(t)^{-1}.
\end{equation*}
We recall that $[\cdot\,,\,\cdot\}$ is the graded
commutator defined in \eqref{EF}.
Using this commutation relation, the commutation of the Cartan currents with
the total currents $F_p(t)$, and the definition of deformed symmetrization \r{sym}
we have
\begin{equation}\label{gc10}
[\EE^+_{p,p+1}(z),\F_p(\bar t^p)\}\ \sim_{K}\
\frac{(-)^{(r_p-1)\delta_{p,m}}}{(\rr_p-1)!}\ \tSym_{\, \bar t^p } \left[
g_{\prt{p+1}}(t^p_{\rr_p},z)\ \F_{p}(\bar t^p_{\rr_p})\ \psi^+_{p}(t^p_{\rr_p})  \right].
\end{equation}

Let us explain the appearance of the phase factor $(-)^{(r_m-1)}$ in \r{gc10} at $p=m$.
Using the definition of the graded commutator in \r{EF}, the commutativity
$\psi^+_m(t)F_m(t')=F_m(t')\psi^+_m(t)$, and the anti-commutativity of the currents $F_m(t)$ we conclude that
\begin{equation*}
\begin{split}
[\EE^+_{m,m+1}(z),\F_m(\bar t^m)\}\ &\!\sim_{K}\ \sum_{\el=1}^{\rr_m}
(-)^{\el-1}g(z,t^m_\el) F_m(t^m_{1})\cdots F_m(t^m_{\el-1})\psi^+_{m}(t^m_\el)F_m(t^m_{\el+1})\cdots
F_m(t^m_{\rr_m})\\
&\!\sim_{K}\  \frac{(-)^{(\rr_m-1)}}{(\rr_m-1)!}\ {\rm ASym}_{\ \bar t^m}\sk{
g(z,t^m_{\rr_m})F_m(t^m_{1})\cdots F_m(t^m_{\rr_m-1})\psi_m^+(t^m_{\rr_m})},
\end{split}
\end{equation*}
where symbol ${\rm ASym}_{\ \bar t^m}(\cdot)$ stands for an anti-symmetrization over the set of the
variables $\bar t^m$. It coincides with deformed symmetrization ${\tSym}_{\ \bar t^m}(\cdot)$
(see equation \eqref{sym}), over the same set.

Within the product of  screening operators  $\Sc_{E_{i-1}\mode0}\cdots \Sc_{E_{p+1}\mode0}$
entering the formula \r{Ezm2} for the Gauss coordinate $\EE^+_{p,i}(z)$ only
the screening operator $\Sc_{E_{p+1}\mode0}$ does not commute with the Cartan current
 $k^+_{p+1}(t^p_{\rr_p})$:
 \begin{equation*}
 \Sc_{E_{p+1}\mode0}\sk{k^+_{p+1}(t^p_{\rr_p})}=-\cci{p+1}k^+_{p+1}(t^p_{\rr_p})
 \EE^+_{p+1,p+2}(t^p_{\rr_p})
 \end{equation*}
 which can be obtained from the commutation relation \r{kiE}.
Using again relation \r{Ezm2}  we obtain that
\begin{equation}\label{gc12}
[\EE^+_{p,i}(z),\F_p(\bar t^p)\}\ \sim_{K}\
\frac{(-)^{(r_p-1)\delta_{p,m}}}{(\rr_p-1)!}\ \tSym_{\, \bar t^p } \left[
g_{\prt{p+1}}(t^p_{\rr_p},z)\ \F_{p}(\bar t^p_{\rr_p})\ \psi^+_{p}(t^p_{\rr_p})
 \EE^+_{p+1,i}(t^p_{\rr_p}) \right].
\end{equation}

Due to
\begin{equation*}
\EE^+_{p,i}(z)\cdot \F_{p+1}\cdots\F_{i-1}\ \sim_{J}\ 0
\end{equation*}
we can rewrite  \r{gc12} as an action of the Gauss coordinate
$\EE^+_{p,i}(z)$ onto the product of  currents $\F_p\cdots\F_{i-1}$ modulo
elements in the ideals $K$ and $J$
\begin{equation}\label{gc14}
\begin{split}
\EE^+_{p,i}(z)\cdot\F_p(\bar t^p)\cdots \F_{i}(\bar t^{i})
&\sim_{K,J}\
\frac{(-)^{(r_p-1)\delta_{p,m}}}{(\rr_p-1)!}\ \tSym_{\, \bar t^p } \left[
g_{\prt{p+1}}(t^p_{\rr_p},z)\ \F_{p}(\bar t^p_{\rr_p})\ \psi^+_{p}(t^p_{\rr_p})
\right. \\
&\qquad\qquad \times \left.
 \EE^+_{p+1,i}(t^p_{\rr_p})\cdot \F_{p+1}(\bar t^{p+1})\cdots \F_{i}(\bar t^{i})
 \right].
\end{split}
\end{equation}
In the last line of \r{gc14} we can use again relation \r{gc12} and  iterating the calculations,   we obtain finally
\begin{equation}\label{gc15}
\begin{split}
&\EE^+_{p,i}(z)\cdot\F_p(\bar t^p)\cdots  \F_{i-1}(\bar t^{i-1})
\F_{i}(\bar t^{i}) \ \sim_{K,J}\ \epsilon_p\ \prod_{s=p}^{i-1} (-)^{(r_s-1)\delta_{s,m}} \\
&\qquad\times
\tSym_{\,\bar t^p,\ldots,\bar t^{i-1}}\left[\comb_p^{i-1}\  \hat{\ZZ}_{i}^{p}(z;\bar t)
\F_p(\bar t^p_{\rr_p})\cdots \F_{i-1}(\bar t^{i-1}_{\rr_{i-1}})\F_i(\bar t^i)
\prod_{s=p}^{i-1}
k^+_{s+1}(t^s_{\rr_s})k^+_{s}(t^s_{\rr_s})^{-1} \right],
\end{split}
\end{equation}
where $\epsilon_p$ is the sign factor
\begin{equation}\label{eps}
\epsilon_i=1\quad\mbox{and}\quad \epsilon_p=(-)^{1+\prt{i}}\quad
\mbox{for}\quad p=1,2,\ldots,i-1.
\end{equation}
Recall that the rational function $\hat{\ZZ}_{i}^{p}(z;\bar t)$ is defined by
\r{zz11} and \r{ZZ-fun}.

Similarly, taking into account that the Gauss coordinate
$\tEE^+_{j,q}(z)$ does not commute only with the product of  currents
$\ticF_{q-1}(\bar t^{q-1})\cdots \ticF_j(\bar t^j)$  in the product \r{gc88} we find
\begin{equation}\label{gc16}
\begin{split}
&(-)^{(\prt{q}+\prt{p})(\prt{q}+\prt{j})}
\tEE^+_{j,q}(z) \cdot \ticF_{q-1}(\bar t^{q-1})\cdots \ticF_j(\bar t^j)
\ticF_{j-1}(\bar t^{j-1}) \ \sim_{\hat K,\hat J}\ \hat\epsilon_q\ \prod_{s=j}^{q-1} (-)^{(r_s-1)\delta_{s,m}} \\
&\times
\tSym_{\,\bar t^{j},\ldots,\bar t^{q-1}}\left[\comb^{q-1}_j\ {\ZZ}_{j}^{q}(z;\bar t)
\ticF_{q-1}(\bar t^{q-1}_1)\cdots \ticF_{j}(\bar t^{j}_{1})\ticF_{j-1}(\bar t^{j-1})
\prod_{s=j}^{q-1}
\tk^+_{s}(t^s_{1})\tk^+_{s+1}(t^s_{1})^{-1} \right],
\end{split}
\end{equation}
where $\hat\epsilon_q$ is the sign factor
\begin{equation}\label{heps}
\hat\epsilon_j=1\quad\mbox{and}\quad \hat\epsilon_q=
(-)^{(\prt{j}+\prt{p})\prt{q}+\prt{j}\prt{p}} \quad
\mbox{for}\quad q=j+1,j+2,\ldots,N,
\end{equation}
and the rational function ${\ZZ}_{j}^{q}(z;\bar t)$ is defined by the formulas \r{zz11} and
\r{ZZ-fun}.

The Gauss coordinates $\FF^+_{q,p}(z)$ and $\tFF^+_{q,p}(z)$ entering
formulas \r{gde1} and \r{gde2} can be
replaced by the product of the corresponding currents (see formulas \r{cFr}, \r{Fzm2}
and \r{tcFr}, \r{tFzm1} respectively)
\begin{equation}\label{gc17}
\FF^+_{q,p}(z)\ \sim_{I}\ \left.\prod_{s=p}^{q-2} f_{\prt{s+1}}(z_{s+1},z_s)^{-1}\
F_p(z_p)\cdots F_{q-1}(z_{q-1})\right|_{z_p=\cdots=z_{q-1}=z},
\end{equation}
\begin{equation}\label{gc18}
\tFF^+_{q,p}(z)\ \sim_{\hat I}\ \left.\prod_{s=p}^{q-2} f_{\prt{s+1}}(z_{s+1},z_s)^{-1}\
\tiF_{q-1}(z_{q-1})\cdots \tiF_{p}(z_{p})\right|_{z_p=\cdots=z_{q-1}=z},
\end{equation}
where we interchanged the orders in the products of  currents and introduced
an auxiliary set of  variables $\bar z=\{z_p,\ldots,z_{q-1}\}$, which at the
end should be all equal to the parameter $z$.

Combining  formula \r{gc44}, the Gauss decomposition
\r{gde1}, the action \r{gc15} of the Gauss coordinates
$\EE^+_{p,i}(z)$  and  formula
\r{gc17} we can calculate the action formulas of the monodromy matrix element
$\LL^+_{i,j}(z)$ onto the unnormalized Bethe vector
\begin{equation*}
\cB(\bar{t})=\Pfp\sk{\F(\bar t)}\prod_{\el=1}^N \lambda_\el(\bar t^\el) \rvec,
\end{equation*}
where the ordered product of the simple root currents $\F(\bar t)$ is given by the formula \r{FF-ff}. We have
\begin{equation}\label{gc19}
\begin{split}
&\LL^+_{i,j}(z)\cdot \cB(\bar{t})=\sum_{p=1}^i\sum_{q=j}^{N+1}
\phi_q\ \epsilon_p\ \comb_{p}^{i-1}\ \comb^{q-1}_j \ \prod_{s=p}^{i-1} (-)^{(r_s-1)\delta_{s,m}} \\
&\quad\times \tSym_{\,\bar t^p,\ldots,\bar t^{i-1},\bar t^{j},\ldots,\bar t^{q-1}}
\left[\frac{
\hat\ZZ_{i}^p(z;\bar t^{p},\ldots,\bar t^i)\ \ZZ_j^{q}(z;\bar t^{j-1},\ldots,
\bar t^{q-1})}{\XX(\bar z;\bar t^p,\ldots,\bar t^{q-1})}
\right.\\[2mm]
&\quad\times
\cB(\bar t^1,\ldots,\bar t^{p-1},
\{z_p,\bar t^{p}_{\rr_{p}} \},\ldots,\{z_{i-1},\bar t^{i-1}_{\rr_{i-1}} \},
\{z_i,\bar t^{i} \},\ldots,\{z_{j-1},\bar t^{j-1} \},\\
&\qquad\qquad\qquad
\{z_j,\bar t^{j}_{1} \},\ldots,\{z_{q-1},\bar t^{q-1}_1 \},
\bar t^q,\ldots,\bar t^N)\\[2mm]
&\quad\qquad\qquad\times \left.\left.
\frac{ \lambda_{p+1} (t^p_{\rr_{p}}) \cdots \lambda_{i} (t^{i-1}_{\rr_{i-1}})
\lambda_{j}(t^j_{1})\cdots \lambda_{q-1}(t^{q-1}_{1}) }
{\lambda_{p}(z_{p})\cdots \lambda_{q-1}(z_{q-1})}
\lambda_{p}(z) \right]\right|_{z_p=\cdots=z_{q-1}=z},
\end{split}
\end{equation}
where we introduce one more rational function $\XX(\bar z,\bar t^p,\ldots,\bar t^{q-1})$
depending on the auxiliary set $\bar z$ and the Bethe parameters
\begin{equation}\label{gc20}
\begin{split}
&\XX(\bar z;\bar t^p,\ldots,\bar t^{q-1})=
\prod_{s=p}^{i-1} f_{\prt{s+1}}(z_{s+1},\{z_s,\bar t^s_{\rr_s}\})\\
&\quad\times \prod_{s=i}^{j-1} f_{\prt{s+1}}(z_{s+1},\{z_s,\bar t^s\})
\prod_{s=j}^{q-2} f_{\prt{s+1}}(z_{s+1},\{z_s,\bar t^s_{1}\})
\ f_{\prt{p}}(\bar t^p_{\rr_p},z_p)^{-1}.
\end{split}
\end{equation}

Similarly, using \r{gc66}, the Gauss decomposition
\r{gde2}, the action \r{gc16} of the Gauss coordinates
$\tEE^+_{j,q}(z)$,  and formula
 \r{gc18}, we can
calculate the action formula of the monodromy matrix element
$\LL^+_{i,j}(z)$ onto the unnormalized Bethe vector
\begin{equation*}
\hcB(\bar{t})=\hPfp\sk{\hat\F(\bar t)}\prod_{\el=1}^N \tk^+_{\el+1}(\bar t^\el) \rvec,
\end{equation*}
where the ordered product of the currents $\hat\F(\bar t)$ is given by \r{tFF-EE}. We obtain
\begin{equation}\label{gc21}
\begin{split}
&\LL^+_{i,j}(z)\cdot \hcB(\bar{t})=\sum_{p=1}^i\sum_{q=j}^{N+1}
\hat\phi_p\ \hat\epsilon_q
\ \comb_p^{i-1}\ \comb^{q-1}_j\ \prod_{s=j}^{q-1} (-)^{(r_s-1)\delta_{s,m}} \\
&\quad\times \tSym_{\,\bar t^p,\ldots,\bar t^{i-1},\bar t^{j},\ldots,\bar t^{q-1}}
\left[\frac{
\hat\ZZ_{i}^p(z;\bar t^{p},\ldots,\bar t^i)\ \ZZ_j^{q}(z;\bar t^{j-1},\ldots,
\bar t^{q-1})}{\hat{\XX}(\bar z;\bar t^p,\ldots,\bar t^{q-1})}
\right.\\[2mm]
&\quad\times
\hcB(\bar t^1,\ldots,\bar t^{p-1},
\{z_p,\bar t^{p}_{\rr_{p}} \},\ldots,\{z_{i-1},\bar t^{i-1}_{\rr_{i-1}} \},
\{z_i,\bar t^{i} \},\ldots,\{z_{j-1},\bar t^{j-1} \},\\
&\qquad\qquad\qquad
\{z_j,\bar t^{j}_{1} \},\ldots,\{z_{q-1},\bar t^{q-1}_1 \},
\bar t^q,\ldots,\bar t^N)\\[2mm]
&\quad\qquad\qquad\times \left.\left.
\frac{ \lambda_{p+1} (t^p_{\rr_{p}})  \cdots \lambda_{i} (t^{i-1}_{\rr_{i-1}})
\lambda_{j}(t^j_{1})\cdots \lambda_{q-1}(t^{q-1}_{1}) }
{\lambda_{p+1}(z_{p})\cdots \lambda_{q}(z_{q-1})}
 \lambda_{q}(z)\right]\right|_{z_p=\cdots=z_{q-1}=z}\;,
\end{split}
\end{equation}
where we introduce another rational function $\hat\XX(\bar z,\bar t^p,\ldots,\bar t^{q-1})$
depending on the auxiliary set $\bar z$ and the Bethe parameters
\begin{equation}\label{gc22}
\begin{split}
&\hat\XX(\bar z,\bar t^p,\ldots,\bar t^{q-1})=
\prod_{s=p}^{i-2} f_{\prt{s+1}}(\{z_{s+1},\bar t^{s+1}_{\rr_{s+1}}\},z_{s})\\
&\quad\times \prod_{s=i-1}^{j-2} f_{\prt{s+1}}(\{z_{s+1},\bar t^{s+1}\},z_{s})
\prod_{s=j-1}^{q-2} f_{\prt{s+1}}(\{z_{s+1},\bar t^{s+1}_{1}\},z_{s})
\ f_{\prt{q}}(z_{q-1},\bar t^{q-1}_{1})^{-1}.
\end{split}
\end{equation}

Let us compare the phase factors entering the first rows of \r{gc19} and
\r{gc21}. Using definitions of these factors given by the formulas
\r{phases}, \r{eps} and \r{heps} we observe that $\hat\phi_p=\epsilon_p$ for
$p=1,\ldots,i$. On the other hand,
\begin{equation}\label{gc23}
\phi_q=(-)^{\prt{q}\prt{j}+(\prt{q}+\prt{j})\prt{i}}
\end{equation}
seems to be different from
 \begin{equation}\label{gc24}
\hat\epsilon_q=(-)^{\prt{q}\prt{j}+(\prt{q}+\prt{j})\prt{p}}.
\end{equation}
However, this is not true
because of the restrictions between on $p$, $i$, $j$, and $q$.
If the parities of the indices $\prt{p}$ and $\prt{i}$ coincide, then the factors
\r{gc23} and \r{gc24} coincide.
Let now consider the case where the  parities of the indices $\prt{p}$ and $\prt{i}$ are different.
Recall that $p\leq i$. Due to the definition of the grading (see \r{grade}), this means that  $\prt{p}=0$
and $\prt{i}=1$. But we consider in this section the action of the diagonal and
the upper triangular  monodromy matrix elements $\LL^+_{i,j}(z)$ onto Bethe vectors. It means that
there is a restriction $p\leq i\leq j\leq q$ so that if   $\prt{p}\not=\prt{i}$
then $\prt{j}=\prt{q}=1$ and both factors in \r{gc23} and \r{gc24} are equals to $-1$.
Further on we  denote these  phase factors
\begin{equation}\label{upf}
\phi_q\epsilon_p=\hat\phi_p
\hat\epsilon_q=\varphi_{p,q}.
\end{equation}

Now we can restore the normalizations of the Bethe vectors \r{uBV1} and
\r{bv} and observe that the actions of the diagonal and the upper triangular
monodromy matrix elements onto these Bethe vectors yield the same recurrence
relations. This  means that the Bethe vectors given by the formulas
\r{uBV1} and  \r{bv} coincide.

We start to restore the normalization of the Bethe vectors \r{gc19} using
 equation \r{FFnor}.
Note that the deformed symmetrization in the formula of the action \r{gc21}  turns
into the usual symmetrization in \r{gc25} due to the property \r{sy-p}.
Using explicit expressions for the rational functions
\r{zz1}, \r{zz2}, and \r{gc20} we obtain
\begin{equation}\label{gc25}
\begin{split}
\LL^+_{i,j}(z)\cdot \BBB(\bar{t})&=\sum_{p=1}^i\sum_{q=j}^{N+1}
\varphi_{p,q}\ \comb^{i-1}_p\ \comb^{q-1}_j  \\
&\quad\times \Sym_{\,\bar t^p,\ldots,\bar t^{i-1},\bar t^{j},\ldots,\bar t^{q-1}}
\Big[\DD(\bar t)\
\YY(z,\bar t)\ \Lambda(z;\bar t)\ \BBB(\{z,\bar t\}')\Big],
\end{split}
\end{equation}
where
the sign factor $\varphi_{p,q}$ is given by  \r{upf},
and the Bethe vector $\BBB(\{z,\bar t\}')$ in the right hand side of this
equation depends on the following set of
parameters:
\begin{equation*}
\begin{split}
\{z,\bar t\}' &=\{\bar t^1,\ldots,\bar t^{p-1},
\{z,\bar t^{p}_{\rr_{p}} \},\ldots,\{z,\bar t^{i-1}_{\rr_{i-1}} \},
\{z,\bar t^{i} \},\ldots,\{z,\bar t^{j-1} \},\\
&\qquad\qquad\qquad \{z,\bar t^{j}_{1} \},\ldots,\{z,\bar t^{q-1}_1 \},
\bar t^q,\ldots,\bar t^N\}\,.
\end{split}
\end{equation*}

The rational function $\DD(\bar t)$ is  given by the product
\begin{equation}\label{DD1}
\DD(\bar t)=
\prod_{s=p}^{i-1}\frac{f_{\prt{s}}(t^s_{\rr_s},\bar t^s_{\rr_s})}
{[(-)^{(r_s-1)}h(t^s_{\rr_s},\bar t^s_{\rr_s})]^{\delta_{s,m}}}
 \prod_{s=j}^{q-1}\frac{f_{\prt{s}}(\bar t^s_{1},t^s_{1}) }
{h(\bar t^s_{1},t^s_1)^{\delta_{s,m}}}.
\end{equation}
 The form of the two other rational functions $\YY(z,\bar t)$ and $\Lambda(z;\bar t)$ strongly depends on the values
of $p$ and $q$.  For $p<i$ and $q>j$
\begin{equation*}
\begin{split}
\YY(z,\bar t)&=  f_{\prt{p}}(z,\bar t^{p-1})\ f_{\prt{q}}(\bar t^{q},z)
\prod_{s=p}^{i-1}h(\bar t^s_{\rr_s},z)^{\delta_{s,m}}
\prod_{s=i}^{j-1}h(\bar t^s,z)^{\delta_{s,m}}
\prod_{s=j}^{q-1}h(\bar t^s_{1},z)^{\delta_{s,m}}\\
&\quad\times
\frac{g(z,t^p_{\rr_p})\prod_{s=p}^{i-2}  g_{\prt{s+1}}(t^{s+1}_{\rr_{s+1}},t^{s}_{\rr_{s}})}
{\prod_{s=p-1}^{i-2} f_{\prt{s+1}}(t^{s+1}_{\rr_{s+1}},\bar t^{s}) }\
\frac{g(z,t^{q-1}_1) \prod_{s=j}^{q-2}  g_{\prt{s+1}}(t^{s+1}_{1},t^{s}_{1}) }
{\prod_{s=j}^{q-1}  f_{\prt{s+1}}(\bar t^{s+1}, t^{s}_1)}\,,
\end{split}
\end{equation*}
\begin{equation*}
\Lambda(z;\bar t)=\frac{ \lambda_{p+1} (t^p_{\rr_{p}}) \cdots  \lambda_{i} (t^{i-1}_{\rr_{i-1}})
 \lambda_{j}(t^j_{1})\cdots  \lambda_{q-1}(t^{q-1}_{1}) }
{ \lambda_{p+1}(z)  \cdots  \lambda_{q-1}(z)},
\end{equation*}
for $p=i$ and $q>j$
\begin{equation*}
\YY(z,\bar t)=  f_{\prt{i}}(z,\bar t^{i-1})\ f_{\prt{q}}(\bar t^{q},z)
\prod_{s=i}^{j-1}h(\bar t^s,z)^{\delta_{s,m}}
\prod_{s=j}^{q-1}h(\bar t^s_{1},z)^{\delta_{s,m}}\
\frac{g(z,t^{q-1}_1) \prod_{s=j}^{q-2}  g_{\prt{s+1}}(t^{s+1}_{1},t^{s}_{1}) }
{\prod_{s=j}^{q-1}  f_{\prt{s+1}}(\bar t^{s+1}, t^{s}_1)}\,,
\end{equation*}
\begin{equation*}
\Lambda(z;\bar t)=\frac{  \lambda_{j}(t^j_{1})\cdots  \lambda_{q-1}(t^{q-1}_{1}) }
{( \lambda_{i+1}(z)  \cdots  \lambda_{q-1}(z))^{\theta_{i+1,q-1}}},
\end{equation*}
for $p<i$ and $q=j$
\begin{equation*}
\YY(z,\bar t)=  f_{\prt{p}}(z,\bar t^{p-1})\ f_{\prt{j}}(\bar t^{j},z)
\prod_{s=p}^{i-1}h(\bar t^s_{\rr_s},z)^{\delta_{s,m}}
\prod_{s=i}^{j-1}h(\bar t^s,z)^{\delta_{s,m}}\
\frac{g(z,t^p_{\rr_p})\prod_{s=p}^{i-2}  g_{\prt{s+1}}(t^{s+1}_{\rr_{s+1}},t^{s}_{\rr_{s}})}
{\prod_{s=p-1}^{i-2} f_{\prt{s+1}}(t^{s+1}_{\rr_{s+1}},\bar t^{s}) },
\end{equation*}
\begin{equation*}
\Lambda(z;\bar t)=\frac{ \lambda_{p+1} (t^p_{\rr_{p}}) \cdots  \lambda_{i} (t^{i-1}_{\rr_{i-1}}) }
{ (\lambda_{p+1}(z)  \cdots  \lambda_{j-1}(z))^{\theta_{p+1,j-1}}},
\end{equation*}
 and finally for $p=i$ and $q=j$
\begin{equation*}
\YY(z,\bar t)=  f_{\prt{i}}(z,\bar t^{i-1})\ f_{\prt{j}}(\bar t^{j},z)
\prod_{s=i}^{j-1}h(\bar t^s,z)^{\delta_{s,m}},
\end{equation*}
\begin{equation*}
\Lambda(z;\bar t)= \frac{\lambda_i(z)^{\delta_{ij}}}
{(\lambda_{i+1}(z)  \cdots  \lambda_{j-1}(z))^{\theta_{i+1,j-1}}},
\end{equation*}
where $\theta_{i,j}$ is the step function
\begin{equation*}
\theta_{i,j}=\begin{cases}
1,& i\leq j,\\
0,& i>j.
\end{cases}
\end{equation*}

Now we restore the normalization of the Bethe vectors \r{gc21} using equation  \r{bv}.
Again, using the explicit expressions for the rational functions
\r{zz1}, \r{zz2}, and \r{gc22} we obtain
\begin{equation}\label{gc28}
\LL^+_{i,j}(z)\cdot \hat\BBB(\bar{t})=\sum_{p=1}^i\sum_{q=j}^{N+1}
\varphi_{p,q}\ \comb^{i-1}_p\ \comb^{q-1}_j \,
 \Sym_{\,\bar t^p,\ldots,\bar t^{i-1},\bar t^{j},\ldots,\bar t^{q-1}}
\Big[\hat\DD(\bar t)\
\YY(z,\bar t)\ \Lambda(z;\bar t)\ \hat\BBB(\{z,\bar t\}')\Big],
\end{equation}
where the only difference compared to the
action formula \r{gc25}  is  the function $\DD(\bar t)$
which is replaced by
\begin{equation}\label{DD2}
\hat\DD(\bar t)=
\prod_{s=p}^{i-1}\frac{f_{\prt{s+1}}(t^s_{\rr_s},\bar t^s_{\rr_s})}{
h(\bar t^s_{\rr_s},t^s_{\rr_s})^{\delta_{s,m}}}
\prod_{s=j}^{q-1} \frac{f_{\prt{s+1}}(\bar t^s_{1},t^s_{1})} {
[(-)^{(r_s-1)}h(t^s_{1},\bar t^s_1)]^{\delta_{s,m}}}.
\end{equation}

Comparing the action formulas \r{gc25} and \r{gc28} we can prove
 Proposition~\ref{main-prop}, if we  prove that
the functions $\DD(\bar t)$ and $\hat\DD(\bar t)$ actually coincide.
First of all we recall that for $s\not=m$ the rational functions $f_{\prt{s}} (u,v)$ and
$f_{\prt{s+1}} (u,v)$ entering the definitions of the  functions \r{DD1} and
\r{DD2} coincide. The difference may occur only in the case when
$s=m$, since by definition
\begin{equation*}
f_{\prt{m}} (u,v)=\frac{u-v+c}{u-v}\quad\mbox{and}\quad
f_{\prt{m+1}} (u,v)=\frac{u-v-c}{u-v}.
\end{equation*}

Assume first that $m\not\in\{p,\ldots,i-1\}$ and $m\not\in\{j,\ldots,q-1\}$. Then the functions
\r{DD1} and \r{DD2} coincide. If $m\in\{p,\ldots,i-1\}$ then the factors in the functions
$\DD(\bar t)$ and $\hat\DD(\bar t)$ depending on the Bethe parameters $\bar t^m$
are both equal to $g(t^m_{\rr_m},\bar t^m_{\rr_m})$. Analogously, if
$m\in\{j,\ldots,q-1\}$, then these factors are equal to $g(\bar t^m_{1},t^m_{1})$.
This means
that  in the Yangian double
$DY(\mathfrak{gl}(m|n))$,
the Bethe vectors constructed via the first current realization  given by the formula
\r{uBV1} coincide with
the Bethe vectors constructed using the second current realization  \r{bv}.

This concludes the proof of the main statement formulated in  Proposition~\ref{main-prop}.
\hfill\qed

\subsection{Actions of the diagonal elements and Bethe equations}

In this section we consider the action of the universal transfer matrix $\trf(z)$ \eqref{strace} onto Bethe vectors.
For this we should find the action of the diagonal monodromy matrix elements. Hence, we should set
$i=j$ in the right hand side of the action formula \r{gc25}.
Since the action formulas \r{gc25} and \r{gc28} coincide,
we  use the first of these relations. We have
\begin{equation}\label{be0}
\begin{split}
\trf(z)\cdot \BBB(\bar{t})&=\sum_{i=1}^{N+1}(-)^{\prt{i}}\sum_{p=1}^i\sum_{q=i}^{N+1}
\varphi_{p,q}\ \comb_p^{i-1}\comb_i^{q-1}\\
&\quad\times \Sym_{\,\bar t^p,\ldots,\bar t^{q-1}}
\Big[\DD(\bar t)\
\YY(z,\bar t)\ \Lambda(z;\bar t)\ \BBB(\{z,\bar t\}')\Big],
\end{split}
\end{equation}
where
\begin{equation*}
\{z,\bar t\}' =\{\bar t^1,\ldots,\bar t^{p-1},
\{z,\bar t^{p}_{\rr_{p}} \},\ldots,\{z,\bar t^{i-1}_{\rr_{i-1}} \},
\{z,\bar t^{i}_{1} \},\ldots,\{z,\bar t^{q-1}_1 \},
\bar t^q,\ldots,\bar t^N\}\,,
\end{equation*}
and we recall that $N=m+n-1$.

Among all the terms in the right hand side of \r{be0} there are
so called 'wanted' terms corresponding to the values  $p=q=i$. One can easily see that they are equal to
\begin{equation*}
\sum_{i=1}^{N+1}(-)^{\prt{i}}\lambda_i(z)f_{\prt{i}}(z,\bar t^{i-1})f_{\prt{i}}(\bar t^{i},z)\ \BBB(\bar t)\,.
\end{equation*}

Let us compare the terms in \r{be0} coming from the action of the
monodromy matrix elements $\LL_{i,i}(z)$ and $\LL_{i+1,i+1}(z)$. They
correspond to the terms in the sums over $p$ and $q$ in the right
hand side of \r{gc25} for $p=i$ and $q=i+1$ in both cases.
For the action of the matrix element $(-)^{\prt{i}}\LL_{i,i}(z)$ these terms are
\begin{equation}\label{be2}
\begin{split}
&\frac{1}{(\rr_i-1)!}\ \Sym_{\,\bar t^i}\left[
\frac{\lambda_i(t^i_1)}{f_{\prt{i+1}}(\bar t^{i+1},t^i_1)  }\
\frac{f_{\prt{i}}(\bar t^{i}_1,t^i_1) }{h(\bar t^{i}_1,t^i_1)^{\delta_{i,m}} }
\BBB(\bar t^1,\ldots,\bar t^{i-1},\{z,\bar t^i_1\},\bar t^{i+1},\ldots,\bar t^N)\right.\\
&\qquad\qquad
\left.\phantom{\frac{\lambda_i(t^i_1)}{f_{\prt{i+1}}(\bar t^{i+1},t^i_1)  }}
\times g(z,t^i_1)f_{\prt{i}}(z,\bar t^{i-1})f_{\prt{i+1}}(\bar t^{i+1},z)
h(\bar t^i_1,z)^{\delta_{i,m}}
\right].
\end{split}
\end{equation}
For the action of the matrix element $(-)^{\prt{i+1}}\LL_{i+1,i+1}(z)$ similar terms are
\begin{equation}\label{be3}
\begin{split}
&\frac{(-)^{1+(\rr_i-1)\delta_{i,m}}}{(\rr_i-1)!}\ \Sym_{\,\bar t^i}\left[
\frac{\lambda_{i+1}(t^i_{\rr_i})}{f_{\prt{i}}(t^i_{\rr_i}, \bar t^{i-1})  }\
\frac{f_{\prt{i}}(t^{i}_{\rr_i},\bar t^i_{\rr_i}) }{h(t^{i}_{\rr_i},\bar t^i_{\rr_i})^{\delta_{i,m}} }
\BBB(\bar t^1,\ldots,\bar t^{i-1},\{z,\bar t^i_{\rr_i}\},\bar t^{i+1},\ldots,\bar t^N)\right.\\
&\qquad\qquad
\left.\phantom{\frac{\lambda_i(t^i_1)}{f_{\prt{i+1}}(\bar t^{i+1},t^i_1)  }}
\times g(z,t^i_{\rr_i})f_{\prt{i}}(z,\bar t^{i-1})f_{\prt{i+1}}(\bar t^{i+1},z)
h(\bar t^i_{\rr_i},z)^{\delta_{i,m}}
\right].
\end{split}
\end{equation}
Symmetrizations in the formulas \r{be2} and \r{be3} can be replaced by summation over
$\el=1,\ldots,\rr_i$
\begin{equation}\label{be22}
\begin{split}
&\sum_{\el=1}^{\rr_i}\left[
\frac{\lambda_i(t^i_\el)}{f_{\prt{i+1}}(\bar t^{i+1},t^i_\el)  }\
\frac{f_{\prt{i}}(\bar t^{i}_\el,t^i_\el) }{h(\bar t^{i}_\el,t^i_\el)^{\delta_{i,m}} }
\BBB(\bar t^1,\ldots,\bar t^{i-1},\{z,\bar t^i_\el\},\bar t^{i+1},\ldots,\bar t^N)\right.\\
&\qquad\qquad
\left.\phantom{\frac{\lambda_i(t^i_\el)}{f_{\prt{i+1}}(\bar t^{i+1},t^i_\el)  }}
\times g(z,t^i_\el)f_{\prt{i}}(z,\bar t^{i-1})f_{\prt{i+1}}(\bar t^{i+1},z)
h(\bar t^i_\el,z)^{\delta_{i,m}}
\right]
\end{split}
\end{equation}
and
\begin{equation}\label{be33}
\begin{split}
&-(-)^{(\rr_i-1)\delta_{i,m}} \sum_{\el=1}^{\rr_i}\left[
\frac{\lambda_{i+1}(t^i_{\el})}{f_{\prt{i}}(t^i_{\el}, \bar t^{i-1})  }\
\frac{f_{\prt{i}}(t^{i}_{\el},\bar t^i_{\el}) }{h(t^{i}_{\el},\bar t^i_{\el})^{\delta_{i,m}} }
\BBB(\bar t^1,\ldots,\bar t^{i-1},\{z,\bar t^i_{\el}\},\bar t^{i+1},\ldots,\bar t^N)\right.\\
&\qquad\qquad
\left.\phantom{\frac{\lambda_i(t^i_1)}{f_{\prt{i+1}}(\bar t^{i+1},t^i_1)  }}
\times g(z,t^i_{\el})f_{\prt{i}}(z,\bar t^{i-1})f_{\prt{i+1}}(\bar t^{i+1},z)
h(\bar t^i_{\el},z)^{\delta_{i,m}}
\right].
\end{split}
\end{equation}
If the set of the Bethe parameters $\bar t$ satisfy a system of equations
\begin{equation}\label{be4}
\frac{\lambda_{i+1}(t^i_{\el})}{\lambda_{i}(t^i_{\el})}=(-)^{(\rr_i-1)\delta_{i,m}}
\frac{f_{\prt{i}}(\bar t^{i}_{\el},t^i_{\el}) }{h(\bar t^{i}_{\el},t^i_{\el})^{\delta_{i,m}} }\
\frac{h(t^{i}_\el,\bar t^i_\el)^{\delta_{i,m}} }{f_{\prt{i}}(t^{i}_\el,\bar t^i_\el) }\
\frac{f_{\prt{i}}(t^i_{\el}, \bar t^{i-1})  }{f_{\prt{i+1}}(\bar t^{i+1},t^i_\el)  },
\end{equation}
then the terms in \r{be22} and \r{be33}  cancel each other. If $i\not=m$, then  equations \r{be4} become the standard Bethe equations
similar to those of $\mathfrak{gl}(N+1)$
\begin{equation}\label{be5}
\frac{\lambda_{i+1}(t^i_{\el})}{\lambda_{i}(t^i_{\el})}=
\frac{f_{\prt{i}}(\bar t^{i}_{\el},t^i_{\el}) }{f_{\prt{i}}(t^{i}_\el,\bar t^i_\el) }\
\frac{f_{\prt{i}}(t^i_{\el}, \bar t^{i-1})  }{f_{\prt{i+1}}(\bar t^{i+1},t^i_\el)  }\;.
\end{equation}
For $i=m$  Bethe equations \r{be4} simplify to
\begin{equation}\label{be6}
\frac{\lambda_{m+1}(t^m_{\el})}{\lambda_{m}(t^m_{\el})}=
\frac{f(t^m_{\el}, \bar t^{m-1})  }{f(t^m_\el,\bar t^{m+1})  }\,.
\end{equation}
This simplified
form of Bethe equations is typical for the models of free fermions, however, one should remember that in the
case under consideration the parameters $t^m_{\el}$ are coupled through equations \eqref{be5} with $i=m\pm1$.

If the Bethe equations are satisfied, then the Bethe vector becomes an eigenvector
of the transfer matrix \r{strace}
\begin{equation*}
\trf(z)\cdot \BBB(\bar t)=\tau(z;\bar t)\ \BBB(\bar t)
\end{equation*}
with the eigenvalue
\begin{equation}\label{be8}
\tau(z;\bar t)=\sum_{i=1}^{N+1} (-)^{\prt{i}}
 \lambda_i(z) f_{\prt{i}}(z,\bar t^{i-1})f_{\prt{i}}(\bar t^{i},z).
\end{equation}
In this case we call $\BBB(\bar t)$ an on-shell Bethe vector.
Note that the Bethe equations \r{be5} and \r{be6} can be considered as conditions
of cancelation of the poles in the eigenvalue \eqref{be8} at $z=t^i_\el$.

Let us verify that  all the remaining 'unwanted' terms in the
action of the transfer matrix \r{strace} onto the on-shell Bethe vector  vanish.
To do this we  calculate the combined coefficient  of the Bethe vector
\begin{equation}\label{be9}
\BBB(\bar t^1,\ldots,\bar t^{i-1},\{z,\bar t^i_{\el_i}\},\ldots, \{z,\bar t^{i+a}_{\el_{i+a}}\},\bar t^{i+a+1},\ldots,\bar t^N)
\end{equation}
for fixed $i$ and \footnote{The case $a=0$ was considered above to obtain Bethe equations.} $a>0$
under summation over Bethe parameters $t^b_{\el_b}$ for $b=i,\ldots,i+a$. These sums
appear from the symmetrization in \r{be0}. One can see that the vector with
Bethe parameters as in \r{be9} may appear only from the actions of the diagonal monodromy
matrix elements $\LL^+_{b,b}(z)$ for $b=i,\ldots,i+a+1$. This occurs if we take in the sums over $p$ and $q$ in \r{gc25}
the only term for $p=i$ and $q=i+a+1$. Restoring definition of the phase factor $\varphi_{i,i+a+1}$
\r{upf} for each $b=i,\ldots,i+a+1$ and denoting it as $\varphi_{i,i+a+1}(b)$ we obtain
\begin{equation*}
(-)^{\prt{b}}\varphi_{i,i+a+1}(b)=\begin{cases}
1&\quad\mbox{for}\quad b=i,\\
(-)^{1+\prt{b}}&\quad\mbox{for}\quad b=i+1,\ldots,i+a,\\
-1&\quad\mbox{for}\quad b=i+a+1.
\end{cases}
\end{equation*}
Using explicitly Bethe equation in the function $\Lambda(z;\bar t)$ we obtain that the coefficient
of the Bethe vector \r{be9} in the right hand side of the action formula \r{be0} is proportional
to the sum
\begin{equation*}
g(z,t^i_{\el_i})^{-1}-\sum_{b=i+1}^{i+a}g(t^b_{\el_b},t^{b-1}_{\el_{b-1}})^{-1}-g(z,t^{i+a}_{\el_{i+a}})^{-1}=0,
\end{equation*}
which obviously vanishes. Note that the same trivial identity was used in the paper
\cite{FKPR} (see unnumbered formula on the page 29 of this paper)
   to prove that the universal off-shell Bethe vectors
 become on-shell provided the Bethe equations are satisfied.

\section{Explicit formulas for the universal Bethe vectors}
\label{BV-calc}

\subsection{Hierarchical relations for the Bethe vector   $\BBB(\bar t)$}

Calculating  the 'positive' projection in the formula for the Bethe vector $\BBB(\bar t)$ \eqref{uBV1} we can obtain
the hierarchical recurrence relation which relates the Bethe vectors constructed for
the Yangian double $DY(\mathfrak{gl}(m|n))$ with the Bethe vectors for the Yangian double
$DY(\mathfrak{gl}(m-1|n))$. Let us separate the product of currents $\F_1(\bar t^1)=F_1(t^1_1)\cdots
F_1(t^1_{\rr_1})$ from the product of the other currents $\F_\el(\bar t^\el)$, $\el=2,\ldots,N$ and apply to the latter
product the normal ordering rule \r{FFpar}. It is obvious from this rule that in order to obtain
desired hierarchical relations for the Bethe vectors (see formula \r{hr3} below)  it is sufficient to calculate
the projection
\begin{equation}\label{gca1}
\Pfp\sk{\F_1(\bar t^1)\cdot \Pfm\sk {\F_2(\bar t^2_{\so}) \F_3(\bar t^3_{\so} )\cdots \F_N(\bar t^N_{\so})}}.
\end{equation}
Using the property  $\Pfm(\F\cdot \Pfp(\F'))=0$ for any elements $\F,\F'\in\oU_F$
such that $\coun(\F')=0$, we reduce the problem to the calculation of the projections
\begin{equation}\label{gca2}
\Pfp\sk{\F_1(\bar t^1)\cdot \Pfm (\F_2(\bar t^2_{\so})\cdot \Pfm(\F_3(\bar t^3_{\so} )
  \cdots \Pfm( \F_N(\bar t^N_{\so})) \cdots )     )}\,.
\end{equation}
The calculation of the projection in \r{gca2} is given in  Appendix~\ref{appC}.
There it is shown that this calculation  yields an answer in the form of a sum over partitions of the sets
$\bar t^1$ and $\bar t^\el_{\so}$,  $\el=2,\ldots,N$
of the Bethe parameters entering in the expression \r{gca2}.

To obtain the hierarchical relations for the Bethe vectors in the framework of this approach we
use formula \r{FFpar} to rewrite the Bethe vector \r{uBV1} as a sum over
partitions of the Bethe parameters
\begin{equation*}
\bar t'=\{\bar t^2,\ldots,\bar t^N\}\Ra \bar t'_{\so} \cup \bar t'_{\st},
\end{equation*}
where
\begin{equation*}
\bar t'_{\so}=\{\bar t^2_{\so},\ldots,\bar t^N_{\so}\}\quad\mbox{and}\quad
\bar t'_{\st}=\{\bar t^2_{\st},\ldots,\bar t^N_{\st}\}.
\end{equation*}
The primed set of Bethe parameters $\bar t'$ differs from the full set $\bar t$
of these parameters \r{Bpar} by excluding the first type of Bethe parameters $\bar t^1$.
It follows from \r{FFpar} and properties of the projections that
\begin{equation}\label{hr3}
\begin{split}
\BBB^{(m|n)}(\bar t)&=
\sum_{\bar t'\Ra \bar t'_{\so} \cup \bar t'_{\st}}
\frac{\Fli{1}(\bar t^1)}{f_{\prt{2}}(\bar t^2_{\so},\bar t^1)}
\Pfp\sk{F_{2,1}(t^1_1)\cdots F_{2,1}(t^1_{r_1}) \Pfm\sk{\Fp(\bar t'_{\so})}}k^+_1(\bar t^1)\\
&\qquad\times
\frac{1}{f_{\prt{2}}(\bar t^2_{\st},\bar t^1)}\ \frac{\prod_ {s=2}^{N}  \Fli{s}(\bar t^s_{\st},\bar t^s_{\so})  }
{ \prod_ {s=2}^{N-1} f_{\prt{s+1}}(\bar t^{s+1}_{\st},\bar t^s_{\so})}\
 \BBB^{(m-1|n)}(\bar t'_{\st})\prod_{s=2}^N \lambda_s(\bar t^s_{\so}),
\end{split}
\end{equation}
where we have identified $\bar t^1_{\so}$ with $\bar t^1$
and used the fact that the Cartan currents $k^+_1(z)$ commute with all the
currents
$F_s(t')$, $s=2,\ldots,N$. Denote
\begin{equation}\label{hr33}
{\cal X}(\bar t)=\frac{\Fli{1}(\bar t^1)}{f_{\prt{2}}(\bar t^2,\bar t^1)}
\Pfp\sk{F_{2,1}(t^1_1)\cdots F_{2,1}(t^1_{r_1}) \Pfm\sk{\Fp(\bar t')}}k^+_1(\bar t^1)
\end{equation}
with $\bar t'=\{\bar t^2,\ldots,\bar t^N\}$.
Then the first line of the right hand side of \r{hr3} is equal to
\begin{equation}\label{hr333}
{\cal X}(\bar t^1,\bar t'_{\so}) .
\end{equation}

To calculate the 'positive' projection from the product of the currents
$F_{2,1}(t^1_1)\cdots F_{2,1}(t^1_{r_1})$ and the 'negative' projection
$\Pfm\sk{\Fp(\bar t')}$ in \r{hr33}
we  use formulas \r{dc10} and \r{dc14} for different
$i$ starting from $i=N$ until $i=2$. We  use the first formula \r{dc10}
for $i=m+1,\ldots,N$ starting from $i=N$ until $i=m+1$, and the
second formula \r{dc14} for $i=2,\ldots,m$ starting from $i=m$ until $i=2$.

The results of Appendix~\ref{appC} show that
the sets $\bar t^\el$ will be further divided into subsets. To describe this division we introduce
for each subset $\bar t^\el$, $\el=1,\ldots,N$ the subdivision
\begin{equation}\label{hr4}
\bar t^\el\Ra\{\bar t^\el_{\el},\bar t^\el_{\el+1},\ldots,\bar t^\el_{N}\}
\end{equation}
such that the following constraints for the cardinalities of the subsets hold:
\begin{equation}\label{hr5}
\#\bar t^\el_q = \# \bar t^{\el'}_q\quad\mbox{for all}\quad \el\not=\el'\quad\mbox{and}\quad q=
\max(\el,\el'),\ldots,N \,.
\end{equation}
Moreover, in order to get a nontrivial result in the calculation of the 'positive' projection in
\r{hr33} we have to impose the following restrictions for the cardinalities of the subsets $\bar t^s$:
\begin{equation*}
\#\bar t^1\geq \#\bar t^2\geq \cdots \geq \#\bar t^N\geq 0.
\end{equation*}

In \r{hr4} and \r{hr5} the superscripts of the subsets describe the type of Bethe
parameter, while the subscripts count the subsets in the division \r{hr4}. One should not confuse this notation
with the notation $\bar t^\el_i=\bar t^\el\setminus \{t^\el_i\}$ used in the previous section~\ref{gen-act}.

Appendix~\ref{appC} demonstrates how the Izergin determinant \cite{Ize-det}
$K_{\prt{i}}(\bar y|\bar x)$ (see \r{Izer})
appears in the calculation of the projections. It also shows how the result of these calculations
can be rewritten in the form of sums over partitions of the set
of Bethe parameters. Let us denote
\begin{equation*}
\begin{split}
 K_{0}(\bar y|\bar x)&=K_{\prt{i}}(\bar y|\bar x)\quad\mbox{for}
\quad i=1,\ldots,m,\\
 K_{1}(\bar y|\bar x)&=K_{\prt{i}}(\bar y|\bar x)=K_{0}(\bar x|\bar y)\quad\mbox{for}
\quad i=m+1,\ldots,N.
\end{split}
\end{equation*}
It is also convenient to introduce for any sets $\bar y$ and $\bar x$ of the same
cardinalities the following product of rational functions:
\begin{equation}\label{hr66}
\Cres(\bar y|\bar x)=g(\bar y,\bar x)h(\bar x,\bar x)\,.
\end{equation}

Let us give more details on the calculation of the projections using the results obtained in
Appendix~\ref{appC}.
The division of the sets \r{hr4} can be presented by the following table
\begin{equation}\label{table}
\!\!\!\!\!\!\!
\begin{array}{ccccccccccccccccccc}
&&\bar t^1_1&\cup&\bar t^1_2&\cup&\cdots&\cup&\bar t^1_{m-1}&\cup&\bar t^1_{m}&\cup&
\bar t^1_{m+1}&\cup&\cdots&\cup&\bar t^1_{N-1}&\cup&\bar t^1_{N}\\[3mm]
&&&&\bar t^2_2&\cup&\cdots&\cup&\bar t^{2}_{m-1}&\cup&\bar t^{2}_{m}&\cup&
\bar t^{2}_{m+1}&\cup&\cdots&\cup&\bar t^{2}_{N-1}&\cup&\bar t^{2}_{N}\\[3mm]
&&&&&&\ddots&&\vdots&&\vdots&& \vdots&&\vdots&&\vdots&&\vdots\\[3mm]
&&&&&&&&\bar t^{m-1}_{m-1}&\cup&\bar t^{m-1}_{m}&\cup&
\bar t^{m-1}_{m+1}&\cup&\cdots&\cup&\bar t^{m-1}_{N-1}&\cup&\bar t^{m-1}_{N}\\[3mm]
&&&&&&&&&&\bar t^{m}_{m}&\cup&
\bar t^{m}_{m+1}&\cup&\cdots&\cup&\bar t^{m}_{N-1}&\cup&\bar t^{m}_{N}\\[3mm]
&&&&&&&&&&&&
\bar t^{m+1}_{m+1}&\cup&\cdots&\cup&\bar t^{m+1}_{N-1}&\cup&\bar t^{m+1}_{N}\\[3mm]
&&&&&&&&&&&& &&\ddots&&\vdots&&\vdots\\[3mm]
&&&&&&&&&&&&&&&&\bar t^{N-1}_{N-1}&\cup&\bar t^{N-1}_{N}\\[3mm]
&&&&&&&&&&&&&&&&&&\bar t^{N}_{N}
\end{array}
\end{equation}
where the cardinalities of the subsets entering the same column are all equal.

For any set $\bar w$ of  cardinality $\#\bar w=d$
we introduce the following ordered product of composed (or simple when $j=i+1$) currents:
\begin{equation}\label{gca6}
\gF_{j,i}(\bar w)= F_{j,i}(w_1)\cdot F_{j,i}(w_2)\cdots F_{j,i}(w_d).
\end{equation}
Then the 'negative' projection $ \Pfm\sk{\Fp(\bar t')}$ entering  the definition of the element \r{hr33} can be written in the form
\begin{equation*}
\frac{\prod_{s=2}^N \Fli{s}(\bar t^s)}{\prod_{s=2}^{N-1} f_{\prt{s+1}}(\bar t^{s+1},\bar t^s)}
\Pfm \sk{\gF_{3,2}(\bar t^2)\Pfm(
  \cdots \Pfm( \gF_{N,N-1}(\bar t^{N-1})\cdot \Pfm( \gF_{N+1,N}(\bar t^N))) \cdots ) }  ,
\end{equation*}
and at the first step we have to calculate
\begin{equation}\label{np12}
\frac{\Fli{N-1}(\bar t^{N-1})\Fli{N}(\bar t^N)}{ f_{\prt{N}}(\bar t^{N},\bar t^{N-1})}
\Pfm \sk{ \gF_{N,N-1}(\bar t^{N-1})\cdot \Pfm\sk{ \gF_{N+1,N}(\bar t^N)}}
\end{equation}
using either formula \r{dc10} or \r{dc14} depending on the relation between $m$ and $N$.

If $m<N$, and hence, $\prt{N}=1$, then we have to use  \r{dc10} to obtain for the element \r{np12}
the sum over partitions of the set $\bar t^{N-1}\Ra \{\bar t^{N-1}_{N-1} \cup \bar t^{N-1}_{N}\}$
such that $\# \bar t^{N-1}_{N}=\#\bar t^N_N$ (see the second to last line in the table above), where
we identify the sets $\bar t^N_N\equiv \bar t^N$
\begin{equation*}
\begin{split}
&(-c)^{-\#\bar t^N_N}\Fli{N-1}(\bar t^{N-1}) \sum_{\bar t^{N-1}\Ra \{\bar t^{N-1}_{N-1}\, \cup\, \bar t^{N-1}_{N}\}}
\frac{f_1(\bar t^{N-1}_N,\bar t^{N-1}_{N-1}) K_1(\bar t^{N}_{N}|\bar t^{N-1}_N) }
{f_1(\bar t^{N}_N,\bar t^{N-1}_{N-1})f_1(\bar t^{N}_N,\bar t^{N-1}_{N})}\\
&\qquad\times\Pfm \sk{ \gF_{N+1,N-1}(\bar t^{N-1}_N)\cdot  \gF_{N,N-1}(\bar t^{N-1}_{N-1})}
\end{split}
\end{equation*}

On the other hand, if $m=N$, and hence, $\prt{N}=0$ (this case corresponds to the algebra
$\mathfrak{gl}(m|1)$),  we have to use  \r{dc14}
to obtain the element \r{np12}  again as the sum over the same partition of the set $\bar t^{N-1}$
\begin{equation*}
\begin{split}
&c^{-\#\bar t^N_N}\Fli{N-1}(\bar t^{N-1}) \sum_{\bar t^{N-1}\Ra \{\bar t^{N-1}_{N-1}\, \cup\, \bar t^{N-1}_{N}\}}
\frac{f_0(\bar t^{N-1}_N,\bar t^{N-1}_{N-1}) \Cres(\bar t^{N}_N|\bar t^{N-1}_{N}) }
{f_0(\bar t^{N}_N,\bar t^{N-1}_{N-1})f_0(\bar t^{N}_N,\bar t^{N-1}_{N})}\\
&\qquad\times\Delta_h(\bar t^{N-1}_N)^{-1}
\Pfm \sk{ \gF_{N+1,N-1}(\bar t^{N-1}_N)\cdot  \gF_{N,N-1}(\bar t^{N-1}_{N-1})}.
\end{split}
\end{equation*}

The next step is to calculate the projection
\begin{equation*}
\frac{\Fli{N-2}(\bar t^{N-2})\Fli{N-1}(\bar t^{N-1})}{ f_{\prt{N-1}}(\bar t^{N-1},\bar t^{N-2})}
\Pfm \sk{ \gF_{N-1,N-2}(\bar t^{N-2})\cdot \Pfm\sk{
\gF_{N+1,N-1}(\bar t^{N-1}_N)\cdot  \gF_{N,N-1}(\bar t^{N-1}_{N-1})}}
\end{equation*}
using equation \r{dc10} for $m<N-1$ and \r{dc14} for $m=N-1$.
Continuing the calculation of the element \r{hr33} using the first formula \r{dc10} and then \r{dc14}
we obtain eventually for this element
\begin{equation}\label{np15}
\begin{split}
&{\cal X}(\bar t)=\sum_{\substack{\bar t^\el\Ra\{\bar t^\el_{\el},\bar t^\el_{\el+1},\ldots,\bar t^\el_{N}\}
\\ \el=1,\ldots,N}}
\prod_{\el=1}^{N-1}\prod_{\el\leq q\leq q'\leq N}^N
f_{\prt{\el+1}}(\bar t^{\el+1}_{q'},\bar t^{\el}_{q})^{-1}
\prod_{\el=1}^{N}\prod_{\el\leq q< q'\leq N}
\frac{f_{\prt{\el}}(\bar t^{\el}_{q'},\bar t^\el_{q})}{h_{\prt{\el}}(\bar t^{\el}_{q'},\bar t^\el_{q})^{\delta_{\el,m}}}\\
&\quad\times\prod_{q=2}^{m-1}\prod_{\el=2}^{q} K_{\prt{\el}}(\bar t^{\el}_q|\bar t^{\el-1}_q)
\prod_{q=m+1}^{N}\prod_{\el=m+1}^{q} K_{\prt{\el}}(\bar t^{\el}_q|\bar t^{\el-1}_q)
\prod_{q=m}^N\prod_{\el=2}^{m} \Cres(\bar t^{\el}_q|\bar t^{\el-1}_q)
\prod_{q=m}^N  \Delta_h(\bar t^1_q)^{-1}\\
&\quad\times    \Fli{1}(\bar t^1)\
\Pfp\sk{\gF_{N+1,1}(\bar t^1_N)
\cdots \gF_{m+1,1}(\bar t^1_m)\cdot
\gF_{m,1}(\bar t^1_{m-1})\cdots \gF_{2,1}(\bar t^1_1)}k^+_1(\bar t^1).
\end{split}
\end{equation}

The projection  in the last line of \r{np15}
can be calculated using methods of the paper \cite{KhP-Kyoto}.
Being multiplied from the right by the product of the Cartan currents $k^+_1(\bar t^1)$ it
can be expressed through an ordered product of the matrix elements of the monodromy
operators $\LL_{1,\el}(t)$, $\el=2,\ldots,N+1$.
This demonstrates that the hierarchical
relations which we are resolving by the calculation of the projection in \r{hr3} are
compatible with the embedding of $DY(\mathfrak{gl}(m-1|n))$ into
$DY(\mathfrak{gl}(m|n))$.

Finally, the element \r{hr33} is given as a multiple sum over partitions
\begin{equation}\label{hr7}
\begin{split}
{\cal X}(\bar t)&=\sum_{\substack{\bar t^\el\Ra\{\bar t^\el_{\el},\bar t^\el_{\el+1},\ldots,\bar t^\el_{N}\}
\\ \el=1,\ldots,N}}
\prod_{\el=1}^{N-1}\prod_{\el\leq q\leq q'\leq N}^N
f_{\prt{\el+1}}(\bar t^{\el+1}_{q'},\bar t^\el_{q})^{-1}
\prod_{\el=1}^{N}\prod_{\el\leq q< q'\leq N}
\frac{f_{\prt{\el}}(\bar t^{\el}_{q'},\bar t^\el_{q})}{h_{\prt{\el}}(\bar t^{\el}_{q'},\bar t^\el_{q})^{\delta_{\el,m}}}\\
&\quad\times\prod_{q=2}^{m-1}\prod_{\el=2}^{q} K_{\prt{\el}}(\bar t^{\el}_q|\bar t^{\el-1}_q)
\prod_{q=m+1}^{N}\prod_{\el=m+1}^{q} K_{\prt{\el}}(\bar t^\el_q|\bar t^{\el-1}_q)
\prod_{q=m}^N\prod_{\el=2}^{m} \Cres(\bar t^{\el}_q|\bar t^{\el-1}_q)\\
&\qquad\times\mathbb{T}_{1,N+1}(\bar t^1_N)
\mathbb{T}_{1,N}(\bar t^1_{N-1})\cdots \mathbb{T}_{1,m+1}(\bar t^1_m)\cdot
{\LL}_{1,m}(\bar t^1_{m-1})\cdots {\LL}_{1,2}(\bar t^1_1).
\end{split}
\end{equation}
Here the symbol $\mathbb{T}_{i,j}(\bar w)$  means
\begin{equation}\label{T-odd}
\mathbb{T}_{i,j}(\bar w)= \Delta_h(\bar w)^{-1}\LL_{i,j}(w_1)\LL_{i,j}(w_2)\cdots
\LL_{i,j}(w_{d-1})\LL_{i,j}(w_d).
\end{equation}
for any set $\bar w$ of cardinality $\#\bar w=d$ and $\prt{i}+\prt{j}=1$.
Obviously, due to the commutation relation \r{NCR-7}, this product of odd matrix elements is symmetric with respect
to the permutations of the parameters $w_i$.

\subsection{Bethe vectors $\BBB(\bar t)$}

The element \r{hr7} for the subsets \r{hr333} should be substituted in the hierarchical relations
\r{hr3}
and the same procedure
should be repeated for the Bethe vector $ \BBB^{(m-1|n)}(\bar t'_{\st})$ in the second line of
\r{hr3}. Finally, we will obtain an explicit expression for the Bethe vector
$ \BBB^{(m|n)}(\bar t)$ as a sum over multiple partitions of the set of Bethe parameters.  Each term
of this sum appears to be a rational coefficient multiplied with a product of
symmetric products of the monodromy matrix elements.
To describe this expression it is necessary to introduce a more convenient indexing of the
multiple partitions.

For all $\el=1,\ldots,N$ we introduce the partition of the sets of Bethe parameters
\begin{equation}\label{hr8}
\bar t^\el=\bigcup_{q=1}^\el\bigcup_{q'=\el}^N \bar t^\el_{q,q'}
\end{equation}
indexed by a pair of positive integers
\begin{equation*}
1\leq q\leq\el\leq q'\leq N\,.
\end{equation*}
We also introduce ordering rules '$\prec$' and  '$\preccurlyeq$' for these pairs according to the following
convention
\begin{equation}\label{hr10}
q,q' \prec p,p'\quad\mbox{if}\quad q<p, \forall q',p'
\quad\mbox{or}\quad q=p, q'<p'
\end{equation}
and
\begin{equation*}
q,q' \preccurlyeq p,p'\quad\mbox{if}\quad q<p, \forall q',p'
\quad\mbox{or}\quad q=p, q'<p'
\quad\mbox{or}\quad q=p,q'=p'.
\end{equation*}

Using this notation and combining \r{hr3} with \r{hr7} we obtain the following expression for the
Bethe vector:
\begin{equation*}
\BBB(\bar t)=\Bp(\bar t)\rvec,
\end{equation*}
where the pre-Bethe vector $\Bp(\bar t)$ is given by the sum over partitions \r{hr8}
\begin{equation}\label{hr77}
\begin{split}
\Bp(\bar t)&=\sum_{\rm part}
\prod_{q,q'\,\preccurlyeq\, p,p'}\prod_{\el=1}^{N-1}
f_{\prt{\el+1}}(\bar t^{\el+1}_{p,p'},\bar t^\el_{q,q'})^{-1}
 \prod_{q,q'\,\prec\, p,p'} g(\bar t^{m}_{p,p'},\bar t^m_{q,q'}) \prod_{\substack{\el=1\\ \el\not=m}}^{N}
 f_{\prt{\el}}(\bar t^{\el}_{p,p'},\bar t^\el_{q,q'})\\
&\quad\times
\prod_{q=1}^{m-2}\prod_{q'=q+1}^{m-1}\prod_{\el=q+1}^{m-1} K_{\prt{\el}}(\bar t^{\el}_{q,q'}|\bar t^{\el-1}_{q,q'})
\quad
\prod_{q=1}^{m-1}\prod_{q'=m}^{N}\prod_{\el=q+1}^{m} \Cres(\bar t^{\el}_{q,q'}|\bar t^{\el-1}_{q,q'})\\
&\quad\times
\prod_{q=1}^{m-1}\prod_{q'=m+1}^{N}\prod_{\el=m+1}^{q'} K_{\prt{\el}}(\bar t^\el_{q,q'}|\bar t^{\el-1}_{q,q'})
\quad
\prod_{q=m}^{N-1}\prod_{q'=q+1}^{N}\prod_{\el=q+1}^{q'} K_{\prt{\el}}(\bar t^{\el}_{q,q'}|\bar t^{\el-1}_{q,q'})
\\
&\quad\times
\prod^{\longrightarrow}_{1\leq q\leq m}
\sk{\prod^{\longleftarrow}_{N+1\geq q'\geq m+1}
\mathbb{T}_{q,q'}(\bar t^q_{q,q'-1})
\prod^{\longleftarrow}_{m\geq q'\geq q+1}
{\LL}_{q,q'}(\bar t^q_{q,q'-1})}\\
&\quad\times
\prod^{\longrightarrow}_{m+1\leq q\leq N}
\sk{\prod^{\longleftarrow}_{N+1\geq q'\geq {q+1}}
{\LL}_{q,q'}(\bar t^q_{q,q'-1})}
\prod_{\el=2}^N {\prod_{q=1}^{\el-1}\prod_{q'=\el}^{N}  } \LL_{\el,\el}(\bar t^\el_{q,q'}).
\end{split}
\end{equation}

The partitions of the Bethe parameters can be pictured by an  ordered table, which is the following union of diagrams
similar to \r{table}
\begin{equation}\label{col-tab}
\mathop{\bigcup}\limits^{\longrightarrow}_{\el=1,\ldots,N} \quad
\begin{array}{ccccccccc}
\bar t^{\el}_{\el,\el}&\cup&\bar t^{\el}_{\el,\el+1}&\cup&\cdots&\cup&\bar t^{\el}_{\el,N-1}&\cup&\bar t^{\el}_{\el,N}\\[2mm]
&&\bar t^{\el+1}_{\el,\el+1}&\cup&\cdots&\cup&\bar t^{\el+1}_{\el,N-1}&\cup&\bar t^{\el+1}_{\el,N}\\[2mm]
&&&&\ddots&&\vdots&&\vdots\\[2mm]
&&&&&&\bar t^{N-1}_{\el,N-1}&\cup&\bar t^{N-1}_{\el,N}\\[2mm]
&&&&&&&&\bar t^{N}_{\el,N}
\end{array}
\end{equation}
The ordering means that  if $\ell'<\ell$, then the diagram associated to $\el'$ in \r{col-tab} is on the left of the diagram corresponding to $\el$.
The ordering rules  \r{hr10} mean literally that if $q,q' \prec p,p'$,  then
the subset $\bar t^{\el'}_{q,q'}$ from the $\el'$-th row is located to the left from the subset
$\bar t^{\el}_{p,p'}$ placed in the $\el$-th row of the diagram.
The  subsets in a given column have all the same cardinality.
The subsets which describe partitions of the same type of Bethe parameters are aligned along the same row of the
diagram (see examples of such tables in \r{hr144}, \r{hr15} and \r{hr1515}).

\subsubsection*{Examples}
Let us consider formula \r{hr77} in some particular cases of small $m$ and $n$.
\paragraph{The case $m=2$ and $n=1$.} In this case $N=m+n-1=2$
and the partitions of the sets $\bar t^1$ and $\bar t^2$ can be pictured by the following
union of two diagrams:
\begin{equation}\label{hr144}
\begin{array}{ccccccccccccc}
\bar t^1&:&\bar t^1_{1,1}&\cup&\bar t^1_{1,2}&\cr
\bar t^2&:&&&\bar t^2_{1,2}&\cup&\bar t^2_{2,2}\cr
\end{array}
\end{equation}
Formula
\r{hr77} in this case simplifies:
\begin{equation}\label{hr13}
\begin{split}
\Bp^{(2|1)}(\bar t^1,\bar t^2)&=\sum_{\rm part}
f(\bar t^2,\bar t^1)^{-1} f(\bar t^1_{1,2},\bar t^1_{1,1})g(\bar t^2_{2,2},\bar t^2_{1,2})
\Cres(\bar t^2_{1,2}|\bar t^1_{1,2})\\
&\times \mathbb{T}_{1,3}(\bar t^1_{1,2})\  \LL_{1,2}(\bar t^1_{1,1})\
\mathbb{T}_{2,3}(\bar t^2_{2,2})\ \LL_{2,2}(\bar t^2_{1,2}).
\end{split}
\end{equation}
After the identifications $\bar t^1_{1,1}\equiv \bar u_{\st}$,   $\bar t^1_{1,2}\equiv \bar u_{\so}$,
 $\bar t^2_{2,2}\equiv \bar v_{\st}$, $\bar t^2_{1,2}\equiv \bar v_{\so}$, we recover from formula
 \r{hr13} the expression \r{cain8} for the Bethe vector.

\paragraph{The case $m=2$ and $n=2$.} Partitions
\r{hr8} can be described by the following table
\begin{equation}\label{hr15}
\begin{array}{ccccccccccccc}
\bar t^1&:&\bar t^1_{1,1}&\cup&\bar t^1_{1,2}&\cup&\bar t^1_{1,3}&&&&&&\\[2mm]
\bar t^2&:&&&\bar t^2_{1,2}&\cup&\bar t^2_{1,3}&\cup&\bar t^2_{2,2}&\cup&\bar t^2_{2,3}&&\\[2mm]
\bar t^3&:&&&&&\bar t^3_{1,3}&&\cup&&\bar t^3_{2,3}&\cup&\bar t^3_{3,3}
\end{array}
\end{equation}
It corresponds to the union of 3 diagrams of type \r{table}.
 With this notation, the
formula \r{hr77} takes the form
\begin{equation}\label{hr16}
\begin{split}
\Bp^{(2|2)}(\bar t^1,\bar t^2,\bar t^3)=\sum_{\rm part}\
&\phantom{\times}f_0(\bar t^2_{1,2},\{\bar t^1_{1,1}\cup \bar t^1_{1,2} \})^{-1}
f_0(\{\bar t^2_{1,3}\cup\bar t^2_{2,2}\cup\bar t^2_{2,3}\} ,\bar t^1)^{-1}\\
&\times f_1(\bar t^3_{1,3},\{\bar t^2_{1,2}\cup\bar t^2_{1,3}\})^{-1}
f_1(\{\bar t^3_{2,3}\cup \bar t^3_{3,3} \},\bar t^2)^{-1}\\
&\times f_0(\bar t^1_{1,3},\{\bar t^1_{1,2} \cup \bar t^1_{1,1}  \}) f_0(\bar t^1_{1,2},\bar t^1_{1,1})\\
&\times g(\bar t^2_{2,3},\{\bar t^2_{2,2} \cup \bar t^2_{1,3}\cup \bar t^2_{1,2}  \})
g(\bar t^2_{2,2} ,\{ \bar t^2_{1,3}\cup \bar t^2_{1,2}  \})  g(\bar  t^2_{1,3}, \bar t^2_{1,2} ) \\
&\times f_1(\bar t^3_{3,3},\{\bar t^3_{1,3}\cup \bar t^3_{2,3}\}) f_1(\bar t^3_{2,3},\bar t^3_{1,3})\\
&\times \Cres(\bar t^2_{1,2}|\bar t^1_{1,2})\Cres(\bar t^2_{1,3}|\bar t^1_{1,3})
K_1(\bar t^3_{1,3}|\bar t^2_{1,3})K_1(\bar t^3_{2,3}|\bar t^2_{2,3}) \\
&\times \mathbb{T}_{1,4}(\bar t^1_{1,3})\mathbb{T}_{1,3}(\bar t^1_{1,2})\  \LL_{1,2}(\bar t^1_{1,1})\
\mathbb{T}_{2,4}(\bar t^2_{2,3})\ \mathbb{T}_{2,3}(\bar t^2_{2,2})\  \LL_{3,4}(\bar t^3_{3,3})\\
&\times \LL_{2,2}(\bar t^2_{1,2})\ \LL_{2,2}(\bar t^2_{1,3})\ \LL_{3,3}(\bar t^3_{1,3})\ \LL_{3,3}(\bar t^3_{2,3}).
\end{split}
\end{equation}
There is a rule which allows to construct a  pre-Bethe vector from any given table of partitions \r{hr8}.
Let us show this rule for the diagram \r{hr15}.
We consider each line in the formula \r{hr16} and explain all the factors entering into this formula
from the diagram \r{hr15}.
\begin{itemize}
\item
For a given subset $\bar t^\el_{i,j}$ in the $\el$-th row of the diagram \r{hr15}, the first and second
lines in the formula \r{hr16} (corresponding to the values $\el=2,3$)
are the product of inverse functions $f_{[\ell]}(\bar t^\el_{i,j},\bar t^{\el-1}_{k,l})^{-1}$
where the subset $\bar t^{\el-1}_{k,l}$ is either above or on the left from the starting subset $\bar t^{\el}_{i,j}$.
\item
The third, fourth and fifth lines in \r{hr16} correspond to products for each line of the diagram with the following rule.
For the lines corresponding to $\bar t^\ell$ with $\ell<m$ (resp. $\ell=m$, resp. $\ell>m$), we form products of
 the function $f_0(\bar x,\bar y)$  (resp. $g(\bar x,\bar y)$, resp. $f_1(\bar x,\bar y)$).
 In these products, the subset $\bar x$ is on the right of the subset $\bar y$ in each of the lines of the diagram \r{hr15}.
\item
The sixth line in \r{hr16} is the product of either Cauchy or Izergin determinants based on neighboring pairs of
subsets $(\bar t^k_{i,j},\bar t^{k-1}_{i,j})$ belonging to the same column of a diagram associated to some $\ell$.

For $\el=1,\ldots,m-1$ and any pair $(\bar t^k_{i,j},\bar t^{k-1}_{i,j})\equiv (\bar x,\bar y)$, we use
\begin{itemize}
\item the Izergin determinant $K_0(\bar x|\bar y)$ if $\el+1\leq k\leq j\leq m-1$;
\item the normalized Cauchy determinant $\Cres(\bar x|\bar y)$ \r{hr66} if $\el+1\leq k\leq m\leq j\leq N$;
\item the Izergin determinant $K_1(\bar x|\bar y)$ if
$m+1\leq k\leq j\leq N$.
\end{itemize}
For $\el=m,\ldots,N-1$ and any pair $(\bar t^k_{i,j},\bar t^{k-1}_{i,j})\equiv (\bar x,\bar y)$, we use
 the Izergin determinant $K_1(\bar x|\bar y)$ if $\el+1\leq k\leq j\leq N$.

Remark that the asymmetry between the cases $\el<m$ and $\el\geq m$ is due to the hierarchical relation
\eqref{hr3}, which is based on the series of inclusions $\mathfrak{gl}(m|n)\supset\mathfrak{gl}(m-1|n)\supset\dots
\supset\mathfrak{gl}(1|n)\supset\mathfrak{gl}(n)$.

In our example of the diagram \r{hr15} there are four such pairs
$(\bar t^2_{1,2}, \bar t^1_{1,2})$, $(\bar t^2_{1,3}, \bar t^1_{1,3})$, $(\bar t^3_{1,3}, \bar t^2_{1,3})$, and
$(\bar t^3_{2,3}, \bar t^2_{2,3})$.
There is no $K_0(\bar x|\bar y)$ determinant in this example, however they can appear for higher $m$.
For instance, this determinant appears in the Bethe vector for the algebra $\mathfrak{gl}(3|2)$ and
constructed by the diagram given in formula \r{hr1515} for the pair of subsets $(\bar t^2_{1,2},\bar t^1_{1,2})$.
\item
The next line is given by the ordered products of monodromy matrix elements $\LL_{i,j}$ with $i<j$
and depending on the subsets $\bar t^i_{i,j-1}$. It is a usual product for
even matrix elements (i.e. when $\prt{i}+\prt{j}=0\mod2$) and a normalized product
\r{T-odd} otherwise. The order of the product is from top to bottom for the lines and right  to left within a line, as it
should be clear by comparing the penultimate line in \r{hr16}
and the diagram \r{hr15}.
\item Last line in \r{hr16} is the product of the diagonal matrix elements depending on the
remaining subsets of
 Bethe parameters that were not used in the previous line. The index $\LL_{i,i}$ of the diagonal
matrix element coincides with the number of the line in the diagram. The order in their product
is not relevant, because they commute
when the pre-Bethe vector \r{hr16} is acting onto the vacuum vector $\rvec$.
\end{itemize}

\paragraph{The case $m=3$ and $n=2$.}  The Bethe vectors in this case can be constructed
by the rules described above using following table of partitions of the Bethe parameters $\bar t^1$,
$\bar t^2$, $\bar t^3$ and $\bar t^4$    {\small%
\begin{equation}\label{hr1515}
 \begin{array}{ccccccccccccccccccccc}
\bar t^1&:&\bar t^1_{1,1}&\!\!\cup&\!\bar t^1_{1,2}&\cup&\bar t^1_{1,3}&\cup&\bar t^1_{1,4}&&&&&&&&&&&\\[1mm]
\bar t^2&:&&&\bar t^2_{1,2}&\cup&\bar t^2_{1,3}&\cup&\bar t^2_{1,4}&\cup&
\bar t^2_{2,2}&\cup&\bar t^2_{2,3}&\cup&\bar t^2_{2,4}&&&&&&\\[1mm]
\bar t^3&:&&&&&\bar t^3_{1,3}&\cup&\bar t^3_{1,4}&&\cup&&
\bar t^3_{2,3}&\cup&\bar t^3_{2,4}&\cup&\!\!\bar t^3_{2,2}&\!\!\cup&\!\!\bar t^3_{3,4}&&\\[1mm]
\bar t^4&:&&&&&&&\bar t^4_{1,4}&&&\cup&&&
\bar t^4_{2,4}&&\!\!\cup&&\!\!\bar t^4_{3,4}&\!\!\cup&\!\!\bar t^4_{4,4}
\end{array}
\end{equation}
}

\subsection{Bethe vectors $\hat{\BBB}(\bar t)$}

Quite analogously one can obtain a hierarchical relation for the Bethe vector \r{bv}
defined by means of the second current realization of Yangian double $DY(\mathfrak{gl}(m|n))$
which are compatible with the embedding of $DY(\mathfrak{gl}(m|n-1))$ into
$DY(\mathfrak{gl}(m|n))$. Another possibility to obtain this hierarchial relation is to apply a special mapping
to the formulas \eqref{hr3}, \eqref{hr7}. This morphism was discussed in the work \cite{PakRS17}. It maps the Bethe vectors
$\BBB(\bar t)$ of $DY(\mathfrak{gl}(m|n))$  to the Bethe vectors
$\hat\BBB(\bar t)$ of $DY(\mathfrak{gl}(n|m))$,  see \eqref{morphism} and the discusion that follows.
Thus, using this mapping and replacing $m\leftrightarrow n$ we can obtain the explicit
hierarchical relation for the Bethe vector $\hat\BBB(\bar t)$. We do not give here this explicit relation, however, we give
an analog of \eqref{hr77} for the Bethe vector $\hat\BBB(\bar t)$.

Again, for all $\el=1,\ldots,N$ we introduce similarly to \r{hr8} a partition of the sets of the Bethe parameters
\begin{equation}\label{hr8h}
\bar t^\el=\bigcup_{q=1}^\el\bigcup_{q'=\el}^N \bar t^\el_{q',q}\,,
\end{equation}
indexed by a pair of positive integers
\begin{equation*}
1\leq q\leq\el\leq q'\leq N\,.
\end{equation*}
We also introduce ordering rules '$\succ$' and  '$\succcurlyeq$' for these pairs according to the following
conventions
\begin{equation*}
p',p \succ q',q\quad\mbox{if}\quad p'>q',\ \forall p,q
\quad\mbox{or}\quad p'=q', p>q
\end{equation*}
and
\begin{equation*}
p,p'\succcurlyeq q,q'  \quad\mbox{if}\quad p'>q',\ \forall p,q
\quad\mbox{or}\quad p'=q',\  p>q
\quad\mbox{or}\quad p'=q',\ p=q.
\end{equation*}

In this notation  we have the following expression for the
Bethe vector:
\begin{equation*}
\hat{\BBB}(\bar t)=\hat{\Bp}(\bar t)\rvec
\end{equation*}
where pre-Bethe vector $\hat{\Bp}(\bar t)$ is given by the sum
over partitions \r{hr8h}
\begin{equation}\label{hr77h}
\begin{split}
\hat{\Bp}(\bar t)&=\sum_{\rm part}
\prod_{p',p\,\succcurlyeq \, q',q}\prod_{\el=1}^{N-1}
f_{\prt{\el+1}}(\bar t^{\el+1}_{p',p},\bar t^\el_{q',q})^{-1}
\prod_{p',p\,\succ\, q',q}
g(\bar t^m_{q',q},\bar t^{m}_{p',p})
\prod_{\substack{\el=1\\ \el\not=m}}^{N}
f_{\prt{\el+1}}(\bar t^{\el}_{p',p},\bar t^\el_{q',q})\\
&\quad\times
\prod_{q'=m+2}^{N}\prod_{q=m+1}^{q'-1}\prod_{\el={m+1}}^{q'-1}
K_{\prt{\el}}(\bar t^{\el+1}_{q',q}|\bar t^{\el}_{q',q})
\prod_{q'=m+1}^{N}\prod_{q=1}^{m}\prod_{\el=m}^{q'-1} \hat\Cres(\bar t^{\el+1}_{q',q}|\bar t^\el_{q',q})\\
&\quad\times
\prod_{q'=m+1}^{N}\prod_{q=1}^{m-1}\prod_{\el=q}^{m-1} K_{\prt{\el}}(\bar t^{\el+1}_{q',q}|\bar t^\el_{q',q})
\prod_{q'=2}^{m}\prod_{q=1}^{q'-1}\prod_{\el=q}^{q'-1} K_{\prt{\el}}(\bar t^{\el+1}_{q',q}|\bar t^\el_{q',q})
\\
&\quad\times
\prod^{\longleftarrow}_{N\geq q'\geq m}
\sk{\prod^{\longrightarrow}_{1\leq q\leq m}
\mathbb{T}_{q,q'+1}(\bar t^{q'}_{q',q})
\prod^{\longrightarrow}_{m< q\leq q'}
{\LL}_{q,q'+1}(\bar t^{q'}_{q',q})}\\
&\quad\times
\prod^{\longleftarrow}_{m>q'\geq 1}
\sk{\prod^{\longrightarrow}_{1\leq q\leq q'}
{\LL}_{q,q'+1}(\bar t^{q'}_{q',q})}
\prod_{\el=1}^{N-1} \prod_{q'=\el+1}^N\prod_{q=1}^{\el}  \LL_{\el+1,\el+1}(\bar t^{\el}_{q',q}),
\end{split}
\end{equation}
 where in contrast to \r{hr66} we  normalize Cauchy determinant $\hat\Cres(\bar y|\bar x)$ differently
\begin{equation*}
\hat\Cres(\bar y|\bar x)=g(\bar x,\bar y)h(\bar y,\bar y) =\Cres(\bar x|\bar y).
\end{equation*}

 The partitions of the Bethe parameters used in \r{hr77h}  can be also
pictured by the ordered union of diagrams similar to \r{table}
\begin{equation*}
\mathop{\bigcup}\limits^{\longleftarrow}_{\el=N,\ldots,1}\quad
\begin{array}{ccccccccc}
\bar t^{\el}_{\el,\el}&\cup&\bar t^{\el}_{\el,\el-1}&\cup&\cdots&\cup&\bar t^{\el}_{\el,2}&\cup&\bar t^{\el}_{\el,1}\\[2mm]
&&\bar t^{\el-1}_{\el,\el-1}&\cup&\cdots&\cup&\bar t^{\el-1}_{\el,2}&\cup&\bar t^{\el-1}_{\el,1}\\[2mm]
&&&&\ddots&&\vdots&&\vdots\\[2mm]
&&&&&&\bar t^{2}_{\el,2}&\cup&\bar t^{2}_{\el,1}\\[2mm]
&&&&&&&&\bar t^{1}_{\el,1}
\end{array}
\end{equation*}
The ordering here is opposite to the one used in table \r{col-tab}.  This means that the
triangle for smaller $\el$ in \r{col-tab} is on the right of the larger $\el$ triangle.
The  subsets in a given  column have all the same cardinality again.
The subsets which describe partitions of the same type of Bethe parameters are aligned along the same row of the
table (see examples of such tables in \r{hr1414} and \r{hr15hat}).

Remark that the two realizations \r{hr77} and \r{hr77h} are related by the morphism $\vph$
defined by \cite{PakRS17}
\begin{equation}\label{morphism}
\vph:\ \begin{cases}
DY(\mathfrak{gl}(m|n))&\to\qquad DY(\mathfrak{gl}(n|m))\,,\\
\quad T_{ij}(x) &\to\ (-1)^{[i][j]+[j]+1}\,\wt T_{j'i'}(x)\,,\quad k'=m+n+1-k\,.
\end{cases}
\end{equation}
Indeed, starting from the pre-Bethe vector $\Bp(\bar t)\in DY(\mathfrak{gl}(m|n))$ and applying $\vph$ we get
the pre-Bethe vector $(-1)^{\#\bar t-\#\bar t^m}\,\hat{\Bp}(\bar s)\in DY(\mathfrak{gl}(n|m))$,
where the set $\bar t$ is divided in subsets $\bar t^\ell_{i,j}$ obeying \r{hr8}, while the set $\bar s$ is divided into
subsets $\bar s^\ell_{i,j}$ obeying \r{hr8h}. The relation between these partitions is given by
$\bar t^\ell_{i,j}=\bar s^{\ell'-1}_{i'-1,j'-1}$ with $k'=m+n+1-k$, $\forall k$. In particular, when $m=n$,
 we have $\vph\big(\Bp(\bar t)\big) = (-1)^{\#\bar t-\#\bar t^m}\,\hat{\Bp}(\bar s)$, as it can be checked on the example $m=n=2$ given in \r{hr16} and \r{hr16hat}.

\subsubsection*{Examples}
For $m=2$ and $n=1$ the partition  \r{hr8h} can be pictured by the  table
\begin{equation}\label{hr1414}
\begin{array}{ccccccccccccc}
\bar t^2&:&\bar t^2_{2,2}&\cup&\bar t^2_{2,1}&\cr
\bar t^1&:&&&\bar t^1_{2,1}&\cup&\bar t^1_{1,1}\cr
\end{array}
\end{equation}
and formula \r{hr77h} reduces to
\begin{equation}\label{hr13h}
\begin{split}
\hat{\Bp}^{(2|1)}(\bar t^1,\bar t^2)&=\sum_{\rm part}
f(\bar t^2,\bar t^1)^{-1} f(\bar t^1_{2,1},\bar t^1_{1,1})g(\bar t^2_{2,2},\bar t^2_{2,1})
K_0(\bar t^2_{2,1}|\bar t^1_{2,1})\\
&\qquad\times \mathbb{T}_{1,3}(\bar t^2_{2,1})\  \mathbb{T}_{2,3}(\bar t^2_{2,2})\
{\LL}_{1,2}(\bar t^1_{1,1})\ \LL_{2,2}(\bar t^1_{2,1}),
\end{split}
\end{equation}
and yields \r{sbvc17} after identification
$\bar t^1_{1,1}\equiv \bar u_{\st}$,   $\bar t^1_{2,1}\equiv \bar u_{\so}$,
 $\bar t^2_{2,2}\equiv \bar v_{\st}$, $\bar t^2_{2,1}\equiv \bar v_{\so}$.

In the case $m=2$ and $n=2$ the  partitions
\r{hr8h} can be described by the following  union of diagrams
\begin{equation}\label{hr15hat}
\begin{array}{ccccccccccccc}
\bar t^3&:&\bar t^3_{3,3}&\cup&\bar t^3_{3,2}&\cup&\bar t^3_{3,1}&&&&&&\\[2mm]
\bar t^2&:&&&\bar t^2_{3,2}&\cup&\bar t^2_{3,1}&\cup&\bar t^2_{2,2}&\cup&\bar t^2_{2,1}&&\\[2mm]
\bar t^1&:&&&&&\bar t^1_{3,1}&&\cup&&\bar t^1_{2,1}&\cup&\bar t^1_{1,1}
\end{array}
\end{equation}
 According to this table,  formula \r{hr77h} takes the form
\begin{equation}\label{hr16hat}
\begin{split}
\hat\Bp^{(2|2)}(\bar t^1,\bar t^2,\bar t^3)=\sum_{\rm part}\
&\phantom{\times}f_0(\bar t^2,\{\bar t^1_{2,1}\cup \bar t^1_{1,1} \})^{-1}
f_0(\{\bar t^2_{3,2}\cup\bar t^2_{3,1}\} ,\bar t^1_{3,1})^{-1}\\
&\times f_1(\bar t^3_{3,1},\{\bar t^2_{3,1}\cup\bar t^2_{2,2}\cup \bar t^2_{2,1}\})^{-1}
f_1(\{\bar t^3_{3,3}\cup \bar t^3_{3,2} \},\bar t^2)^{-1}\\
&\times f_0(\bar t^1_{3,1},\{\bar t^1_{2,1} \cup \bar t^1_{1,1}  \}) f_0(\bar t^1_{2,1},\bar t^1_{1,1})\\
&\times g(\{\bar t^2_{3,1} \cup \bar t^2_{2,2}\cup \bar t^2_{2,1}  \},\bar t^2_{3,2})
g(\{ \bar t^2_{2,2}\cup \bar t^2_{2,1}  \},\bar t^2_{3,1} )  g( \bar t^2_{2,1},\bar  t^2_{2,2} ) \\
&\times f_1(\{\bar t^3_{3,3}\cup \bar t^3_{3,2}\},\bar t^3_{3,1}) f_1(\bar t^3_{3,2}, \bar t^3_{3,3})\\
&\times K_0(\bar t^2_{2,1}|\bar t^1_{2,1}) K_0(\bar t^2_{3,1}|\bar t^1_{3,1})
 \hat\Cres(\bar t^3_{3,1}|\bar t^2_{3,1})\hat\Cres(\bar t^3_{3,2}|\bar t^2_{3,2}) \\
&\times \mathbb{T}_{1,4}(\bar t^3_{3,1})\mathbb{T}_{2,4}(\bar t^3_{3,2})\  \LL_{3,4}(\bar t^3_{3,3})\
\mathbb{T}_{1,3}(\bar t^2_{2,1})\ \mathbb{T}_{2,3}(\bar t^2_{2,2})\  \LL_{1,2}(\bar t^1_{1,1})\\
&\times \LL_{2,2}(\bar t^1_{3,1})\ \LL_{2,2}(\bar t^1_{2,1})\ \LL_{3,3}(\bar t^2_{3,2})\ \LL_{3,3}(\bar t^2_{3,1}).
\end{split}
\end{equation}
Comparing \r{hr16hat} and  the diagram \r{hr15hat} one can formulate the rules to associate with
a partition diagram the explicit formula for the Bethe vector similar to those formulated in the previous subsection.
We leave this exercise to the interested reader.

\subsection{Dual Bethe vectors and examples for $DY(\mathfrak{gl}(2|1))$}

In order to obtain explicit expressions for the dual Bethe vectors
$\CCC(\bar t)$ and $\hat\CCC(\bar t)$ we have to exploit the definition and
the properties of the antimorphism \r{antimo} and \r{pro-anti}. { It is clear that for
even operators $\Psi(T_{i,j}(\bu))=T_{j,i}(\bu)$.}
Consider an odd monodromy matrix element $\LL_{i,j}(u)$ for $i<j$. Then
it means that $\prt{i}=0$ and $\prt{j}=1$ and it follows from the
commutation relations \r{NCR-7}  that for any set $\bu$ with cardinality
$\#\bu=a$ the product
$\mathbb{T}_{i,j}(\bu)$ given by equality \r{T-odd} is symmetric
with respect to permutations of the parameters $u_i$.

For the odd monodromy  matrix element $\LL_{i,j}(u)$ with $i>j$
and the set $\bu$ we define a product
\begin{equation*}
\mathbb{T}_{i,j}(\bu)=\Delta'_{h}(\bu)^{-1} \LL_{i,j}(u_1)\LL_{i,j}(u_2)
\cdots \LL_{i,j}(u_{a-1})\LL_{i,j}(u_a)\,,
\end{equation*}
which is also symmetric with respect to the permutations in $\bu$ due
to the commutation relations \r{NCR-7}.

Let us apply the antimorphism \r{antimo} to the product $\mathbb{T}_{i,j}(\bu)$ with
$i<j$. Using \r{pro-anti} we obtain for $i<j$
\begin{equation*}
\begin{split}
&\Psi\sk{\mathbb{T}_{i,j}(\bu)}=\Delta_h(\bu)\Psi\sk{\LL_{i,j}(u_1)\LL_{i,j}(u_2)
\cdots \LL_{i,j}(u_{a-1})\LL_{i,j}(u_a)}\\
&\quad=(-)^{a(a-1)/2}\Delta_h(\bu)\Psi\sk{\LL_{i,j}(u_a)}\Psi\sk{\LL_{i,j}(u_{a-1})}\cdots
\Psi\sk{\LL_{i,j}(u_2)}\Psi\sk{\LL_{i,j}(u_1)}\\
&\quad=(-)^{a(a-1)/2}\Delta'_h(\bu)\Psi\sk{\LL_{i,j}(u_1)}\Psi\sk{\LL_{i,j}(u_{2})}\cdots
\Psi\sk{\LL_{i,j}(u_{a-1})}\Psi\sk{\LL_{i,j}(u_a)}\\
&\quad=(-)^{a(a-1)/2}\mathbb{T}_{j,i}(\bu).
\end{split}
\end{equation*}
Analogously we can calculate for $i<j$ that
\begin{equation}\label{PsiT}
\Psi\sk{\mathbb{T}_{j,i}(\bu)}=(-)^{a(a+1)/2}\mathbb{T}_{i,j}(\bu),
\end{equation}
taking into account that in this case
\begin{equation*}
\Psi\sk{\LL_{j,i}(u)}=(-)^{\prt{j}(\prt{i}+1)}\LL_{i,j}(u)=-\LL_{i,j}(u).
\end{equation*}

Relation \r{PsiT} shows that for any $i,j$ such that $\prt{i}+\prt{j}=1$
\begin{equation*}
\Psi\sk{\Psi\sk{\mathbb{T}_{i,j}(\bu)}}=(-)^a\mathbb{T}_{i,j}(\bu),
\end{equation*}
and the antimorphism $\Psi$ is idempotent of the forth order.

Thus, we have described the action of the antimorphism $\Psi$ onto symmetric products of even and odd operators. Applying this action to the  pre-Bethe vectors $\Bp(\bar t)$ \r{hr77} and $\hat\Bp(\bar t)$ \r{hr77h} we respectively obtain explicit expressions for the dual pre-Bethe vectors $\Cp(\bar t)$
and $\hat\Cp(\bar t)$. Up to the sign factor they are still given by \r{hr77} and \r{hr77h} with opposite order of the operator products and the transposition
$T_{i,j}\to T_{j,i}$.
Let us give explicit formulas for a particular case of (dual) Bethe vectors $\BBB(\bar t)$,
$\hat\BBB(\bar t)$, $\CCC(\bar t)$ and
$\hat\CCC(\bar t)$
defined by the formulas \r{hr13}, \r{hr13h} and related to the
Yangian double $DY(\mathfrak{gl}(2|1))$.
 Then, we have two sets of Bethe parameters $\bar t^\el$ with
 cardinalities $\#\bar t^\el=\rr_\el$, $\el=1,2$, which
 we rename as  $\bar t^1\equiv \bu$ and
$\bar t^2\equiv \bv$ with cardinalities $\rr_1=a$ and $\rr_2=b$. Formulas  \r{uBV1}, \r{dual1},
\r{bv} and \r{dbv} for these Bethe vectors
acquire the form
\begin{equation}\label{cain8}
\begin{split}
\BBB_{a,b}(\bu,\bv)&= f(\bv,\bu)^{-1}
\sum g(\bv_{\so},\bu_{\so})
f(\bu_{\so},\bu_{\st})g(\bv_{\st},\bv_{\so})h(\bu_{\so},\bu_{\so})\\
&\quad \times \mathbb{T}_{1,3}(\bu_{\so})\LL_{1,2}(\bu_{\st})
\mathbb{T}_{2,3}(\bv_{\st})\lambda_2(\bv_{\so})  \rvec,
\end{split}
\end{equation}
\begin{equation}\label{sbvc17}
\begin{split}
\hat\BBB_{a,b}(\bu,\bv)&=
 f(\bv,\bu)^{-1} \sum {K_p(\bv_{\so}|\bu_{\so})}
f(\bu_{\so},\bu_{\st})g(\bv_{\st},\bv_{\so})\\
& \times
\mathbb{T}_{1,3}(\bv_{\so})\ \mathbb{T}_{2,3}(\bv_{\st})
\LL_{1,2}(\bu_{\st}) \lambda_2(\bu_{\so})\rvec,
\end{split}
\end{equation}
\begin{equation}\label{ecain9}
\begin{split}
\CCC_{a,b}(\bu,\bv)= (-)^{\rr_2(\rr_2-1)/2} &\ f(\bv,\bu)^{-1}
\sum
g(\bv_{\so},\bu_{\so})\  f(\bu_{\so},\bu_{\st})\
g(\bv_{\st},\bv_{\so})\  h(\bu_{\so},\bu_{\so})\    \\
&\times \lvec\lambda_{2}(\bv_{\so})\
\mathbb{T}_{3,2}(\bv_{\st})\cdot
\LL_{2,1}(\bu_{\st}) \cdot
\mathbb{T}_{3,1}(\bu_{\so}),
\end{split}
\end{equation}
\begin{equation}\label{anau3}
\begin{split}
\hat\CCC_{a,b}(\bu,\bv)=
(-)^{\rr_2(\rr_2-1)/2} &\  f(\bv,\bu)^{-1} \sum
{K_p(\bv_{\so}|\bu_{\so})}
f(\bu_{\so},\bu_{\st})g(\bv_{\st},\bv_{\so})\\
& \times
\lvec \lambda_2(\bu_{\so})\ \LL_{2,1}(\bu_{\st}) \
\mathbb{T}_{3,2}(\bv_{\st})\
\mathbb{T}_{3,1}(\bv_{\so}),
\end{split}
\end{equation}
where the sums run over partitions of the sets
$\bu\Ra\{\bu_{\so},\bu_{\st}\}$ and $\bv\Ra\{\bv_{\so},\bv_{\st}\}$ such that
$\#\bu_{\so}=\#\bv_{\so}=p\leq\min(a,b)$.

Formulas \r{cain8}--\r{anau3} were already used in the series of papers
\cite{HLPRS1,HLPRS2,HLPRS3} to calculate the
form factors of the monodromy matrix elements in the supersymmetric quantum integrable
models associated with the super-Yangian  $Y(\mathfrak{gl}(2|1))$.

\section*{Acknowledgements}

N.A.S. thanks LAPTH in Annecy-le-Vieux for the hospitality and stimulating scientific atmosphere,
and CNRS for partial financial support. The work of A.L. has been funded by the
Russian Academic Excellence Project 5-100 and by joint NASU-CNRS project F14-2016. The
work of S.P. was supported in part by the RFBR grant 14-01-00547-a. N.A.S. was supported
by the grant RFBR-15-31-20484-mol-a-ved.

\appendix

\section{Composed currents and Gauss coordinates}
\label{CCGC}

In the completed algebras $\oU_F$, $\oU_E$, $\hat\oU_F$, and $\hat\oU_E$ a product of the total currents
has specific analytical properties. This means that if one performs the normal ordering of the current generators
in these products, then one can observe the pole structure of this product, which is
encoded into commutation relations of the total currents. This normal ordering procedure demonstrates that
the products $F_i(u)F_{i+1}(v)$, $E_{i+1}(v)E_{i}(u)$, $\tiF_{i+1}(v)\tiF_{i}(u)$, and  $\tiE_i(u)\tiE_{i+1}(v)$  have
simple poles at $u=v$. We define the composed currents $F_{j,i}(u)$, $E_{i,j}(u)$, $\tiF_{j,i}(u)$ and $\tiE_{i,j}(u)$
for $1\leq i<j\leq m+n$ inductively as the residues
\begin{equation}\label{F-com}
F_{j,i}(v)=\mathop{{\rm res}}\limits_{u=v} F_{a,i}(v)F_{j,a}(u)=-\mathop{{\rm res}}\limits_{u=v} F_{a,i}(u)F_{j,a}(v)\,,
\end{equation}
\begin{equation}\label{E-com}
E_{i,j}(v)=\mathop{{\rm res}}\limits_{u=v} E_{a,j}(u)E_{i,a}(v)=-\mathop{{\rm res}}\limits_{u=v} E_{a,j}(v)E_{i,a}(u)\,,
\end{equation}
\begin{equation}\label{tF-com}
\tiF_{j,i}(v)=\mathop{{\rm res}}\limits_{u=v} \tiF_{j,a}(u)\tiF_{a,i}(v)=-\mathop{{\rm res}}\limits_{u=v}
 \tiF_{j,a}(v)\tiF_{a,i}(u)\,,
\end{equation}
\begin{equation}\label{tE-com}
\tiE_{i,j}(v)=\mathop{{\rm res}}\limits_{u=v} \tiE_{i,a}(v)\tiE_{a,j}(u)=
-\mathop{{\rm res}}\limits_{u=v} \tiE_{i,a}(u)\tiE_{a,j}(v)\,,
\end{equation}
where $i<a<j$ and we denoted the simple root currents as follows: $F_i(u)\equiv F_{i+1,i}(u)$, $E_i(u)\equiv E_{i,i+1}(u)$,
$\tiF_i(u)\equiv \tiF_{i+1,i}(u)$, and $\tiE_i(u)\equiv \tiE_{i,i+1}(u)$.

Calculating the residues via the commutation relations \r{FiFii}, \r{EiEii}, \r{tFiFii}, and \r{tEiEii} in \r{F-com}--\r{tE-com}
respectively we obtain
\begin{equation}\label{cFr}
F_{j,i}(v)=\cci{i+1}\cdots \cci{j-1} F_{j,j-1}(v)F_{j-1,j-2}(v)\cdots F_{i+1,i}(v)\,,
\end{equation}
\begin{equation}\label{cEr}
E_{i,j}(v)=\cci{i+1}\cdots \cci{j-1} E_{i,i+1}(v)E_{i+1,i+2}(v)\cdots E_{j-1,j}(v)\,,
\end{equation}
\begin{equation}\label{tcFr}
\tiF_{j,i}(v)=\cci{i+1}\cdots \cci{j-1} \tiF_{i+1,i}(v)\tiF_{i+2,i+1}(v)\cdots \tiF_{j,j-1}(v)\,,
\end{equation}
\begin{equation}\label{tcEr}
\tiE_{i,j}(v)=\cci{i+1}\cdots \cci{j-1} \tiE_{j-1,j}(v)\tiE_{j-2,j-1}(v)\cdots \tiE_{i,i+1}(v)\,.
\end{equation}

Let us prove one of these formulas, namely \r{cFr}.
Consider  \r{F-com} for $j=i+2$ and $a=i+1$.
Since we know that the product $F_{i+1,i}(v)F_{i+2,i+1}(u)$ has a simple
pole at $u=v$, we can calculate
the residue in \r{F-com} as follows
\begin{equation*}
\begin{split}
F_{i+2,i}(v)&=\mathop{{\rm res}}\limits_{u=v} F_{i+1,i}(v)F_{i+2,i+1}(u)=
(u-v)F_{i+1,i}(v)F_{i+2,i+1}(u)\Big|_{u=v}\\
&=(u-v+\cci{i+1})F_{i+2,i+1}(u)F_{i+1,i}(v)\Big|_{u=v}=\cci{i+1}F_{i+2,i+1}(v)F_{i+1,i}(v)\,.
\end{split}
\end{equation*}
Here  we used the commutation relation \r{FiFii} to transform the first line into the second one.
Now we repeat the same calculation to define the current $F_{i+3,i}(v)$ using the simple root current
$F_{i+3,i+2}(u)$ and the composed current $F_{i+2,i}(v)$ that we just calculated.
Due to the commutativity of $F_{i+3,i+2}(u)$ and
$F_{i+1,i}(v)$ we obtain that
\begin{equation*}
F_{i+3,i}(v)=\cci{i+1}\cci{i+2} F_{i+3,i+2}(v)F_{i+2,i+1}(v)F_{i+1,i}(v).
\end{equation*}
 Iterating the calculation, we get formula \r{cFr}. Formulas \r{cEr}--\r{tcEr}
 can be proved completely analogously.

The composed currents are important in calculating the universal Bethe
vectors using formulas \r{uBV1} and \r{bv}.
In this section we show that the projections of the composed
currents discussed in section \ref{gen-act} coincide  with the Gauss coordinates
of the universal monodromy matrix \r{GF2}--\r{GE2} and \r{GF1}--\r{GE1}, up to some unessential prefactors.
To do this we rewrite the defining formulas for the composed currents in an integral form.

Both equations in \r{F-com} can be written through the contour integrals
\begin{equation}\label{iF}
\begin{split}
F_{j,i}(v)&=-\oint_{C_0}du\ F_{a,i}(v)F_{j,a}(u)+\oint_{C_\infty}du\ \frac{u-v+\cci{a}}{(u-v)_{>}}F_{j,a}(u)F_{a,i}(v)\\
&=-\oint_{C_\infty}du\ F_{a,i}(u)F_{j,a}(v)+\oint_{C_0}du\ \frac{u-v-\cci{a}}{(u-v)_{<}}F_{j,a}(v)F_{a,i}(u)
\end{split}
\end{equation}
where $C_0$ and $C_\infty$ are small closed contours around the points 0 and $\infty$ on the complex plane $u$.
The rational functions  $\frac{1}{(u-v)_{\lessgtr}}$ are defined by the series \r{expan}.

For any formal series $G(u)=\sum_{\ell\in\ZZ} G\mode{\ell} u^{-\ell-1}$, we define $G^{(\pm)}(u)$ as
\begin{equation}\label{ghc}
G^{(\pm)}(u)=\pm\sum_{\substack{\ell\geq 0\\ \ell<0}} G\mode{\ell} u^{-\ell-1}\,.
\end{equation}
It is obvious that the half-currents $F^{(\pm)}$ and  $E^{(\pm)}$ coincide with the corresponding projections of the
currents only for the simple root currents $F_i(u)$ and $E_i(u)$. For the composed currents this is not the case, but nevertheless one can prove that
\begin{equation}\label{vanish}
\begin{split}
\Pfp\sk{F^{(-)}_{j,i}(u)\cdot \eF}&=0\,,\qquad \Pfm\sk{\eF\cdot F^{(+)}_{j,i}(u)}=0\,,\\
\Pep\sk{\eE\cdot E^{(-)}_{i,j}(u)}&=0\,,\qquad \Pem\sk{ E^{(+)}_{j,i}(u) \cdot \eE}=0\,,
\end{split}
\end{equation}
for any elements $\eF\in\oU_F$ and $\eE\in\oU_E$. Similar properties can be formulated for
the projections $\hPfpm$ and $\hPepm$.

Using the notation \r{ghc} and calculating the formal contour integrals in \r{iF} as
\begin{equation}\label{in-cal}
\oint_{C_0}du\ G(u)=\oint_{C_\infty}du\ G(u)=G\mode0
\end{equation}
we obtain the following expressions for the composed currents $F_{j,i}(v)$:
\begin{equation}\label{iF1}
\begin{split}
F_{j,i}(v)&= [F_{j,a}\mode0,F_{a,i}(v)]-\cci{a}F^{(-)}_{j,a}(v)F_{a,i}(v)\\
&=[F_{j,a}(v),F_{a,i}\mode0]+\cci{a}F_{j,a}(v)F^{(+)}_{a,i}(v).
\end{split}
\end{equation}

For the composed currents $E_{i,j}(v)$ defined by the formula \r{E-com} we have
\begin{equation}\label{iE}
\begin{split}
E_{i,j}(v)&=-\oint_{C_0}du\ E_{a,j}(u)E_{i,a}(v)+\oint_{C_\infty}du\ \frac{u-v+\cci{a}}{(u-v)_{>}}E_{i,a}(v)E_{a,j}(u)\\
&=-\oint_{C_\infty}du\ E_{a,j}(v)E_{i,a}(u)+\oint_{C_0}du\ \frac{u-v-\cci{a}}{(u-v)_{<}}E_{i,a}(u)E_{a,j}(v),
\end{split}
\end{equation}
or using \r{in-cal} we obtain for these composed currents
\begin{equation}\label{iE1}
\begin{split}
E_{i,j}(v)&= [E_{i,a}(v),E_{a,j}\mode0]-\cci{a}E_{i,a}(v)E^{(-)}_{a,j}(v)\\
&=[E_{i,a}\mode0,E_{a,j}(v)]+\cci{a}E^{(+)}_{i,a}(v)E_{a,j}(v).
\end{split}
\end{equation}

Quite similarly, we have for the currents $\tiF_{j,i}(v)$ defined by the relation \r{tF-com}
\begin{equation}\label{itF}
\begin{split}
\tiF_{j,i}(v)&=\oint_{C_\infty}du\ \tiF_{j,a}(u)\tiF_{a,i}(v)-\oint_{C_0}du\ \frac{u-v+\cci{a}}{(u-v)_{<}}\tiF_{a,i}(v)\tiF_{j,a}(u)\\
&=\oint_{C_0}du\ \tiF_{j,a}(v)\tiF_{a,i}(u)-\oint_{C_\infty}du\ \frac{u-v-\cci{a}}{(u-v)_{>}}\tiF_{a,i}(u)\tiF_{j,a}(v),
\end{split}
\end{equation}
or calculating these formal contour integrals
\begin{equation}\label{itF1}
\begin{split}
\tiF_{j,i}(v)&= [\tiF_{j,a}\mode0,\tiF_{a,i}(v)]+\cci{a}\tiF_{a,i}(v)\tiF^{(+)}_{j,a}(v)\\
&=[\tiF_{j,a}(v),\tiF_{a,i}\mode0]-\cci{a}\tiF^{(-)}_{a,i}(v)\tiF_{j,a}(v).
\end{split}
\end{equation}

Finally, for the composed currents $\tiE_{j,i}(v)$ defined by the relation \r{tE-com} we can calculate
\begin{equation}\label{itE}
\begin{split}
\tiE_{i,j}(v)&=\oint_{C_\infty}du\ \tiE_{i,a}(v)E_{a,j}(u)-\oint_{C_0}du\ \frac{u-v+\cci{a}}{(u-v)_{<}}\tiE_{a,j}(u)\tiE_{i,a}(v)\\
&=\oint_{C_0}du\ \tiE_{i,a}(u)\tiE_{a,j}(v)-\oint_{C_\infty}du\ \frac{u-v-\cci{a}}{(u-v)_{>}}\tiE_{a,j}(v)\tiE_{i,a}(u),
\end{split}
\end{equation}
or
\begin{equation}\label{itE1}
\begin{split}
\tiE_{i,j}(v)&= [{\tiE}_{i,a}(v),{\tiE}_{a,j}\mode0]+\cci{a}\tiE^{(+)}_{a,j}(v)\tiE_{i,a}(v)\\
&=[{\tiE}_{i,a}\mode0,{\tiE}_{a,j}(v)]-\cci{a}\tiE_{a,j}(v)\tiE^{(-)}_{i,a}(v).
\end{split}
\end{equation}

\subsection{Projections of the composed currents}

Formulas \r{iF1}, \r{iE1}, \r{itF1}, and \r{itE1} are convenient for the calculation of the projections
of the composed currents. Indeed, let us consider $a=j-1$ in the first line of \r{iF1} and apply to the both
sides of this equation the 'positive' projection $\Pfp$
defined by the formula \r{proj1}. Analogously, we can consider
the second line in \r{iF1} for $a=i+1$ and apply to this equality the 'negative' projection $\Pfm$.
Using the properties of the projections \r{vanish} we have
\begin{equation}\label{iF2}
\begin{split}
\Pfp\sk{F_{j,i}(v)}&=\left[F_{j,j-1}\mode0,\Pfp\sk{F_{j-1,i}(v)}\right]\,,\\
\Pfm\sk{F_{j,i}(v)}&=\left[\Pfm\sk{F_{j,i+1}(v)},F_{i+1,i}\mode0\right]\,,
\end{split}
\end{equation}
where we have used the  commutativity of the projections with the adjoint action of the
 zero modes of the simple root currents proved in the next appendix.
Then equations \r{iF2} can be easily iterated to obtain
\begin{equation}\label{iF3}
\begin{split}
\Pfp\sk{F_{j,i}(v)}&=\Sc_{F_{j-1}\mode0}\ \Sc_{F_{j-2}\mode0}\cdots
\Sc_{F_{i+1}\mode0}\sk{\FF^+_{i+1,i}(v)}\,,\\
\Pfm\sk{F_{j,i}(v)}&=(-)^{j-i}\Sc_{F_{i}\mode0}\ \Sc_{F_{i+1}\mode0}\cdots
\Sc_{F_{j-2}\mode0}\sk{\FF^-_{j,j-1}(v)}\,,
\end{split}
\end{equation}
where we have used the relation between the projections of the simple root currents
and the Gauss coordinates: $\Pfpm(F_{i+1,i}(v))=\pm\FF^\pm_{i+1,i}(v)$.

Quite analogously we can obtain from the formulas  \r{iE1}, \r{itF1} and \r{itE1}:
\begin{equation}\label{iE3}
\begin{split}
\Pep\sk{E_{i,j}(v)}&=(-)^{j-i-1}
\Sc_{E_{j-1}\mode0}\ \Sc_{E_{j-2}\mode0}\cdots \Sc_{E_{i+1}\mode0}\sk{\EE^+_{i,i+1}(v)}\,,\\
\Pem\sk{E_{i,j}(v)}&=-\Sc_{E_{i}\mode0}\ \Sc_{E_{i+1}\mode0}\cdots
\Sc_{E_{j-2}\mode0}\sk{\EE^-_{j-1,j}(v)}\,,
\end{split}
\end{equation}
\begin{equation}\label{itF3}
\begin{split}
\hPfm\sk{\tiF_{j,i}(v)}&=-\Sc_{\tiF_{j-1}\mode0}\ \Sc_{\tiF_{j-2}\mode0}\cdots
\Sc_{\tiF_{i+1}\mode0}\sk{\tFF^-_{i+1,i}(v)}\,,\\
\hPfp\sk{\tiF_{j,i}(v)}&=(-)^{j-i-1}\Sc_{\tiF_{i}\mode0}\ \Sc_{\tiF_{i+1}\mode0}\cdots \Sc_{\tiF_{j-2}\mode0}
\sk{\tFF^+_{j,j-1}(v)}\,,
\end{split}
\end{equation}
\begin{equation}\label{itE3}
\begin{split}
\hPem\sk{\tiE_{i,j}(v)}&=(-)^{j-i}
\Sc_{\tiE_{j-1}\mode0}\ \Sc_{\tiE_{j-2}\mode0}\cdots \Sc_{\tiE_{i+1}\mode0}\sk{\tEE^-_{i,i+1}(v)}\,,\\
\hPep\sk{\tiE_{i,j}(v)}&=\Sc_{\tiE_{i}\mode0}\ \Sc_{\tiE_{i+1}\mode0}\cdots
\Sc_{\tiE_{j-2}\mode0}\sk{\tEE^+_{j-1,j}(v)}\,.
\end{split}
\end{equation}

In the rest of this section we are going to demonstrate that the 'positive' projections of the composed
currents given by the first lines in the formulas \r{iF3}, \r{iE3} and the second lines in the formulas
\r{itF3}, \r{itE3} coincide with the Gauss coordinate of the universal monodromy operator $\LL^+_{i,j}(v)$.
To do this we consider equation \r{TM-2} for $i\to i$, $j\to j-1$, $k\to j-1$, $l\to j$, and $i<j-1$
\begin{equation}\label{scr1}
[\LL^\pm_{i,j-1}(u), \LL^+_{j-1,j}(v)]=\frac{\cci{j-1}}{u-v}\sk{\LL^\pm_{i,j}(u)\LL^+_{j-1,j-1}(v) -
\LL^+_{i,j}(v)\LL^\pm_{j-1,j-1}(u)}\,.
\end{equation}
To obtain \r{scr1} from \r{TM-1} we take into account that $(-)^{(\prt{i}+\prt{j-1})(\prt{j-1}+\prt{j})}=1$
for any $i,j$ satisfying $i<j-1$ and  the sign factor $(-)^{\prt{j}(\prt{i}+\prt{j-1})+\prt{i}\prt{j-1}}$
is equal to $(-)^{\prt{j-1}}$.

One can easily see from the Gauss decompositions and the modes expansions of the
Gauss coordinates \r{L-oper} that the zero modes of the monodromy matrix elements
\begin{equation}\label{zmme}
\begin{split}
\mathop{\rm res}\limits_{v\to\infty}v\LL^+_{i,i+1}(v)&=
(\LL^+_{i,i+1})\mode0=(\FF^+_{i+1,i})\mode0=(\tFF^+_{i+1,i})\mode0=F_i\mode0=\tiF_i\mode0\,,\\
\mathop{\rm res}\limits_{v\to\infty}v\LL^+_{i+1,i}(v)&=
(\LL^+_{i+1,i})\mode0=(\EE^+_{i,i+1})\mode0=(\tEE^+_{i,i+1})\mode0=E_i\mode0=\tiE_i\mode0\,,
\end{split}
\end{equation}
coincide with the zero modes of the corresponding currents.

Let us multiply equation \r{scr1} by $v$ and send $v\to\infty$. Due to the formulas \r{zmme}
this relation becomes
\begin{equation}\label{Fzm0}
\cci{j-1}\ \LL^\pm_{i,j}(u)=\Sc_{F_{j-1}\mode0}\sk{\LL^\pm_{i,j-1}(u)}\,,
\end{equation}
or
\begin{equation}\label{Fzm}
\cci{j-1}\ \sk{\FF^\pm_{j,i}(u)k_i^\pm(u)+\cdots }=\Sc_{F_{j-1}\mode0}
\sk{\FF^\pm_{j-1,i}(u)k_i^\pm(u)+\cdots},
\end{equation}
where dots denote the terms given by the Gauss decomposition  \r{GF2}.
One can prove using weight arguments that the contribution of these terms vanishes and
due to the commutativity of the Cartan current $k^\pm_i(u)$ with the zero mode $F_{j-1}\mode0$ at $i<j-1$
we obtain from \r{Fzm}
\begin{equation*}
\cci{j-1}\ \FF^\pm_{j,i}(u) =\Sc_{F_{j-1}\mode0}\sk{\FF^\pm_{j-1,i}(u)}.
\end{equation*}
Iterating this relation for the 'positive' Gauss coordinates we obtain
\begin{equation}\label{Fzm2}
c_{[i,j]}\ \FF^+_{j,i}(u) =\Sc_{F_{j-1}\mode0}\cdots \Sc_{F_{i+1}\mode0}
\sk{\FF^+_{i+1,i}(u)}=\Pfp\sk{F_{j,i}(u)},
\end{equation}
according to the first line in the relations \r{iF3}, where we denoted by $c_{[i,j]}$ the product
\begin{equation}\label{ccc}
c_{[i,j]}=\cci{i+1}\cci{i+2}\cdots \cci{j-2}\cci{j-1}\,.
\end{equation}
In particular, we set $c_{[i,i+1]}=1$.

Formula \r{Fzm2} describes the relation between the 'positive' projection of the composed currents
and the 'positive' Gauss coordinates. The relations between the 'negative' projection of the composed currents
and the 'negative' Gauss coordinates are more tricky. To find them we apply the 'negative' projection
to the first equation in \r{iF1} for $a=j-1$ to obtain
\begin{equation}\label{nrca1}
\Pfm\sk{F_{j,i}(u)} = \sk{\Sc_{F_{j-1}\mode0}-\cci{j-1}\FF^-_{j,j-1}(u)}\Pfm\sk{F_{j-1,i}(u)}\,,
\end{equation}
where we have used the equality $F^{(-)}_{j,j-1}(u)=\FF^-_{j,j-1}(u)$ between the 'negative' half-currents
and the 'negative' Gauss coordinates. Iterating \r{nrca1}
we obtain an expression for the 'negative' projection, which uses only the zero modes screening operators
and the 'negative' Gauss coordinates
\begin{equation*}
\begin{split}
\Pfm\sk{F_{j,i}(u)} &= -\sk{\Sc_{F_{j-1}\mode0}-\cci{j-1}\FF^-_{j,j-1}(u)}
\sk{\Sc_{F_{j-2}\mode0}-\cci{j-2}\FF^-_{j-1,j-2}(u)}\cdots\\
&\qquad\qquad\cdots \sk{\Sc_{F_{i+1}\mode0}-\cci{i+1}\FF^-_{i+2,i+1}(u)}
\FF^{-}_{i+1,i}(u)\,,
\end{split}
\end{equation*}
where at the last step we used the relation $\Pfm\sk{F_{i+1,i}(u)}=-\FF^-_{i+1,i}(u)$.
Developing the parentheses in the latter relation we obtain finally
\begin{equation}\label{nrca3}
\Pfm\sk{F_{j,i}(u)} =-c_{[i,j]}\sk{\FF^-_{j,i}(u)+\sum_{\ell=1}^{j-i-1}(-)^\ell
\sum_{j>i_\ell>\cdots>i_1>i} \FF^-_{j,i_\ell}(u) \cdots \FF^-_{i_2,i_1}(u)\FF^-_{i_1,i}(u)  }.
\end{equation}
This expression is very useful for the calculating the action of the monodromy matrix elements
onto Bethe vectors.

On the other hand,
we can relate the projection of the composed current given by the second line in formula  \r{itF3}
with the Gauss coordinates defined by the relation \r{GF1}. To do this we consider relation \r{TM-1} for
$i\to i$, $j=k\to i+1$, $l\to j$, and $i<j-1$, which is equivalent to
\begin{equation}\label{scr2}
[\LL^+_{i,i+1}(u), \LL^+_{i+1,j}(v)]=\frac{\cci{i+1}}{u-v}\sk{\LL^+_{i+1,i+1}(v)\LL^+_{i,j}(u) -
\LL^+_{i+1,i+1}(u)\LL^+_{i,j}(v)}\,.
\end{equation}
Again the factor  $(-)^{(\prt{i}+\prt{i+1})(\prt{i+1}+\prt{j})}=1$ becomes trivial
for any $i,j$ satisfying $i<j-1$ and  the sign factor $(-)^{\prt{i}(\prt{i+1}+\prt{j})+\prt{i+1}\prt{j}}$
is equal to $(-)^{\prt{i+1}}$.    Multiplying equation \r{scr2} by $u$ and sending $u\to\infty$
we  obtain using \r{GF1} a relation between the following Gauss coordinates:
\begin{equation}\label{tFzm1}
\cci{i+1}\ \tFF^+_{j,i}(v) =-\Sc_{\tiF_{i}\mode0}\sk{\tFF^+_{j,i+1}(v)}\,.
\end{equation}
Iterating this equality we find
\begin{equation}\label{tFzm2}
c_{[i,j]}\ \tFF^+_{j,i}(u) =(-)^{j-i-1}\Sc_{\tiF_{i}\mode0}\cdots \Sc_{\tiF_{j-2}\mode0} \sk{\tFF^+_{j,j-1}(u)}=
\hPfp\sk{\tiF_{j,i}(v)}\,.
\end{equation}
For the relation between the 'negative' projections of the composed currents and
the 'negative' Gauss coordinates we obtain
\begin{equation}\label{nrca4}
\hPfm\sk{\tiF_{j,i}(u)} =-c_{[i,j]}\sk{\tFF^-_{j,i}(u)+\sum_{\ell=1}^{j-i-1}(-)^\ell
\sum_{j>i_\ell>\cdots>i_1>i} \tFF^-_{i_1,i}(u)  \tFF^-_{i_2,i_1}(u)\cdots \tFF^-_{j,i_\ell}(u)  }\,.
\end{equation}

Using again the  \r{TM-2} for $i\to i+1$, $j\to i$, $k\to j$, $l\to i+1$, and $i<j-1$, we obtain using
analogous arguments and the Gauss decomposition  \r{GE1} a relation between the Gauss coordinate
$\tEE^+_{i,j}(v)$ and the projection of the composed current $\hPep(\tiE_{i,j}(v)$:
\begin{equation*}
c_{[i,j]}\ \tEE^+_{i,j}(v) =\Sc_{\tiE_{i}\mode0}\cdots \Sc_{\tiE_{j-2}\mode0}
\sk{\tEE^+_{j-1,j}(v)}=\hPep\sk{\tiE_{i,j}(v)}\,.
\end{equation*}
Finally, from the relation \r{TM-1} for $i\to j-1$, $j\to i$, $k\to j$, $l\to j-1$, $i<j-1$, and \r{GE2} we obtain:
\begin{equation}\label{Ezm2}
c_{[i,j]}\ \EE^+_{i,j}(u) =(-)^{j-i-1}\Sc_{E_{j-1}\mode0}\cdots
\Sc_{E_{i+1}\mode0} \sk{\EE^+_{i,i+1}(u)}=\Pep\sk{E_{i,j}(v)}.
\end{equation}

Summarizing the considerations above we conclude that the 'positive' projections of the composed currents coincide with the
corresponding Gauss coordinates of the universal monodromy operator. The formulas for the relation
of the  'negative' projections of the composed currents are a little bit more complicated and one
can obtain formulas similar to \r{nrca3} and \r{nrca4} for the other two types of composed currents
$E_{i,j}(u)$, and $\tiE_{i,j}(u)$.

\section{Commutativity of the projection and screening operators}
\label{comPS}

The adjoint actions by the zero modes of the simple root currents $F_i\mode0$, $E_i\mode0$ and
$\tiF_i\mode0$, $\tiE_i\mode0$ play an important role. For any elements $\F\in U_F$, $\eE\in U_E$,
$\hat\F\in\hat U_F$ and $\hat\eE\in \hat U_E$ we introduce the screening operators
\begin{equation}\label{sc1}
\begin{split}
\Sc_{F_i\mode0}(\F)&\equiv [F_i\mode0,\F]\,,\qquad \Sc_{E_i\mode0}(\eE)\equiv [E_i\mode0,\eE]\,,\\
\Sc_{\hat F_i\mode0}(\hat\F)&\equiv [\hat F_i\mode0,\hat\F]\,,\qquad \Sc_{\hat E_i\mode0}(\hat\eE)
\equiv [\hat E_i\mode0,\hat\eE]\,.
\end{split}
\end{equation}
One can check that the intersections between standard and current Borel subalgebras
 are all stable under respective action
of the screening operators.

Let us check, for example, that the subalgebras $U_F^\pm$ defined by \r{inter} are invariant
under the adjoint action of the screening operator $\Sc_{F_i\mode0}$ for $i=1,\ldots,N$.
It follows from \r{ordF} that any element $\F\in U_F$ can be presented in the normal
ordered form $\F=\sum_\el \F^{(-)}_\el\ot \F^{(+)}_\el$, where by definition the
elements $\F^{(\pm)}_\el\in U^\pm_F$. Then
\begin{equation*}
\Sc_{F_i\mode0}(\F)=\sum_\el \Sc_{F_i\mode0}\sk{\F^{(-)}_\el}\cdot \F^{(+)}_\el+ \sum_\el \F^{(-)}_\el\cdot
\Sc_{F_i\mode0}\sk{\F^{(+)}_\el},
\end{equation*}
and by definition of the projection $\Pfp$ \r{proj1} we have
\begin{equation}\label{sc3}
\Pfp\sk{\Sc_{F_i\mode0}(\F)}=\sum_\el \varepsilon\sk{\Sc_{F_i\mode0}\sk{\F^{(-)}_\el}}\cdot \F^{(+)}_\el+
\sum_\el \varepsilon\sk{\F^{(-)}_\el}\cdot
\Sc_{F_i\mode0}\sk{\F^{(+)}_\el}.
\end{equation}
The first sum in the right hand side of equation \r{sc3} vanishes, because
$\Sc_{F_i\mode0}\sk{\F^{(-)}_\el}\in U^-_F$
if $\varepsilon\sk{\F^{(-)}_\el}=0$. It also vanishes if $\varepsilon\sk{\F^{(-)}_\el}=1$ due to the
definition of the screening operators and the commutation relations
\begin{equation*}
\begin{split}
\Sc_{F_i\mode0}\sk{k^-_i(u)}&=\cci{i}F_i^{(-)}(u)k^-_i(u)\,,\\
\Sc_{F_i\mode0}\sk{k^-_{i+1}(u)}&=-\cci{i+1}F_i^{(-)}(u)k^-_{i+1}(u),
\end{split}
\end{equation*}
which easily follow from \r{kiF}. Since $\varepsilon\sk{\F^{(-)}_\el}\in\CC$ the equation \r{sc3}
can be rewritten in the form
\begin{equation*}
\Pfp\sk{\Sc_{F_i\mode0}(\F)}=
\Sc_{F_i\mode0}\sk{\sum_\el \varepsilon\sk{\F^{(-)}_\el}\cdot\sk{\F^{(+)}_\el}}=
\Sc_{F_i\mode0}\sk{ \Pfp(\F)},
\end{equation*}
which proves the assertion. A commutativity of the projections and the corresponding other screening
operators can be proved analogously.

\section{Calculation of the projection}
\label{appC}

Let  $\bv$ be a
 set of variables with cardinality $\#\bv=b$. Let us  consider a
product of composed currents \r{cFr}
\begin{equation}\label{gca3}
F_{j_1,i}(v_1)\cdot F_{j_2,i}(v_2)\cdots F_{j_{b-1},i}(v_{b-1})\cdot F_{j_{b},i}(v_{b})
\end{equation}
with the following restrictions for the indices of the composed currents
\begin{equation}\label{gca4}
j_1\geq j_{2} \geq \cdots \geq j_{b-1}\geq  j_b\geq i+1\,.
\end{equation}
In the previous papers on the method of projections  these products were called {\it strings}.

For any $\el,\el'=1,\ldots,N$ and $\el\leq\el'$ denote by $U_{\el,\el'}$
the subalgebra of $\overline U_F$ formed by the modes
of the currents $F_\el(t),F_{\el+1}(t),\ldots,F_{\el'}(t)$. Let $U_{\el,\el'}^\coun=U_{\el,\el'}\cap{\rm Ker}\,\coun $
be the corresponding  augmentation ideal.

\begin{proposition}
Using the commutation relations between composed currents one can prove
 the following equation:
\begin{equation}\label{gca11}
\begin{split}
&F_{i,i-1}(u_1)\cdots F_{i,i-1}(u_a)\cdot \Pfm\sk{F_{j_1,i}(v_1)\cdot F_{j_2,i}(v_2)\cdots F_{j_{b-1},i}(v_{b-1})\cdot F_{j_{b},i}(v_{b})} \\
&= \frac{\cci{i}^{-b}}{(a-b)!}\ \tSym_{\,\bu}\left[\prod_{\el=1}^b g_{\prt{i}}(v_\el,u_{\el})
\prod_{1\leq \el<\el'\leq b}f_{\prt{i}}(u_\el,u_{\el'})
\frac{f_{\prt{i}}(v_{\el'},u_{\el})}{f_{\prt{i}}(v_{\el'},v_{\el})}\prod_{\el=1}^b\prod_{\el'=b+1}^a
f(u_\el,u_{\el'})
  \right.\\
 & \left.\phantom{\prod_{\el=1}^b} \times
 F_{j_1,i-1}(u_1)\cdot F_{j_2,i-1}(u_{2})\cdots F_{j_{b},i-1}(u_{b})\cdot
 F_{i,i-1}(u_{b+1})\cdots F_{i,i-1}(u_{a})
 \right],\\
&\qquad\qquad\qquad \quad
\mbox{mod}\quad \Pfm(U^\coun_{i,j_1-1})\cdot U_{i-1,j_1-1}\;.
\end{split}
\end{equation}
\end{proposition}

Below the equalities between elements $\cA_1$ and $\cA_2$ from the subalgebra $\overline U_F$
modulo elements  $\Pfm(U^\coun_{i,j-1})\cdot U_{i-1,j-1}$ will be denoted
by the symbol $\cA_1\sieq{i}{j}\cA_2$.

Let us prove \r{gca11} step by step. First of all we observe that the
'negative' projection of the product of the composed currents \r{gca3} with restrictions \r{gca4}
can be factorized \cite{KhP-Kyoto,OPS}
\begin{equation*}
\begin{split}
&\Pfm\sk{F_{j_1,i}(v_1)\cdot F_{j_2,i}(v_2)\cdots F_{j_{b-1},i}(v_{b-1})\cdot F_{j_{b},i}(v_{b})}\\
&=\Pfm\sk{F_{j_1,i}(v_1;v_2,\ldots,v_b)}\cdot \Pfm\sk{F_{j_2,i}(v_2;v_3,\ldots,v_b)}\cdots
\Pfm\sk{F_{j_b,i}(v_b)},
\end{split}
\end{equation*}
where $F_{j,i}(v_1;v_2,\ldots,v_b)$ is the following linear combination of the composed
currents of the same type
\begin{equation}\label{gca13}
F_{j,i}(v_1;v_2,\ldots,v_b)=F_{j,i}(v_1)-\sum_{\el=2}^b h_{\prt{i}}(v_\el,v_1)^{-1}
\prod_{\substack{\el'=2\\ \el'\not=\el}}^b
\frac{f_{\prt{i}}(v_{\el'},v_\el)}{f_{\prt{i}}(v_{\el'},v_1)}\ F_{j,i}(v_\el).
\end{equation}
Next, we observe that due to the first relation in \r{iF1} for the composed currents
we have
\begin{equation*}
\Pfm\sk{F_{j,i}(v)}+F^{(-)}_{j,i}(v)\sieq{i}{j}\Sc_{F_{j-1}\mode0}\sk{\Pfm\sk{F_{j-1,i}(v)}+F^{(-)}_{j-1,i}(v)}.
\end{equation*}
Iterating this relation we obtain that
\begin{equation*}
\Pfm\sk{F_{j,i}(v)}+F^{(-)}_{j,i}(v)\sieq{i}{j}\Sc_{F_{j-1}\mode0}\cdots \Sc_{F_{i+1}\mode0}
\sk{\Pfm\sk{F_{i+1,i}(v)}+F^{(-)}_{i+1,i}(v)},
\end{equation*}
and since $\Pfm\sk{F_{i+1,i}(v)}+F^{(-)}_{i+1,i}(v)=0$ we arrive at
\begin{equation*}
\Pfm\sk{F_{j,i}(v)}\sieq{i}{j}-F^{(-)}_{j,i}(v).
\end{equation*}
This means that
\begin{equation}\label{gca17}
\begin{split}
&\Pfm\sk{F_{j_1,i}(v_1)\cdot F_{j_2,i}(v_2)\cdots F_{j_{b-1},i}(v_{b-1})\cdot F_{j_{b},i}(v_{b})}\\
&\quad \sieq{i}{j}(-)^bF^{(-)}_{j_1,i}(v_1;v_2,\ldots,v_b)\cdot F^{(-)}_{j_2,i}(v_2;v_3,\ldots,v_b)\cdots
F^{(-)}_{j_b,i}(v_b).
\end{split}
\end{equation}
 Hence, calculating the projection \r{gca2} one can move the terms of the form $\Pfm(U^\coun_{i+1,j-1})$
to the left through the product of the currents $F_1(u),\cdots F_{i-1}(u)$,
where they disappear under the 'positive' projection $\Pfp$.
This fact allows us to replace in the left hand side of \r{gca11} the product of the currents
and 'negative' projection by the product
\begin{equation*}
(-)^b\ F_{i,i-1}(u_1)\cdots F_{i,i-1}(u_a)\cdot
F^{(-)}_{j_1,i}(v_1;v_2,\ldots,v_b)\cdot F^{(-)}_{j_2,i}(v_2;v_3,\ldots,v_b)\cdots
F^{(-)}_{j_b,i}(v_b).
\end{equation*}
The commutation relations between the product of currents $F_{i,i-1}(u)$ and the 'negative' half-currents
$F^{(-)}_{j,i}(v)$ can be calculated via
\begin{equation}\label{gca19}
\begin{split}
F_{i,i-1}(u)F^{(-)}_{j,i}(v)=f_{\prt{i}}(v,u)\Big(F^{(-)}_{j,i}(v)&-h_{\prt{i}}(v,u)^{-1}F^{(-)}_{j,i}(u)\Big)F_{i,i-1}(u)\\
&+
\cci{i}^{-1}g_{\prt{i}}(u,v)F_{j,i-1}(u).
\end{split}
\end{equation}
The latter equation is a consequence of the commutation relations between  the simple root currents and the composed
currents
\begin{equation*}
F_{i,i-1}(u)F_{j,i}(v)=f_{\prt{i}}(v,u)\ F_{j,i}(v)F_{i,i-1}(u)-\delta(u,v)F_{j,i-1}(u)
\end{equation*}
and the definition of the 'negative' half-current
\begin{equation*}
F^{(-)}_{j,i}(v)=-\sum_{p<0}F_{j,i}^{(p)}\ u^{-p-1}\,.
\end{equation*}

Using commutation relation \r{gca19} we obtain that
\begin{equation}\label{gca21}
\begin{split}
&F_{i,i-1}(u_1)\cdots F_{i,i-1}(u_a)\cdot F^{(-)}_{j,i}(v) \\
&\quad= f_{\prt{i}}(v,\bu)\tilde F^{(-)}_{j,i}(v;u_1,\ldots,u_a)\cdot
F_{i,i-1}(u_1)\cdots F_{i,i-1}(u_a)\\
&\quad+\sum_{q=1}^a \cci{i}^{-1}g_{\prt{i}}(u_q,v)\prod_{q'=q+1}^a
\frac{(u_q-u_{q'})\epsilon_{i,m+1}+\cci{i}}{(u_q-u_{q'})\epsilon_{i,m+1}-\cci{i}}\\
&\qquad \times F_{i,i-1}(u_{1})\cdots F_{i,i-1}(u_{q-1})\cdot F_{i,i-1}(u_{q+1})\cdots F_{i,i-1}(u_{a})
\cdot F_{j,i-1}(u_{q}),
\end{split}
\end{equation}
where
\begin{equation}\label{gca22}
\tilde F^{(-)}_{j,i}(v;u_1,\ldots,u_a)=F^{(-)}_{j,i}(v)-\sum_{\el=1}^a
h_{\prt{i}}(v,u_\el)^{-1}\prod_{\substack{q=1\\ q\not=\el}}^a
\frac{f_{\prt{i}}(u_\el,u_q)}{f_{\prt{i}}(v,u_q)}\ F^{(-)}_{j,i}(u_\el).
\end{equation}
The linear combination of the 'negative' half-currents \r{gca22} in the first term of the
right hand side of  \r{gca21}
commutes  with all products of the currents $F_{i-2}(u),\ldots, F_1(u)$.
Therefore, this term eventually disappears under the 'positive' projection in \r{gca2}.
To transform the sum over $q$ in the right hand side of \r{gca21} we move
the composed current to the right using the commutation relation for $i\not=m+1$
\begin{equation} \label{gca25}
F_{j,i-1}(u_2)F_{i,i-1}(u_1)=f_{\prt{i}}(u_1,u_2)^{-1}F_{i,i-1}(u_1)F_{j,i-1}(u_2)\,,
\end{equation}
and for $i=m+1$
\begin{equation*}
F_{j,m}(u_2)F_{m+1,m}(u_1)=-f_{\prt{m+1}}(u_2,u_1)^{-1}F_{m+1,m}(u_1)F_{j,m}(u_2),
\end{equation*}
or, what is the same,
 \begin{equation} \label{gca26}
F_{j,m}(u_2)F_{m+1,m}(u_1)=-f(u_1,u_2)^{-1}F_{m+1,m}(u_1)F_{j,m}(u_2).
\end{equation}
Here we have used the fact that $\prt{m+1}=1$ and $f_1(u_2,u_1)=f(u_1,u_2)$. The both  cases
of $i\not=m+1$ and $i=m+1$ can be combined in one formula, and due
to the definition of the deformed symmetrization \r{sym} the sum in \r{gca21} can be written as
follows
\begin{equation}\label{gca23}
\begin{split}
&F_{i,i-1}(u_1)\cdots F_{i,i-1}(u_a)\cdot F^{(-)}_{j,i}(v)
 \\
&\quad \sieq{i}{j}\frac{\cci{i}^{-1}}{(a-1)!}\ \tSym_{\,\bu}\Big(g_{\prt{i}}(u_a,v)
 F_{i,i-1}(u_{1})\cdots F_{i,i-1}(u_{a-1})
\cdot F_{j,i-1}(u_{a})\Big),
\end{split}
\end{equation}
or
\begin{equation}\label{gca24}
\begin{split}
&F_{i,i-1}(u_1)\cdots F_{i,i-1}(u_a)\cdot F^{(-)}_{j,i}(v) \\
&\quad \sieq{i}{j}\frac{\cci{i}^{-1}}{(a-1)!}\ \tSym_{\,\bu}\Big(g_{\prt{i}}(u_1,v)f_{\prt{i}}(u_1,\bu_1)
F_{j,i-1}(u_{1})\cdot F_{i,i-1}(u_{2})\cdots F_{i,i-1}(u_{a})
\Big).
\end{split}
\end{equation}
Here,  one has to use the commutation relations \r{gca25} and
\r{gca26} in order to obtain \r{gca24} from \r{gca23}.

Using now the definition of the linear combinations of the half-currents \r{gca13}
and summation formula
\begin{equation*}
g_{\prt{i}}(u,v_1)f_{\prt{i}}(\bv_1,u)=g_{\prt{i}}(u,v_1)f_{\prt{i}}(\bv_1,v_1)
+\sum_{\el=2}^b g_{\prt{i}}(u,v_\el)g_{\prt{i}}(v_1,v_\el)
\prod_{\substack{\el'=2\\ \el'\not=\el}}^bf_{\prt{i}}(v_{\el'},v_\el),
\end{equation*}
the relation \r{gca24} can be rewritten in the form
\begin{equation*}
\begin{split}
&F_{i,i-1}(u_1)\cdots F_{i,i-1}(u_a)\cdot F^{(-)}_{j,i}(v_1,v_2,\ldots,v_b) \sieq{i}{j}\frac{\cci{i}^{-1}}{(a-1)!}
\\
&\quad \times \tSym_{\,\bu}\Big(g_{\prt{i}}(u_1,v_1)f_{\prt{i}}(u_1,\bu_1)\frac{f_{\prt{i}}(\bv_1,u_1)}
{f_{\prt{i}}(\bv_1,v_1)}
F_{j,i-1}(u_{1})\cdot F_{i,i-1}(u_{2})\cdots F_{i,i-1}(u_{a})
\Big).
\end{split}
\end{equation*}
Now we can use this relation for calculating the commutation of the product
of  currents $F_{i,i-1}(u_1)\cdots F_{i,i-1}(u_a)$ with the 'negative' projection
\r{gca17} modulo terms which  vanish under 'positive' projection in \r{gca1}.
The result gives us the proof of the relation \r{gca11}. Note that the deformed
symmetrization $\tSym_{\,\bu}$ over the set $\bu$ becomes a usual antisymmetrization over this
set for $i=m+1$. \hfill\qed

Let us stress the meaning of the relation \r{gca11}. Moving the 'negative' projection
of the string \r{gca3} through the product of currents $F_{i,i-1}(u_1)\cdots F_{i,i-1}(u_a)$
we are obtaining  linear combinations of  analogous strings
\begin{equation}\label{gca29}
F_{j_1,i-1}(u_{1})\cdot F_{j_2,i-1}(u_{2})\cdots F_{j_a,i-1}(u_{a})
\end{equation}
modulo elements which are irrelevant for the calculation the 'positive' projection in the
definition of the Bethe vector \r{uBV1}
with restrictions
\begin{equation}\label{dc00}
j_1\geq j_2\geq \cdots \geq j_{a-1}\geq j_a\geq i\,,
\end{equation}
such that the first $b$ indices $j_\el$, $\el=1,\ldots,b$ in the string \r{gca29} coincide with
the corresponding indices of the string \r{gca3} and the rest indices $j_{b+1}=\ldots=j_a=i$.

This linear combination is given by the deformed symmetrization over the set $\bu$
which can be reduced to the sum over partitions of this set. Let us describe these partitions.

Let $p_1$  be the number of equal indices $j_\el$ starting from $j_1$. Then, let $p_2$
be the number of equal indices $j_\el$ starting from $j_{p_1+1}$ and so on. Assume that
the whole set of  indices $j_\el$ is divided into  $s$ subsets of  identical indices
with cardinalities $p_l$, $l=1,\ldots,s$ and all $p_l>0$.
The number $s$ counts the number of groups of
the same type of composed currents in the string \r{gca3}. It is clear that this number is in the
interval $1\leq s\leq b$ including  the cases when all currents are the same ($s=1$) or
when all currents are different ($s=b$).
The restriction for the indices in the product of composed currents \r{gca3}
induces a natural decomposition of the set $\bv$
\begin{equation}\label{dc0}
\bv=\{v_1,v_2,\ldots,v_{b-1},v_b\}\Ra\{\bv^1,\ldots,\bv^s\}
\end{equation}
 into $s$ non-intersecting subsets
with  cardinalities $\#\bv^q=p_q$, $q=1,\ldots,s$.
Here we were forced to use a superscript to count these
subsets and this superscript should not be confused with the index which characterizes
the type of  Bethe parameters.

Assume that $a\geq b$.
Let us decompose the set $\bu$ into $s+1$ non-intersecting subsets
\begin{equation}\label{dc1}
\bu=\{u_1,u_2,\ldots,u_{a-1},u_a\}\Ra\{\bu^1,\ldots,\bu^s,\bu^{s+1}\}
\end{equation}
such that
\begin{equation*}
\#\bu^q=p_q>0\quad\mbox{and}\quad \#\bu^{s+1}=a-b.
\end{equation*}
The last subset $\bu^{s+1}$ can be empty for the terms in \r{gca11} with $a=b$.
According to the definition of the sizes of the subsets $\bu^q$, $q=1,\ldots,s$ we have
\begin{equation*}
j_1=\ldots=j_{p_1}>j_{p_1+1}=\ldots=j_{p_2}> \cdots >
j_{p_{s-1}+1}=\ldots=j_{p_s}>i\, .
\end{equation*}
Let
\begin{equation*}
j_{p_{\el-1}+1}=\ldots=j_{p_\el}=j'_\el
\end{equation*}
for $\el=1,\ldots,s$.
Using the definition of the ordered  product of the same type of composed or simple currents given by the
formula \r{gca6}
and dividing   the initial set of variables $\bv$ \r{dc0} into subsets
$\bv^q$, $q=1,\ldots,s$  we can transform the string \r{gca3}  to
\begin{equation}\label{gca7}
\begin{split}
&F_{j_1,i}(v_1)\cdot F_{j_2,i}(v_2)\cdots F_{j_{b-1},i}(v_{b-1})\cdot F_{j_{b},i}(v_{b})\to\\
&\quad \to \gF_{j'_1,i}(\bv^1)\cdot \gF_{j'_2,i}(\bv^2)\cdots \gF_{j'_{s-1},i}(\bv^{s-1})\cdot  \gF_{j'_{s},i}(\bv^{s}).
\end{split}
\end{equation}
Denote the ordered product of currents in the right hand side of \r{gca7} as
\begin{equation*}
\gF_{\bar\jmath',i}(\bv)=
\gF_{j'_1,i}(\bv^1)\cdot \gF_{j'_2,i}(\bv^2)\cdots \gF_{j'_{s-1},i}(\bv^{s-1})\cdot  \gF_{j'_{s},i}(\bv^{s}),
\end{equation*}
where the set $\bar\jmath'=\{j'_1,\cdots,j'_s\}$ such that $j'_1>j'_2>\cdots> j'_s>i$.

Similarly, dividing the set $\bu$ into subsets \r{dc1} the string \r{gca29} transforms into
\begin{equation}\label{gca88}
\gF_{\bar\jmath,i-1}(\bu)=
\gF_{j'_1,i-1}(\bu^1)\cdot \gF_{j'_2,i-1}(\bu^2)\cdots \gF_{j'_{s},i-1}(\bu^{s})\cdot  \gF_{i,i-1}(\bu^{s+1}),
\end{equation}
where $\bar\jmath=\{j'_1,\cdots,j'_s,i\}$.

In order to rewrite the summation over permutations over the elements of the set
$\bu$ in the right hand side of \r{gca11} we multiply both sides of this equation
by the rational function $\Delta_{f_{\prt{i}}}(\bu)\Delta_{h_{\prt{i}}}(\bu)^{-\delta_{i,m+1}}$.
Then, using the fact that the deformed symmetrization (or antisymmetrization in the case
$i=m+1$) can be transformed into the
usual symmetrization over the set $\bu$
\begin{equation*}
\frac{\Delta_{f_{\prt{i}}}(\bu)}{\Delta_{h_{\prt{i}}}(\bu)^{\delta_{i,m+1}}} \ \tSym_{\,\bu}(G(\bu))=
 \Sym_{\,\bu}
 \sk{\frac{\Delta_{f_{\prt{i}}}(\bu)}{\Delta_{h_{\prt{i}}}(\bu)^{\delta_{i,m+1}}}\cdot G(\bu)}
\end{equation*}
for any formal series  $G(\bu)$,   we can replace it into the sum over partitions
\r{dc1} and symmetrizations inside the partial subsets
\begin{equation*}
 \Sym_{\,\bu}\ (\cdot)=\sum_{\bu\Ra\{\bu^1,\ldots,\bu^s,\bu^{s+1}\}}\
 \Sym_{\,\bu^1} \Sym_{\,\bu^2}\cdots  \Sym_{\,\bu^{s}}  \Sym_{\,\bu^{s+1}} (\cdot)\;.
\end{equation*}

Below we use the fact that after multiplication of both sides of \r{gca11} by the rational function
$\Delta_{f_{\prt{i}}}(\bu)\Delta_{h_{\prt{i}}}(\bu)^{-\delta_{i,m+1}}$ we can perform summations
over symmetrization in each of the non-intersecting subsets
$\bu^q$, $q=1,\ldots,s+1$ in the right hand side of \r{gca11}.

For any composed current $F_{j,i}(u)$, $j>i$ we introduce its parity $\cpr_{i,j}$ defined as
\begin{equation*}
\cpr_{i,j}=\prt{i}+\prt{j}=\begin{cases}
1,\quad i\leq m\leq j-1,\\
0,\quad i>m\quad\mbox{or}\quad m>j-1.
\end{cases}
\end{equation*}
We refer to the composed currents with the parity 1 as {\it odd} and with the parity 0 as {\it even}.
Using the commutation relations between simple root currents one can check that the commutation
relations between even composed currents are the same as for even simple root currents, while
the odd composed currents anticommute
\begin{equation}\label{dc6}
\begin{split}
(u-v-\cci{i})F_{j,i}(u)F_{j,i}(v)&=(u-v+\cci{i})F_{j,i}(v)F_{j,i}(u),\quad\mbox{for}\quad \cpr_{i,j}=0,\\
F_{j,i}(u)F_{j,i}(v)&=-F_{j,i}(v)F_{j,i}(u)\quad\mbox{for}\quad \cpr_{i,j}=1.
\end{split}
\end{equation}

If $m+1< i\leq N$, then it is clear from the restrictions \r{gca4} and \r{dc00}
that only even currents (both simple and composed) enter both sides of the
equation \r{gca11}. Otherwise, for  $i=m+1$ all the currents (again both simple and composed)
in the right hand side of \r{gca11} are odd. When $1< i\leq m$, then
there are simultaneously odd and even currents in the right hand side of \r{gca11}, and according to the structure of the initial
string \r{gca3} all odd currents are placed  to the left of all even currents.
In this case there are $s'$ ($1\leq s'< s$) factors in the string which are
products of the same odd currents. Due to the
commutation relations of the composed currents \r{dc6} the symmetrizations over sets $\bu^q$,
$q=1,\ldots,s'$ and $q=s'+1,\ldots,s+1$ will be performed differently.
For $m+1\leq i\leq N$  all symmetrizations
over sets $\bu^q$ for all $q=1,\ldots,s+1$ are the same.
The number $s'$ can be calculated as follows
\begin{equation}\label{dc88}
s'=\sum_{\el=1}^s \cpr_{i,j'_\el}\,.
\end{equation}

Let us first consider the case $m+1\leq i\leq N$.
Multiplying both sides of \r{gca11} by the function $\Fli{i-1}(\bu)$ we obtain
\begin{equation}\label{dc8}
\begin{split}
&\Fli{i-1}(\bu)\gF_{i,i-1}(\bu)\cdot \Pfm\sk{\gF_{\bar\jmath',i}(\bv)}\\
&\quad \sieq{i}{j_1} \frac{\cci{i}^{-b}}{\Tpf{f}{i}{\bv}}\
\sum_{\bu\Ra\{\bu^1,\ldots,\bu^s,\bu^{s+1}\}}\
\prod_{q<q'}^{s+1} f_{\prt{i}}(\bu^{q},\bu^{q'})
\prod_{q<q'}^{s} f_{\prt{i}}(\bv^{q'},\bu^{q})\Fli{i-1}(\bu)
 \gF_{\bar\jmath,i-1}(\bu)\\
&\qquad
\times\prod_{q=1}^s \Sym_{\,\bu^q}\left[ \Tpfp{f}{i}{\bu^q}
\prod_{\el} g_{\prt{i}}(v_{\el},u_{\el})
\prod_{\el<\el'}f_{\prt{i}}(v_{\el'},u_{\el})
\right]_{\substack{v_{\el},v_{\el'}\in \bv^q\\ u_{\el},u_{\el'}\in \bu^q}}\;,
\end{split}
\end{equation}
where we used the fact that the product of function $\Fli{i-1}(\bu)$ and
the string \r{gca88} is symmetric with respect to  permutations within every
subset $\bu^q$. In particular, this symmetry allows one to remove the symmetrization
over the set $\bu^{s+1}$ and to cancel the combinatorial factor $(a-b)!^{-1}$ in \r{gca11}.
Note that if $i=m+1$, then all the currents in the product
$\gF_{\bar\jmath,m}(\bu)$ become odd and the symmetry
 over permutations of the variables  in each set $\bu^q$ is provided by the
 function $\Fli{m}(\bu)=\Tpf{g}{m}{\bu}$.

The remaining symmetrization over each subsets $\bu^q$, $q=1,\ldots,s$ is
a famous Izergin determinant \cite{Ize-det} defined for two sets $\bar y$ and $\bar x$ with the same
cardinality $\#\bar y=\#\bar x=p$ as follows
\begin{equation}\label{Izer}
\begin{split}
K_{\prt{i}}(\bar y|\bar x)&=\Sym_{\,\bar x}\left[\Tpfp{f}{i}{\bar x}
\prod_{\el=1}^{p} g_{\prt{i}}(y_{\el},x_{\el})
\prod_{\el<\el'}^{p} f_{\prt{i}}(y_{\el'},x_{\el})
\right]\\
&=\Tpf{g}{i}{\bar y} \Tpfp{g}{i}{\bar x} h_{\prt{i}}(\bar y,\bar x)
\det\left[\frac{g_{\prt{i}}(y_{\el},x_{\el'})}{h_{\prt{i}}(y_{\el},x_{\el'})}\right]_{\el,\el'=1,\ldots,p}\;.
\end{split}
\end{equation}
Thus, we conclude that if the index $i$ belongs to the interval  $m+1\leq i\leq N$,
then equation \r{gca11} can be rewritten as a sum over partitions of the set $\bu$
which is defined by the string $\gF_{\bar\jmath,i}(\bv)$
\begin{equation}\label{dc10}
\begin{split}
&\Fli{i-1}(\bu)\gF_{i,i-1}(\bu)\cdot \Pfm\sk{\gF_{\bar\jmath',i}(\bv)} \\
&\quad\sieq{i}{j_1} \frac{\cci{i}^{-b}}{\Tpf{f}{i}{\bv}}\
\sum_{\bu\Ra\{\bu^1,\ldots,\bu^s,\bu^{s+1}\}}\ \prod_{q<q'}^{s+1} f_{\prt{i}}(\bu^{q},\bu^{q'})
\prod_{q<q'}^{s} f_{\prt{i}}(\bv^{q'},\bu^{q})\\
 &\qquad \times  \prod_{q=1}^s
K_{\prt{i}}(\bv^q|\bu^q)\ \  \Fli{i-1}(\bu)
 \gF_{\bar\jmath,i-1}(\bu).
\end{split}
\end{equation}

Let us consider now the case when $1<i\leq m$. As it was mentioned above, in this case
the product of currents $\gF_{\bar\jmath,i-1}(\bu)$  contains both odd and even composed
currents. Therefore, in order to perform symmetrization over subsets $\bu^q$ we have to use different
approaches for odd and even currents.

Let $s'$, $1\leq s'\leq s$ be the number  of
 products of the same odd currents in the right hand side of \r{gca11} defined by  \r{dc88}.
Then the symmetrization over the sets $\bu^q$ for $s'<q\leq s+1$ in \r{dc8} is
the same as described above. It  leads to the appearance of  Izergin
determinants depending on the corresponding sets of variables.
Since variables from the subsets $\bu^q$ for $1 \leq q\leq s'$ become the
arguments of the odd anticommuting currents, the relation \r{gca11} after
multiplication by the function \r{dc9}  takes the form
\begin{equation}\label{dc11}
\begin{split}
&\Fli{i-1}(\bu)\gF_{i,i-1}(\bu)\cdot \Pfm\sk{\gF_{\bar\jmath',i}(\bv)} \\
&\quad\sieq{i}{j_1} \frac{\cci{i}^{-b}}{\Tpf{f}{i}{\bv}}\
\sum_{\bu\Ra\{\bu^1,\ldots,\bu^s,\bu^{s+1}\}}\
\prod_{q<q'}^{s+1} f_{\prt{i}}(\bu^{q},\bu^{q'})
\prod_{q<q'}^{s} f_{\prt{i}}(\bv^{q'},\bu^{q})\\
&\quad
\times \prod_{q=s'+1}^s
K_{\prt{i}}(\bv^q|\bu^q)
 \prod_{q=1}^{s'} \Sym_{\,\bu^q}\left[ \Tpfp{g}{i}{\bu^q}
\prod_{\el} g_{\prt{i}}(v_{\el},u_{\el})
\prod_{\el<\el'}f_{\prt{i}}(v_{\el'},u_{\el})
\right]_{\substack{v_{\el},v_{\el'}\in \bv^q\\ u_{\el},u_{\el'}\in \bu^q}}
\\
 &\quad \times \Fli{i-1}(\bu)  \prod_{q=1}^{s'}\Tpfp{h}{i}{\bu^q}\
 \gF_{\bar\jmath,i-1}(\bu),
\end{split}
\end{equation}
where we have used factorization $\Tpfp{f}{i}{\bu^q}=\Tpfp{g}{i}{\bu^q}\Tpfp{h}{i}{\bu^q}$.

The fact that products of odd currents in the right hand side of \r{dc11}
can be extracted from the symmetrization over the sets $\bu^q$, $q=1,\ldots,s'$
follows from the observation that the function $\Fli{i-1}(\bu)=\Tpf{f}{i-1}{\bu}$
for $1<i\leq m$ contains the factors $\Tpf{h}{i-1}{\bu^q}$ and $\Tpf{g}{i-1}{\bu^q}$.
The first factor together with the function $\Tpfp{h}{i}{\bu^q}$ produces a symmetric
function of the variables of the set $\bu^q$
\begin{equation*}
\Tpf{h}{i-1}{\bu^q}\Tpfp{h}{i}{\bu^q}=\Tpf{h}{i}{\bu^q}\Tpfp{h}{i}{\bu^q}=
h_{\prt{i}}(\bu^q,\bu^q)\quad\mbox{for}\quad 1<i\leq m\,,
\end{equation*}
while the second factor $\Tpf{g}{i-1}{\bu^q}$ makes symmetric the product of the odd
currents depending on the set of variables $\bu^q$.

Let us denote the normalized symmetrization in the third line of \r{dc11} as
$\Cau_{\prt{i}}(\bv|\bu)$
\begin{equation*}
\Cau_{\prt{i}}(\bv|\bu)=\Tpfp{h}{i}{\bu}\
\Sym_{\,\bu}\left[ \Tpfp{g}{i}{\bu^q}
\prod_{\el} g_{\prt{i}}(v_{\el},u_{\el})
\prod_{\el<\el'}f_{\prt{i}}(v_{\el'},u_{\el})
\right]_{\substack{v_{\el},v_{\el'}\in \bv\\ u_{\el},u_{\el'}\in \bu}}\;.
\end{equation*}
This function is proportional to the Cauchy determinant according to the chain of
equalities
\begin{equation*}
\begin{split}
\Cau_{\prt{i}}(\bv|\bu)&=\Tpfp{h}{i}{\bu} \Tpfp{g}{i}{\bu}\
{\rm ASym}_{\,\bu}\left[
\prod_{\el} g_{\prt{i}}(v_{\el},u_{\el})
\prod_{\el<\el'}f_{\prt{i}}(v_{\el'},u_{\el})
\right]_{v_{\el},v_{\el'}\in \bv;\  u_{\el}\in \bu}\\
&=\Tpfp{f}{i}{\bu}\Tpf{f}{i}{\bv}\  {\rm ASym}_{\,\bu}\left[
\prod_{\el} g_{\prt{i}}(v_{\el},u_{\el})
\right]_{v_{\el}\in \bv;\ u_{\el}\in \bu}\\
&=\frac{\Tpfp{f}{i}{\bu}\Tpf{f}{i}{\bv}}{\Tpfp{g}{i}{\bu}\Tpf{g}{i}{\bv}}\
g_{\prt{i}}(\bv,\bu)=\Tpfp{h}{i}{\bu}\Tpf{h}{i}{\bv}g_{\prt{i}}(\bv,\bu),
\end{split}
\end{equation*}
where the symbol ${\rm ASym}_{\,\bu}$ means antisymmetrization with respect
to the set $\bu$.

Thus, the relation \r{gca11} for $1<i\leq m$  can be presented as the following
 sum over partitions
 \begin{equation}\label{dc14}
\begin{split}
&\Fli{i-1}(\bu) \gF_{i,i-1}(\bu)\cdot \Pfm\sk{\gF_{\bar\jmath',i}(\bv)} \\
&\quad\sieq{i}{j_1} \frac{\cci{i}^{-b}}{\Tpf{f}{i}{\bv}}\
\sum_{\bu\Ra\{\bu^1,\ldots,\bu^s,\bu^{s+1}\}}\ \prod_{q<q'}^{s+1} f_{\prt{i}}(\bu^{q},\bu^{q'})
\prod_{q<q'}^{s} f_{\prt{i}}(\bv^{q'},\bu^{q})\\
&\qquad
\times\prod_{q=1}^{s'}
\Cau_{\prt{i}}(\bv^q|\bu^q)
\prod_{q=s'+1}^s
K_{\prt{i}}(\bv^q|\bu^q)\ \ \Fli{i-1}(\bu) \gF_{\bar\jmath,i-1}(\bu),
\end{split}
\end{equation}
where $s'$ is given by \r{dc88}.

Now we  apply \r{dc10} and \r{dc14} for the calculation of the projection
\r{gca2} to obtain the recursion for the Bethe vectors given by \r{uBV1}.

We should add the rule for the ordering of the sets $\bu^q$ in the formulas \r{dc10} and \r{dc14}.
As it was shown in the definition of the string \r{gca88} the sets marked by the smaller indices enter
more complicated composed current placed on the left of the string \r{gca88}.

\end{document}